\documentclass[10pt]{article}%
\usepackage[a4paper,top=2.8cm,bottom=2.8cm,left=3cm,right=3cm]{geometry} % 
\usepackage{sgame}
\usepackage{amsmath}%
\usepackage{amsthm}%
\usepackage{amsfonts}%
\usepackage{amssymb}%
\usepackage{mathrsfs}%				% to get more rounded calligraphic with \mathscr{}
\usepackage{braket}					% to get set notation, e.g. $ A = \Set { x | x \in X } $
\usepackage{mathtools}
\usepackage{enumitem}
\usepackage{xcolor}%
\usepackage{graphicx}%
\usepackage{bm}%					% to get boldface greek symbol \bm{}
\usepackage{tikz}%
\usetikzlibrary{calc}%
\usetikzlibrary{cd}%
\usepackage{float}%
\usepackage{etoolbox}
\usepackage{changepage}
\usepackage{natbib}          
\usepackage[english]{babel}
\usepackage[utf8]{inputenc}
\usepackage[T1]{fontenc}
\usepackage{lmodern} % To switch to Latin Modern
\rmfamily % To load Latin Modern Roman and enable the following NFSS declarations.
\DeclareFontShape{T1}{lmr}{b}{sc}{<->ssub*cmr/bx/sc}{}
\DeclareFontShape{T1}{lmr}{bx}{sc}{<->ssub*cmr/bx/sc}{}
\usepackage{epigraph}
\makeatletter
\patchcmd{\epigraph}{\@epitext{#1}}{\itshape\@epitext{#1}}{}{}
\makeatother
\usepackage{titlesec}
\titleformat{\section}
  {\normalfont\sc\centering}
  {\thesection.}{1em}{}
\titleformat{\subsection}
  {\normalfont\centering}
  {\normalfont\sc\centering\thesubsection}{1em}{}{}
\titleformat{\subsubsection}
  {\normalfont\slshape\centering}
  {\normalfont\sc\centering\thesubsubsection}{1em}{}{}
\titleformat{\paragraph}
{\normalfont\sc\slshape\normalsize\centering}{\theparagraph}{1em}{}
\titlespacing*{\paragraph}
{0pt}{3.25ex plus 1ex minus .2ex}{1.5ex plus .2ex}
\usepackage{xparse}
\AtBeginDocument%{\let\mathbb\varmathbb}

\DeclareFontFamily{U}{ntxmia}{}
\DeclareFontShape{U}{ntxmia}{m}{it}{<-> ntxmia }{}
\DeclareFontShape{U}{ntxmia}{b}{it}{<-> ntxbmia }{}
\DeclareSymbolFont{lettersA}{U}{ntxmia}{m}{it}
\SetSymbolFont{lettersA}{bold}{U}{ntxmia}{b}{it}

\ExplSyntaxOn
\NewDocumentCommand{\varmathbb}{m}
 {
  \tl_map_inline:nn { #1 }
   {
    \use:c { varbb##1 }
   }
 }
\tl_map_inline:nn { ABCDEFGHIJKLMNOPQRSTUVWXYZ }
 {
  \exp_args:Nc \DeclareMathSymbol{varbb#1}{\mathord}{lettersA}{\int_eval:n { `#1+64 }}
 }
\exp_args:Nc \DeclareMathSymbol{varbbk}{\mathord}{lettersA}{169}
\ExplSyntaxOff
\makeatletter
\g@addto@macro\@floatboxreset\centering
\makeatother
\makeatletter
\def\bbordermatrix#1{\begingroup \m@th
  \@tempdima 4.75\p@
  \setbox\z@\vbox{%
    \def\cr{\crcr\noalign{\kern2\p@\global\let\cr\endline}}%
    \ialign{$##$\hfil\kern2\p@\kern\@tempdima&\thinspace\hfil$##$\hfil
      &&\quad\hfil$##$\hfil\crcr
      \omit\strut\hfil\crcr\noalign{\kern-\baselineskip}%
      #1\crcr\omit\strut\cr}}%
  \setbox\tw@\vbox{\unvcopy\z@\global\setbox\@ne\lastbox}%
  \setbox\tw@\hbox{\unhbox\@ne\unskip\global\setbox\@ne\lastbox}%
  \setbox\tw@\hbox{$\kern\wd\@ne\kern-\@tempdima\left[\kern-\wd\@ne
    \global\setbox\@ne\vbox{\box\@ne\kern2\p@}%
    \vcenter{\kern-\ht\@ne\unvbox\z@\kern-\baselineskip}\,\right]$}%
  \null\;\vbox{\kern\ht\@ne\box\tw@}\endgroup}
\makeatother
\makeatletter
\newcommand{\mathleft}{\@fleqntrue}
\newcommand{\mathcenter}{\@fleqnfalse}
\makeatother
\DeclareMathOperator{\supp}{supp}
\DeclareMathOperator{\proj}{proj}

\makeatletter
\def\th@plain{%
  \thm@notefont{}% same as heading font
  \itshape % body font
}
\def\th@definition{%
  \thm@notefont{}% same as heading font
  \normalfont % body font
}
\makeatother
\numberwithin{equation}{section}
\usepackage{stmaryrd}

\newcommand{\mcal}{\mathcal}
\newcommand{\mscr}{\mathscr}
\newcommand{\mbf}{\mathbf}
\newcommand{\mbb}{\mathbb}
\newcommand{\bN}{\mathbb{N}}

\newcommand{\varep}{\varepsilon}
\newcommand{\Llra}{\Longleftrightarrow}

\newcommand{\Lra}{\Longrightarrow}

\newcommand{\la}{\langle}
\newcommand{\ra}{\rangle}
\newcommand{\abs}[1]{\left\lvert#1\right\rvert}				%absolute value
\newcommand{\cto}{\rightrightarrows}					%Correspondence mapping
\newcommand{\sAlg}{\Sigma}
\renewcommand{\Game}{\Gamma}
\newcommand{\Ann}{\mathrm{a}}
\newcommand{\Bob}{\mathrm{b}}

\newcommand{\Utilities}{\mcal{U}}
\newcommand{\Acts}{\mcal{A}}

\newcommand{\States}{\Theta}
\newcommand{\state}{\theta}
\newcommand{\NonNull}{\Omega}
\newcommand{\Outcomes}{\mathcal{Z}}
\newcommand{\bd}{\mathsf{bd}}
\newcommand{\cbd}{\mathsf{cbd}}
\newcommand{\md}{\mathsf{md}}

\newcommand{\Burnt}{\mathbb{B}}
\newcommand{\Und}{\mbb{U}}
\newcommand{\OSR}{\mbb{S}}
\newcommand{\ICD}{\mbb{M}}
\newcommand{\SR}{\mbb{R}}
\newcommand{\Ad}{\mathsf{A}}
\newcommand{\IA}{\mbf{A}}

\newcommand{\prefeq}{\bm{\succsim}}
\newcommand{\pref}{\bm{\succ}}
\newcommand{\outprefeq}{\,\widehat{\bm{\succsim}}\,}
\newcommand{\outpref}{\,\widehat{\bm{\succ}}\,}
\newcommand{\outprefeqE}{\, {\widehat{\bm{\succsim}}}{\vphantom{\bm{\succsim}}}^{E} \,}

\newcommand{\CDTeq}{\prefeq^{h}}

\newcommand{\CMsim}{{\overset{\scriptscriptstyle CM}{\prefeq_{i}}}}
\newcommand{\CSEUsimR}{{\overset{\scriptscriptstyle CSEU}{{\prefeq}_i|_R}}}
\newcommand{\prefeqR}{\prefeq_i\!\!|_{R}}
\newcommand{\prefeqhR}{\prefeq^{h}_i\!\!|_{R}\,}
\newcommand{\SEs}{\succcurlyeq}
\newcommand{\Coalitions}{\mcal{I}}
\newcommand{\effrel}[1][]{\overset{#1}{\rightarrow}} 
\def\SetVertic{\egroup\;\hspace{-0.1cm}\middle|\hspace{-0.1cm}\;\bgroup}
{\catcode`\|=\active
  \xdef\Sets{\protect\expandafter\noexpand\csname Sets \endcsname}
  \expandafter\gdef\csname Sets \endcsname#1%
  	{\hspace{-0.05cm}\left\{\hspace{-0.08cm}%
     		\:{
     			\mathcode`\|32768\let|\SetVertic
     		#1}
     		\:\hspace{-0.08cm}\right\}}
}

{\catcode`\|=\active
  \xdef\Round{\protect\expandafter\noexpand\csname Round \endcsname}
  \expandafter\gdef\csname Round \endcsname#1{\left(%
     \:{
     \mathcode`\|32768\let|\SetVertic
     #1}\:\right)}
}

{\catcode`\|=\active
  \xdef\Rounds{\protect\expandafter\noexpand\csname Rounds \endcsname}
  \expandafter\gdef\csname Rounds \endcsname#1%
  	{\hspace{-0.05cm}\left(\hspace{-0.08cm}%
     		\:{
     			\mathcode`\|32768\let|\SetVertic
     		#1}
     		\:\hspace{-0.08cm}\right)}
}

{\catcode`\|=\active
  \xdef\Square{\protect\expandafter\noexpand\csname Square \endcsname}
  \expandafter\gdef\csname Square \endcsname#1{\left[%
     \:{
     \mathcode`\|32768\let|\SetVertic
     #1}\:\right]}
}

{\catcode`\|=\active
  \xdef\Squares{\protect\expandafter\noexpand\csname Squares \endcsname}
  \expandafter\gdef\csname Squares \endcsname#1%
  	{\hspace{-0.05cm}\left[\hspace{-0.08cm}%
     		\:{
     			\mathcode`\|32768\let|\SetVertic
     		#1}
     		\:\hspace{-0.08cm}\right]}
}

{\catcode`\|=\active
  \xdef\Angle{\protect\expandafter\noexpand\csname Angle \endcsname}
  \expandafter\gdef\csname Angle \endcsname#1{\left\langle%
     \:{
     \mathcode`\|32768\let|\SetVertic
     #1}\:\right\rangle}
}

{\catcode`\|=\active
  \xdef\Angles{\protect\expandafter\noexpand\csname Angles \endcsname}
  \expandafter\gdef\csname Angles \endcsname#1%
  	{\hspace{-0.05cm}\left\langle\hspace{-0.08cm}%
     		\:{
     			\mathcode`\|32768\let|\SetVertic
     		#1}
     		\:\hspace{-0.08cm}\right\rangle}
}
\usepackage{varioref}
\usepackage[pagebackref,hypertexnames=false]{hyperref}
\usepackage[noabbrev,nameinlink]{cleveref}
\Crefname{paragraph}{Section}{Sections}
\definecolor{egtgreen}{RGB}{75, 155, 8}
\definecolor{egtpurple}{RGB}{10, 33, 128}
\definecolor{egtred}{RGB}{103, 25, 15}
\definecolor{burgundy}{rgb}{0.5, 0.0, 0.13}
\definecolor{cyanp}{RGB}{45,129,173}
\hypersetup{colorlinks=true,linkcolor={egtred},citecolor=green!50!black,bookmarksnumbered=true}
\hypersetup{%
  pdfcreator={},
  pdfsubject={},%
  pdfproducer={},%
pdfinfo={%
  Title={Revealing Sequential Rationality and Forward Induction},%
  Author={Pf. Guarino},%
  ModDate={ },%
  CreationDate={ }%
}%
}
\newcommand{\Mref}[2][cyanp]{%
\hypersetup{linkcolor=#1}%
\Cref{#2}%
\hypersetup{linkcolor=burgundy}%
}
\newcommand{\Sref}[2][burgundy]{%
\hypersetup{linkcolor=#1}%
\Cref{#2}%
\hypersetup{linkcolor=burgundy}%
}
\usepackage{xpatch}\xpatchbibmacro{pageref}{parens}{brackets}{}{}
\makeatletter
\patchcmd{\BR@backref}{\newblock}{\newblock[}{}{}
\patchcmd{\BR@backref}{\par}{]\par}{}{}
\makeatother
\theoremstyle{plain}
\newtheorem{theorem}{Theorem}
\newtheorem{definition}{Definition}[section]
\newtheorem{proposition}{Proposition}
\newtheorem{lemma}{Lemma}[section]
\newtheorem{corollary}{Corollary}

\newtheorem{algorithm}{Algorithm}
\newtheorem{remark}{Remark}[section]
\theoremstyle{definition}
\newtheorem{notation}{Notation}

\theoremstyle{remark}

\newtheorem{case}{Case}

\makeatletter
\def\thm@space@setup{%
  \thm@preskip=0.5cm %plus 0.5cm %minus 2cm
  \thm@postskip=\thm@preskip % or whatever, if you don't want them to be equal
}
\makeatother
\newtheoremstyle{myclaimstyle} % name
    {\topsep}                    % Space above
    {\topsep}                    % Space below
    {\itshape}                   % Body font
    {}                           % Indent amount
    {\itshape}                   % Theorem head font
    {.}                          % Punctuation after theorem head
    {.5em}                       % Space after theorem head
    {}  % Theorem head spec (can be left empty, meaning ?normal?)

\theoremstyle{myclaimstyle}

\newtheorem{claim}{Claim}

\makeatletter
\preto\claim{%
  \patchcmd\cref@thmnoarg
    {\trivlist}
    {\list{}{\leftmargin\parindent\rightmargin\parindent}}
    {}{}%
  \patchcmd\cref@thmoptarg
    {\trivlist}
    {\list{}{\leftmargin\parindent\rightmargin\parindent}}
    {}{}%
  \patchcmd\thmt@original@endclaim{\endtrivlist}{\endlist}{}{}%
}
\makeatother

\newtheoremstyle{named}{}{}{\itshape}{}{\bfseries}{.}{.5em}{\thmnote{#3}}
\theoremstyle{named}
\newtheorem*{namedtheorem}{Theorem}

\makeatletter
\newtheoremstyle{axioms}{}{}{\itshape}{}{\bfseries}{.}{.5em}%
{#1 \thmnumber{#2} \@ifnotempty{#3}{ (\thmnote{#3})}}
\theoremstyle{axioms}

\newtheorem{GhirardatoAxiom}{Axiom}

\makeatletter
\newtheoremstyle{case}
  {5pt}% space before
  {5pt}% space after
  {\addtolength{\@totalleftmargin}{0em}
   \addtolength{\linewidth}{-1em}
   \parshape 1 1em \linewidth}% body font
  {}% indent
  {\normalfont}% header font
  {.}% punctuation
  {.5em}% after theorem header
  {}% header specification (empty for default)
\makeatother

\theoremstyle{case}

\AtEndEnvironment{proof}{\setcounter{step}{0}}
\AtEndEnvironment{proof}{\setcounter{claim}{0}}
\AtEndEnvironment{proof}{\setcounter{Cases}{0}}
\makeatletter
\def\thm@space@setup{%
  \thm@preskip=0.5cm %plus 0.5cm %minus 2cm
  \thm@postskip=\thm@preskip % or whatever, if you don't want them to be equal
}
\makeatother
\usepackage{thmtools}

\declaretheorem[style=definition, name=Example, qed=$\diamond$]{example}
\renewcommand\thmcontinues[1]{Continued}
\declaretheoremstyle[
  spaceabove=3pt, spacebelow=6pt,
  headfont=\normalfont\itshape,
  notefont=\mdseries,
  bodyfont=\normalfont,
  postheadspace=.5em
]{innerproof}

\renewcommand\thmcontinues[1]{}

\declaretheorem[style= innerproof, numbered=unless unique, name=Proof of Claim, qed=$\square$]{proofclaim}
\renewcommand\thmcontinues[1]{}
\makeatletter
\preto\proofclaim{%
  \patchcmd\cref@thmnoarg
    {\trivlist}
    {\list{}{\leftmargin\parindent\rightmargin\parindent}}
    {}{}%
  \patchcmd\cref@thmoptarg
    {\trivlist}
    {\list{}{\leftmargin\parindent\rightmargin\parindent}}
    {}{}%
  \patchcmd\thmt@original@endproofclaim{\endtrivlist}{\endlist}{}{}%
}
\makeatother

\usepackage{datetime}
\makeatletter
\newcommand\Romanmonth{\@Roman{\month}}
\makeatother
\newdateformat{mydate}{\THEDAY.\Romanmonth.\THEYEAR}
\title{\vspace{-1.3cm}Revealing Sequential Rationality and Forward Induction%\vspace{0.5cm} \Large{Note}% 
\thanks{I am grateful to Pierpaolo Battigalli, Emiliano Catonini, Francesco De Sinopoli, Shuige Liu, Andr\'{e}s Perea, Elias Tsakas, Gabriel Ziegler, and Peio Zuazo-Garin for most useful comments and conversations. Also, I would like to thank the audiences of the GAMES2020 conference, the ESEM2022 conference, the  Conference in Honor of Edi Karni, and the GRASS XVII workshop. Some results contained in this paper previously circulated---in a considerable different form---in papers under different titles. Of course, all errors are mine. Finally, I thankfully acknowledge financial support from the Austrian Science Fund (FWF) (P31248-G27) and from MIUR under the PRIN 2017 program (grant number 2017K8ANN4).}%
\\
}
\author{%
Pierfrancesco Guarino\thanks{%
University of Udine (Department of Economics and Statistics -- DIES). %
\textit{E-mails:} \texttt{pf.guarino@hotmail.com} \& %
\texttt{pierfrancesco.guarino@uniud.it}.}
}
\date{\mydate\today}
\begin{document}
\renewcommand\thmcontinues[1]{Continued}

\maketitle

%%%%%%%%%%%%%%%%%%%%%%%%%%%%%%%%%%%%%%%%%
%%%%%%%%%%%%%%%%%%%%%%%%%%%%%%%%%%%%%%%%%
%%%%%%%%%%%%%%%%%%%%%%%%%%%%%%%%%%%%%%%%%
%%%%%%%% ABSTRACT %%%%%%%%%%%%%%%%%%%%%%%%%
%%%%%%%%%%%%%%%%%%%%%%%%%%%%%%%%%%%%%%%%%
%%%%%%%%%%%%%%%%%%%%%%%%%%%%%%%%%%%%%%%%%
%%%%%%%%%%%%%%%%%%%%%%%%%%%%%%%%%%%%%%%%%

\begin{abstract}
\noindent Given a dynamic ordinal game, we deem a strategy sequentially rational if there exist a Bernoulli utility function and a conditional probability system with respect to which the strategy is a maximizer. We establish a complete class theorem by characterizing sequential rationality via the new Conditional B-Dominance. Building on this notion, we introduce Iterative Conditional B-Dominance, which is an iterative elimination procedure that characterizes the implications of forward induction in the class of games under scrutiny and selects the unique backward induction outcome in dynamic ordinal games with perfect information satisfying a genericity condition.  Additionally, we show that Iterative Conditional B-Dominance, as a `forward induction reasoning' solution concept, captures: %
\emph{(i)} the unique backward induction outcome obtained via sophisticated voting in binary agendas with sequential majority voting; %
\emph{(ii)}  farsightedness in dynamic ordinal games derived from social environments; %
\emph{(iii)} a unique outcome in ordinal Money-Burning Games. 

\bigskip


\end{abstract}

\epigraph{``[\dots] the theory of choice is based on the premise that choice is, or ought to be, governed by ordinal preference relations on the set of alternatives [\dots]''}%
{---\citet[Chapter 1.1.2, p.4]{Karni_2014}}

\vspace{-0.5cm}

\epigraph{```Look, I had the opportunity to get $2$ for sure, and nevertheless I decided to play in this subgame [\dots]. So think now well and make your decision.{'}''}%
{---\citet[Section 2.6, p.1013]{Kohlberg_Mertens_1986}}

%\clearpage

\tableofcontents

\clearpage

%%%%%%%%%%%%%%%%%%%%%%%%%%%%%%%%%%%%%%%%%
%%%%%%%%%%%%%%%%%%%%%%%%%%%%%%%%%%%%%%%%%
%%%%%%%%%%%%%%%%%%%%%%%%%%%%%%%%%%%%%%%%%
%%%%%%%%%%%%%%%%%%%%%%%%%%%%%%%%%%%%%%%%%
%%%%%%%%%%%%%%%%%%%%%%%%%%%%%%%%%%%%%%%%%
%%%%%%%% SECTION %%%%%%%%%%%%%%%%%%%%%%%%%
%%%%%%%%%%%%%%%%%%%%%%%%%%%%%%%%%%%%%%%%%
%%%%%%%%%%%%%%%%%%%%%%%%%%%%%%%%%%%%%%%%%
%%%%%%%%%%%%%%%%%%%%%%%%%%%%%%%%%%%%%%%%%
%%%%%%%%%%%%%%%%%%%%%%%%%%%%%%%%%%%%%%%%%
%%%%%%%%%%%%%%%%%%%%%%%%%%%%%%%%%%%%%%%%%
\section{Introduction}
\label{sec:introduction}

%%%%%%%%%%%%%%%%%%%%%%%%%%%%%%%%%%%%%%%%%%
%%%%%%%%%%%%%%%%%%%%%%%%%%%%%%%%%%%%%%%%%%
%%%%%%%%%%%%%%%%%%%%%%%%%%%%%%%%%%%%%%%%%%
%%%%%%%%% SUB-Section %%%%%%%%%%%%%%%%%%%%%%%%%%%%%%%%%%%%%%%%%%
%%%%%%%%%%%%%%%%%%%%%%%%%%%%%%%%%%%%%%%%%%
%%%%%%%%%%%%%%%%%%%%%%%%%%%%%%%%%%%%%%%%%%
\subsection{Motivation \& Results}
\label{subsec:motivation_results}

Brought to the attention of the game-theoretical community at large in \citet[Section 2.6, p.1013]{Kohlberg_Mertens_1986},\footnote{Forward induction has been identified for the first time by Elon Kohlberg, who first spelled out publicly the idea behind it in \cite{Kohlberg_1981}, which is an unpublished contribution to a conference (we are extremely grateful to the author for having explained the story behind this reference). To the best of our knowledge, the first mention in press to this idea is in \citet[Section 8]{Kreps_Wilson_1982} (with the related Figure 14 at page 884), which revolves around a private communication the authors had with Kohlberg back in 1980 concerning sensible restrictions on beliefs in sequential equilibria.} \emph{forward induction} is a powerful notion which informally revolves around the basic idea that players should take seriously into account what has happened in the past to draw opportune inferences from it that should then guide their choices.\footnote{See \citet[Section 2.4]{Kohlberg_1990}. This can also be captured by the \emph{Best-Rationalization Principle} of \citet[Section 1.1, p.180]{Battigalli_1996}  (see also the \emph{Rationalization Principle} stated in
\citet[Section 6, p.50]{Stalnaker_1998}).} Given how natural this idea is, it has been extensively studied in the game-theoretical literature and it has been proven most useful in both theoretical\footnote{See, for example, \cite{Ponssard_1991}, \cite{Ben-Porath_Dekel_1992}, \cite{Bagwell_Ramey_1996}, \cite{Friedenberg_2019}, \cite{Pomatto_2022}.} and experimental\footnote{See, for example, \cite{Cooper_et_al_1993}, \cite{vanHuyck_et_al_1993}, \cite{Cachon_Camerer_1996}, \cite{Huck_Mueller_2005}, 
\cite{Brandts_et_al_2007}.} applications.

In approaching the topic from a purely theoretical standpoint, we can identify two main streams of literature:\footnote{Of course, with the understanding that there have been works that---prioritizing a certain approach---have established strong connections between these two streams of literature (e.g., \cite{Sobel_et_al_1990}, \cite{Battigalli_Siniscalchi_2002}, \cite{Catonini_2019}, and \cite{Battigalli_Catonini_2023}).}
\begin{itemize}[leftmargin=*]
\item the \emph{equilibrium-based approach}, which---as exemplified in \citet[Sections 11, 13.3, \& 13.4]{Hillas_Kohlberg_2002} and starting from  \cite{Kohlberg_Mertens_1986}---has given to this idea a central role for equilibrium refinements purposes and that has focused on properly formalizing it in an equilibrium setting, with \cite{Govindan_Wilson_2009} and \cite{Man_2012} as two recent examples of these endeavours explicitly focusing on this notion;\footnote{See also \cite{McLennan_1985} (which does not explicitly refer to this idea with the term ``forward induction''), \cite{Cho_1987}, \cite{Banks_Sobel_1987}, \cite{Cho_Kreps_1987}, \cite{Cho_Sobel_1990}, \cite{Sobel_et_al_1990}, and \cite{Reny_1992}. Notably, \cite{vanDamme_1989}, \cite{Gul_Pearce_1996}, \cite{Govindan_Robson_1998} discuss potential shortcomings (and related solutions) behind this notion.}

\item the \emph{non-equilibrium based approach}, which---grounded on Rationalizability-style solution concepts for dynamic games and tightly linked with the epistemic\footnote{Regarding the epistemic approach to game theory, see \cite{Perea_2012}, \cite{Dekel_Siniscalchi_2015}, and \cite{Bonanno_2015}, where this last work approaches the topic by focusing on games with ordinal utilities.} game-theoretical literature---has built on the notion of Strong Rationalizability\footnote{\label{foot:SR}A common name for this solution concept---only hinted in \cite{Pearce_1984}, but explicitly adopted in \cite{Battigalli_1997} and subsequent work---is ``Extensive-Form Rationalizability''. Here, we employ a terminology introduced in \cite{Battigalli_Siniscalchi_1999b}, which distinguishes this solution concept from other forms of Rationalizability that can be formalized for the analysis of dynamic games in their extensive form (i.e., Initial (or Weak) Rationalizability \emph{\`a la} \cite{Ben-Porath_1997}, Backward Rationalizability \emph{\`a la} \cite{Penta_2015}, and Forward and Backward Rationalizability \emph{\`a la} \cite{Meier_Perea_2023}).} as introduced in \citet[Definition 9, p.1042]{Pearce_1984}  and redefined in \citet[Definition 2, p.46]{Battigalli_1997}.\footnote{See also \cite{Battigalli_Siniscalchi_1999b}, \cite{Shimoji_2002}, \cite{Battigalli_2003}, \cite{Battigalli_Siniscalchi_2003}, \cite{Shimoji_2004}, \cite{Battigalli_Friedenberg_2012}, \cite{Battigalli_Prestipino_2013}, \cite{Arieli_Aumann_2015}, \cite{Catonini_2021}.}
\end{itemize}
Thus, it is important to observe that both the aforementioned approaches are entrenched in standard game-theoretical assumptions and practice, with players' preferences over outcomes of a game represented via Bernoulli utility functions that are unique up to positive affine transformation and that are assumed to be transparent\footnote{That is, common knowledge in the informal sense of the expression.} between the players.

In this paper, we take a third route that starts from a natural question that is in the spirit of the revealed preference approach: 
\begin{quote}
\emph{assuming that we are outside observers of a given dynamic strategic interactions, what are the strategies---and consequently the outcomes---that we should expect if we assume that players are reasoning according to forward induction?}
\end{quote}

Thus, in order to answer this question, we depart from standard game-theoretical practice by simply assuming transparency of the players' \emph{ordinal} preferences over outcomes in line with a choice-theoretic approach to the problem.\footnote{See, for example, \citet[Section 1.1.1, pp.2--3]{Karni_2014}.}  As a  result we are forced to focus on ordinal games, i.e., games where the players' preferences can be represented via utility functions that are only unique up to monotone transformations, where it should be observed that we always employ the term ``ordinal'' throughout this work as a shortcut for the expression \emph{``representable via an ordinal (i.e., unique up to monotone transformation) utility function''}.\footnote{In the spirit of this point, we also employ the expression ``ordinal preference'', even if---for example---also when we deliver a von Neumann-Morgenstern utility function we still actually work with ordinal preferences (with the obvious caveat that in this latter case they are over the domain of lotteries).} Thus, by zeroing in on ordinal games, we actually work in the spirit of the Wilson's doctrine as stated in \cite{Wilson_1987} via---taking the perspective of outside observers---a relaxation of common knowledge assumptions regarding the players' risk attitudes.

To answer the problem set forth above, first of all, we capture what would be the behavioral implications of having players acting according to sequential rationality given our `ordinal' assumptions. Thus, with respect to this point, we introduce in \Mref{def:rational} a notion of \emph{sequential rationality}:
\begin{itemize}[leftmargin=*]
\item that is in the spirit of the Bayesian paradigm as set forth in \cite{Savage_1972} from a decision-theoretic standpoint, 

\item whereas it is in line with the notion of sequential rationality as set forth in \citet[Section 4, p.872]{Kreps_Wilson_1982} from a game-theoretic standpoint.\footnote{\label{foot:weak_sequential_rationality} Actually, the notion we introduce  is in the spirit of a weakening of the one introduced in  \cite{Kreps_Wilson_1982}  that goes back to \citet[Definition, p.631]{Reny_1992}, where it is called ``Weak Sequential Rationality''.}
\end{itemize}
In particular, we deem a strategy sequentially rational if there exist a---unique up to positive affine transformation and state independent---Bernoulli utility function and a conditional probability system (i.e., a family of probability measures well-behaved with respect to the presence of conditioning events) such that the strategy is a subjective expected utility maximizer. In order to capture the behavioral implications we are interested in, we introduce in \Mref{def:conditional_B-dominance} a dominance notion, called ``Conditional B-Dominance'', that works on ordinal preferences and that we show in \Mref{th:rationality} characterizes our notion of sequential rationality. In particular, we also provide a revealed preference interpretation of the aforementioned result in \Mref{prop:DT_take}, by establishing a complete class theorem in the spirit of the seminal \citet[Theorem 2.2, p.183]{Wald_1949} and \citet[Lemma 3, p.1048]{Pearce_1984}.\footnote{See also \cite{Battigalli_et_al_2016}.}

Now, one point is in order regarding what we just described: whereas the notions of sequential rationality and Conditional B-Dominance along with the corresponding---game-theoretic and decision-theoretic---characterization results can be seen as extensions of those introduced for static settings in \cite{Borgers_1993} and \cite{Lo_2000} and---as such---they could be considered hardly surprising from a theoretical standpoint, this should not diminish the relevancy of what we obtain. Indeed, bypassing the fact that the results are not trivial and to be delivered need an extensive usage of the notion of normal form information set introduced in \citet[Definition 2, p.329]{Mailath_et_al_1993}, our \Mref{th:rationality} and \Mref{prop:DT_take} have far-reaching implications, as exemplified by their potential usage in experimental settings employing the strategy method.\footnote{Introduced in \cite{Selten_1967}. See \cite{Brandts_Charness_2011} for a survey of the results obtained with this tool.} 
In particular, an experimenter that would like to infer the rationality of their subjects---in the spirit of \cite{Kneeland_2015}---in an experiment based on  a dynamic strategic interaction could rely on the aforementioned results, dropping all at once possible assumptions regarding the players' risk attitudes. Now, regarding this point, it should also be emphasized that, even if players' risk attitudes can be elicited with standard decision-making tasks, from a conceptual standpoint what has been elicited in that kind of game against nature should not necessarily automatically translate in an actual game-theoretical environment.\footnote{With respect to this point, see in particular
\citet[Section 1.1, pp.185--186]{Gilboa_Schmeidler_2003}, where they explicitly refer to this issue with respect to the problem of dealing with the potential presence of other-regarding preferences. A similar point is made in a different context in \citet[Section 3.4, p.12]{Daley_Sadowsky_2023}.}

Going back to our undertaking, having an opportune notion of sequential rationality along with a dominance notion that can capture its behavioral implications, we can tackle the problem set forth  above by introducing in \Mref{def:ICBD} an iterative elimination procedure based on our Conditional B-Dominance, that we call ``Iterative Conditional B-Dominance'' and that we show satisfies nonemptiness in \Mref{prop:nonemptiness}. Additionally, we introduce in \Mref{def:ordinal_strong_rationalizability} a solution concept built on our notion of sequential rationality that we call ``Ordinal Strong Rationalizability'' that is algorithmically characterized by Iterative Conditional B-Dominance, as we establish in \Mref{prop:ICBD_SR}. Once more, the solution concepts introduced with related results could look hardly surprising, since they could be seen as extensions of Iterated Conditional Dominance of \citet[Section 3]{Shimoji_Watson_1998} and of Strong Rationalizability in its version formalized in \citet[Definition 2, p.46]{Battigalli_1997} with our notions of Conditional B-Dominance and of sequential rationality in place. However, two points make what we obtain particularly relevant.

First of all, from a methodological standpoint, we show in \Mref{th:ICD_general}, with the related \Mref{prop:ICBD_DT_take}, that Iterative Conditional B-Dominance captures the revealed preference implications of forward induction reasoning by linking it to Strong Rationalizability: indeed, once more having in mind an experiment employing the strategy method, the strategy profiles we observe with related outcomes given the assumed ordinal preferences that are compatible with Iterative Conditional B-Dominance allow us to infer the existence of a Bernoulli utility function and opportune beliefs for which the strategies are expected utility maximizers.\footnote{In doing so, incidentally, we provide a characterization of Strong Rationalizability which is true to the spirit of the name of this very solution concept. Indeed, Rationalizability owes its name to a suggestion made to Douglas Bernheim  by Kevin Roberts---during a visiting of the latter to MIT back in 1981---in light of the conceptual similarities between the aforementioned solution concept and the notion of rationalizability in abstract choice theory (we are extremely grateful to Kevin Roberts for having shared his memories regarding this event).}

In second place, what we obtain, beyond its potential usage in experimental settings, has immediate theoretical applications. Indeed, it turns out that---really as it should be---Iterative Conditional B-Dominance delivers the unique backward induction outcome in dynamic games satisfying a genericity condition known as the ``No-Relevant Ties'' condition, as we show in \Mref{prop:relevant_ties}. In the standard game-theoretical setting with Bernoulli utility functions, the result that shows that backward and forward induction are \emph{outcome} equivalent in this class of games is known as Battigalli's theorem and it has been established  in \citet[Theorem 4, p.53]{Battigalli_1997}.\footnote{See \cite{Perea_2010} for a discussion of backward and forward induction reasoning.} Now, interestingly, even if risk attitudes should not actually play any role in dynamic games satisfying the ``No-Relevant Ties'' condition,\footnote{As pointed out in \citet[Section 4, p.52]{Battigalli_1997}.} this result has never been established before in an ordinal setting, which should actually be its most natural environment. Thus, in order to deliver \Mref{prop:relevant_ties}, we exploit tools introduced in \cite{Marx_Swinkels_1997} and, in doing so, we link Iterative Conditional B-Dominance and Iterated Admissibility.\footnote{See also \cite{Brandenburger_Friedenberg_2014} and \cite{Catonini_2023} with respect to the relation between Iterated Admissibility, backward induction, and forward induction.}

Our framework, along with the results we establish, turns out to be particularly appealing in a variety of contexts and has immediate applications. Indeed, the ordinal setting we work is a rather natural one for voting applications\footnote{See for example \citet[Footnote 3, p.1338]{Moulin_1979} and all the related literature reviewed in \Sref{subsec:binary_agendas}.} and as a result, by introducing a framework appropriate to capture binary agendas with sequential majority voting, we show in \Mref{prop:sophisticated_forward} how our \Mref{prop:relevant_ties} happens to link sophisticated voting \emph{{\`a} la} \cite{Farquharson_1969} to forward induction reasoning.\footnote{Concerning forward induction and voting, see the analysis of \cite{De_Sinopoli_2004} in the context of the model of representative democracy of \cite{Besley_Coate_1997}.} Also, we show in  \Mref{prop:social_environments} how Iterative Conditional B-Dominance can be used to capture farsighted behavior in the spirit of the notion of Social Rationalizability of \cite{Herings_et_al_2004} in social environments \emph{{\`a} la} \cite{Chwe_1994}, which are typically based on ordinal preferences. Finally, in \Mref{prop:money-burning} we establish that the cutting power of forward induction identified in `standard' money-burning games by \cite{Ben-Porath_Dekel_1992} can be replicated via Iterative Conditional B-Dominance in `ordinal' money-burning games. As a matter of fact, we build on this latter result in \Sref{subsubsec:abstract_GT_choice_theory} by arguing that Iterative Conditional B-Dominance, in particular in light of the decision-theoretic take concerning it that we describe in \Sref{subsubsec:ICBD_DT_take},\footnote{With respect to this point, see \citet[Section 5]{Chambers_et_al_2017}.} can be the building block to operationalize forward induction to pursue endeavours in the spirit of the revealed preference approach to game theory along the lines of \cite{Sprumont_2000}.\footnote{See also \cite{Galambos_2005}, \cite{Demuynck_Lauwers_2009}, \cite{Carvajal_et_al_2013}, and \cite{Cherchye_et_al_2013} and---more in general---the references  in \Sref{subsubsec:abstract_GT_choice_theory}.} In particular, this could be a key approach to disentangle backward and forward induction reasoning,\footnote{Regarding backward induction, see \cite{Ray_Zhou_2001}, \cite{Ray_Snyder_2013}, and \cite{Schenone_2020}. Also, \cite{Brandenburger_Friedenberg_2009} is a different take on this issue.}  a challenging problem that has been also attracted the attention of a recent stream of experimental literature.\footnote{See \cite{Berninghaus_et_al_2014}, \cite{Ghosh_et_al_2015},  \cite{Balkenborg_Nagel_2016}, \cite{Chlass_Perea_2016},  and \cite{Evdokimov_Rustichini_2016}.}

%%%%%%%%%%%%%%%%%%%%%%%%%%%%%%%%%%%%%%%%%%
%%%%%%%%%%%%%%%%%%%%%%%%%%%%%%%%%%%%%%%%%%
%%%%%%%%%%%%%%%%%%%%%%%%%%%%%%%%%%%%%%%%%%
%%%%%%%%% SUB-Section %%%%%%%%%%%%%%%%%%%%%%%%%%%%%%%%%%%%%%%%%%
%%%%%%%%%%%%%%%%%%%%%%%%%%%%%%%%%%%%%%%%%%
%%%%%%%%%%%%%%%%%%%%%%%%%%%%%%%%%%%%%%%%%%
\subsection{Related Literature}
\label{subsec:related_literature}

This work is related to various papers belonging to different streams of literature.  In its focus on providing a \emph{Bayesian foundation} to the notion of sequential rationality with a marked revealed preference flavour,  it is related to \cite{Borgers_1993}, \cite{Lo_2000}, \cite{Siniscalchi_2022}, and \cite{de_Oliveira_Lamba_2022} (where we expand on the relation between the present work and \cite{Siniscalchi_2022} in \Sref{subsubsec:structural_rationality}). This point is also related to that stream of literature that focuses on the behavioral implications compatible with players acting rationally according to the Bayesian paradigm, in the spirit of \cite{Green_Park_1996} and \cite{Zambrano_2005}.\footnote{See also \cite{Ledyard_1986}.} In focusing on ordinal games, it is related to \cite{Grant_et_al_2016} (with the caveat that this latter work focuses on state-dependent preferences), \cite{Bonanno_Tsakas_2018}, and \cite{Guarino_Ziegler_2022}.\footnote{See also \cite{Morris_Takahashi_2012}.} Clearly, as pointed out in the previous section, it is related to the papers on forward induction previously mentioned with an immediate link in particular to \cite{Shimoji_Watson_1998}, with respect to our introduction of an \emph{iterative procedure} based on the dominance notion we introduce (i.e., Iterative Conditional B-Dominance), and to \cite{Battigalli_1997}, in providing an ordinal version of Strong Rationalizability. Also, there is an immediate link to \cite{Battigalli_1997} with respect to the problem of establishing a relation between backward and forward induction reasoning via the \emph{unique backward induction outcome} in dynamic games with perfect information satisfying the genericity condition known as ``No Relevant Ties''. Finally, this paper is also linked to a growing literature that focuses on B-Dominance such as \cite{Weinstein_2016} and \cite{Alon_et_al_2021} as a tool to avoid assumptions regarding the transparency of players' risk attitudes.

%%%%%%%%%%%%%%%%%%%%%%%%%%%%%%%%%%%%%%%%%%
%%%%%%%%%%%%%%%%%%%%%%%%%%%%%%%%%%%%%%%%%%
%%%%%%%%%%%%%%%%%%%%%%%%%%%%%%%%%%%%%%%%%%
%%%%%%%%% SUB-Section %%%%%%%%%%%%%%%%%%%%%%%%%%%%%%%%%%%%%%%%%%
%%%%%%%%%%%%%%%%%%%%%%%%%%%%%%%%%%%%%%%%%%
%%%%%%%%%%%%%%%%%%%%%%%%%%%%%%%%%%%%%%%%%%
\subsection{Synopsis}
\label{subsec:synopsis}

In \Sref{sec:game-theoretical_framework}, we introduce the primitive objects of our analysis, namely, dynamic games with ordinal preferences, and we recall the definitions of various dominance notions for games in strategic form representation. In \Sref{sec:rationality_dominance}, we set forth our notion of sequential rationality, we formalize the dominance notion for dynamic games in their extensive form representation that we study in this paper, i.e., Conditional B-Dominance, and we show that a strategy that is sequentially rational according to our definition is not conditionally B-dominated and vice versa, by additionally providing a decision-theoretic interpretation of the result as a complete class theorem.  In \Sref{sec:revealed_FI}, we define Iterative Conditional B-Dominance, i.e., the algorithm that iteratively eliminates conditionally B-dominated strategies, and Ordinal Strong Rationalizability, its Rationalizability-type solution concept counterpart, we study the properties they satisfy, and we link them to Iterated Conditional Dominance and Strong Rationalizability. \Sref{sec:relevant_ties} is devoted to investigate the behavioral implications of Iterative Conditional B-Dominance in  dynamic games with perfect information which satisfy a genericity condition called ``No Relevant Ties'', where we show that Iterative Conditional B-Dominance selects the unique backward induction outcome in this class of games, a result that we exploits in some of the applications we describe in \Sref{sec:applications}, where we show how Iterative Conditional B-Dominance can provide a `forward induction reasoning' flavour to outcomes obtained in different  game-theoretical contexts. Finally, in \Sref{sec:discussion}, we address various issues related to our analysis. All the proofs of the results established in the paper are relegated to \Sref{app:proofs}.

%%%%%%%%%%%%%%%%%%%%%%%%%%%%%%%%%%%%%%%%%
%%%%%%%%%%%%%%%%%%%%%%%%%%%%%%%%%%%%%%%%%
%%%%%%%%%%%%%%%%%%%%%%%%%%%%%%%%%%%%%%%%%
%%%%%%%%%%%%%%%%%%%%%%%%%%%%%%%%%%%%%%%%%
%%%%%%%%%%%%%%%%%%%%%%%%%%%%%%%%%%%%%%%%%
%%%%%%%% SECTION %%%%%%%%%%%%%%%%%%%%%%%%%
%%%%%%%%%%%%%%%%%%%%%%%%%%%%%%%%%%%%%%%%%
%%%%%%%%%%%%%%%%%%%%%%%%%%%%%%%%%%%%%%%%%
%%%%%%%%%%%%%%%%%%%%%%%%%%%%%%%%%%%%%%%%%
%%%%%%%%%%%%%%%%%%%%%%%%%%%%%%%%%%%%%%%%%
%%%%%%%%%%%%%%%%%%%%%%%%%%%%%%%%%%%%%%%%%
\section{Game-Theoretical Framework}
\label{sec:game-theoretical_framework}

%%%%%%%%%%%%%%%%%%%%%%%%%%%%%%%%%%%%%%%%%%
%%%%%%%%%%%%%%%%%%%%%%%%%%%%%%%%%%%%%%%%%%
%%%%%%%%%%%%%%%%%%%%%%%%%%%%%%%%%%%%%%%%%%
%%%%%%%%% SUB-Section %%%%%%%%%%%%%%%%%%%%%%%%%%%%%%%%%%%%%%%%%%
%%%%%%%%%%%%%%%%%%%%%%%%%%%%%%%%%%%%%%%%%%
%%%%%%%%%%%%%%%%%%%%%%%%%%%%%%%%%%%%%%%%%%
\subsection{Dynamic Ordinal Games}
\label{subsec:dynamic_ordinal_games}

The primitive object of our analysis is a finite dynamic game with ordinal preferences\footnote{See \cite{Bonanno_2018} for a textbook focusing on games with ordinal preferences.} 
and perfect recall in its extensive form representation
\begin{equation}
\label{eq:dynamic_ordinal_game}
\Game := \Angles { I , (A_j)_{j \in I} , X, Z , (H_j, \mbf{S}_j )_{j \in I}, \zeta , (\succsim_j)_{j \in I} },
\end{equation} 
which we simply refer to as \emph{dynamic game}, where this definition possibly allows for simultaneous moves.\footnote{For a similar definition, see \citet[Section 2]{Battigalli_Friedenberg_2012}, inspired by \citet[Definition 200.1, Chapter 11.1.2]{Osborne_Rubinstein_1994}, with the caveat that the latter does not explicitly allow for simultaneous moves (see  \citet[Chapter 6.3.2]{Osborne_Rubinstein_1994} for the corresponding extension).}  
In \Mref{eq:dynamic_ordinal_game}, $I$ denotes the set of players and $A_i$, with $i \in I$, is the set of \emph{actions} of  player $i$. The set $X$ is the set of \emph{histories}, where a history $x$ is either the empty sequence $\la \varnothing \ra$ (i.e., the initial history), or it is a sequence $(a^1, \dots, a^K)$,  where $a^k := (a^{k}_{i})_{i \in I}$ with $a^{k}_i \in A_i$ for every $i \in I$ and for every $1 \leq k \leq K$. We let $A(x) : = \prod_{i \in I} A_i (x)$ denote the set of actions available to the players at history $x$: given that we let $\abs{Y}$ denote the cardinality of an arbitrary set $Y$, if $\abs{A_i (x)} \geq 2$, then player $i$ is \emph{active} at history $x$, otherwise she is inactive. The set $Z \subseteq X$ is the set of \emph{terminal} histories, alternatively called \emph{outcomes}, i.e., the set of histories such that $\abs{A (z)} = 0$, for every $z \in Z$.  When the set of histories is endowed with the binary relation $\sqsubseteq$ capturing the notion of precedence, the resulting tuple $(X, \sqsubseteq)$ is an \emph{arborescence}.\footnote{It should be observed that---rather often---in the literature it can be alternatively found the term ``tree'' in place of ``arborescence'' (as in  \citet[Chapter 6.1.2, p.92]{Osborne_Rubinstein_1994}). See \citet[Chapter 3.3.1, pp.77-79]{Fudenberg_Tirole_1991} for an explanation of the difference between these two notions.}

We let $H_i$ denote the set of \emph{information sets} of player $i$, i.e., this is the partition of all those non-terminal histories where player $i$ is active, where the elements of $H_i$ satisfy the property that, if $x, x' \in h$, with $h \in H_i$, then $A_i (x) = A_i (x')$. We extend to information sets the notational convention introduced for histories and we let $A_i (h)$ denote the set of actions available to player $i$ at information set $h \in H_i$. Given that $H := \bigcup_{j \in I} H_j$, we let $I_h \subseteq I$ denote the set of players that are active at information set $h \in H$. A dynamic game is said to be with \emph{observable actions} if $h$ is a singleton,  for every $h \in H$. A dynamic game with observable actions is said to be with \emph{perfect information} if $I_h$ is a singleton, for every information set $h \in H$.

A \emph{standard strategy} of player $i$ is a function $\mbf{s}_i : H_i \to \bigcup_{\overline{h} \in H_i} A_i (\overline{h})$ such that $\mbf{s}_i (h) \in A_i (h)$, for every $h \in H_i$, where $\mbf{S}_i$ denotes the set of standard strategies of player $i$, with $\mbf{S} := \prod_{j \in I} \mbf{S}_j$ and $\mbf{S}_{-i} := \prod_{j \in I \setminus \{i\}} \mbf{S}_j$. We let $\zeta \in Z^\mbf{S}$ denote an \emph{outcome function}.

Finally, for every $i \in I$, we let $\succsim_i \subseteq Z \times Z$ be a  \emph{rational preference relation}\footnote{Here, we follow the terminology set forth in \citet[Definition 1.B.1, p.6]{Mas-Colell_et_al_1995} and \citet[Chapter 1.2, p.3]{Chambers_Echenique_2016}.} over the outcomes of the game, that is, a binary relation that is complete and transitive. As it is customary, we let $\succ_i$ (resp., $\sim_i$) denote the asymmetric (resp., symmetric) part of $\succsim_i$ and we let  $\succsim := (\succsim_{j})_{j \in I}$ denote a profile of rational preference relations of the players (where the symmetric and asymmetric parts are defined accordingly).

%%%%%%%%%%%%%%%%%%%%%%%%%%%%%%%%%%%%%%%%%%
%%%%%%%%%%%%%%%%%%%%%%%%%%%%%%%%%%%%%%%%%%
%%%%%%%%%%%%%%%%%%%%%%%%%%%%%%%%%%%%%%%%%%
%%%%%%%%% SUB-Section %%%%%%%%%%%%%%%%%%%%%%%%%%%%%%%%%%%%%%%%%%
%%%%%%%%%%%%%%%%%%%%%%%%%%%%%%%%%%%%%%%%%%
%%%%%%%%%%%%%%%%%%%%%%%%%%%%%%%%%%%%%%%%%%
\subsection{Strategies \& The Reduced Strategic Form}
\label{subsec:strategies_reduced_strategic_form}

For every history $x \in X$ and player $i \in I$, we let $\mbf{S}_i (x)$ denote the set of player $i$'s \emph{standard strategies that reach history} $x$, i.e.,
\begin{equation*}
\mbf{S}_i (x) := \Set { \mbf{s}_i \in \mbf{S}_i | \exists z \in X \ \exists \mbf{s}_{-i} \in \mbf{S}_{-i} : x \sqsubseteq z , \ \zeta (\mbf{s}_i, \mbf{s}_{-i}) = z }, 
\end{equation*} 
with $\mbf{S}_{-i} (x)$ similarly defined. As a result, for every $h \in H$, we let $\mbf{S}_i (h) := \bigcup_{x \in h} \mbf{S}_i (x)$ denote the set of player $i$'s standard strategies that reach information set $h$, with $\mbf{S}_{-i} (h)$ (as before) similarly defined. Now, given that we  let $H (\mbf{s}_i)$ denote the \emph{set of information sets that are allowed by} $\mbf{s}_i$, i.e., 
\begin{equation*}
H (\mbf{s}_i) := \Set { h \in H | \mbf{S}_i (h) \neq \emptyset } ,
\end{equation*} 
two standard strategies $\mbf{s}_i, \mbf{s}'_i \in \mbf{S}_i$ are deemed \emph{behaviorally equivalent} if $H (\mbf{s}_i) = H (\mbf{s}'_i)$ and $\mbf{s}_i (h) = \mbf{s}'_i (h)$ for every $h \in H (\mbf{s}_i) = H (\mbf{s}'_i)$. %

Thus, a \emph{strategy}\footnote{\label{foot:strategy}Alternatively called ``plan of actions'' as in the words of  \citet[Section 2]{Rubinstein_1991} or ``reduced strategy''. We use the---nonstandard---expression ``standard strategy'' to refer to an element of $\mbf{S}_i$, for an arbitrary player $i$, where usually such an object is simply called ``strategy'', since we want to save that term for the actual primitive objects of interests for our analysis, which are the elements of $S_i$.} $s_i$ of player $i$  is a maximal set of behaviorally equivalent standard strategies, where in the remainder of this work the focus is on strategies. We let $S_i$ denote the set of strategies of player $i$ and---following standard notational conventions---we let $S_{-i} := \prod_{j \in I \setminus \{i\}} S_j$, with $S := \prod_{j \in I} S_j$. Given a nonempty subset $R \subseteq S$, $R$ is  a \emph{restriction} if $R = \prod_{j \in I} R_j$, i.e., when $R$ has a product structure.\footnote{Concerning our usage of the term ``restriction'' with respect to the related literature, see \Sref{subsubsec:restrictions}.}

We extend the conventions introduced above for standard strategies to strategies in a natural fashion. Hence, we let $S (h)$ denote the set of strategy profiles that reach information set $h \in H$, where, given that we let $\proj$ denote the projection operator as canonically defined, we set $S_i (h) := \proj_{S_i} S (h)$ and $S_{-i} (h) := \proj_{S_{-i}} S (h)$. Building on this notation, for every $i \in I$ and $h \in H_i$, we let $S^i (h)$ denote  the set of strategy profiles that allow information set $h \in H_i$, called the \emph{conditional problem of player $i$ at $h \in H_i$}, where, clearly, there exists a `natural' bijection between an information set $h \in H_i$ and the conditional problem of player $i$ at $h$, for every $i \in I$.  Two points need to be emphasized concerning $S^i (h)$, for every $i \in I$ and $h \in H_i$: first, by assuming perfect recall, we have that $S^i (h) = S_i (h) \times S_{-i} (h)$; in second place, $S^i (h)$ does not need to have a product structure, which is the case when $S_{-i} (h)$ does not have a product structure.\footnote{\label{foot:Mailath_et_al_1993}See \citet[Section 3, p.283]{Mailath_et_al_1993} for an example based on the ($3$-player) Selten's Horse to see the potential lack of product structure of $S_{-i} (h)$.}  For every $i \in I$ and $h \in H_i$, we call $S_{-i} (h)$ a \emph{conditioning event for player $i$}, with $\mcal{H}_i := \Set { S_{-i} (h) | h \in H_i}$.  Given an arbitrary player $i \in I$ and a restriction $R \subseteq S$, for every $i \in I$ and $h \in H_i$, we let $R_i (h) := R_i \cap S_{i} (h)$ and $R_{-i} (h) := R_{-i} \cap S_{-i} (h)$ (from what is written above, not necessarily a restriction), with $R^i (h) := R_i (h) \times R_{-i} (h)$ being a \emph{restricted conditional problem} of player $i$, where all these sets can be empty. Also, we let $H (s_i)$ denote the set of information sets allowed by $s_i \in S_i$, with $H_i (s_i)$ and $H_{-i} (s_i)$ defined accordingly. Finally, we let $H (R)$ (resp., $Z (R)$) denote the set of information sets (resp., the set of terminal histories) that are allowed by strategy profiles in $R$.

In the following, with an innocuous abuse of notation, given that---as already pointed out---we focus throughout this work on strategies, we actually refer to an outcome function as a function $\zeta \in Z^S$ mapping \emph{strategies} to terminal histories.

In light of the objects and notation set forth above, we can now introduce the notion of reduced strategic form representation,\footnote{Traditionally, this is called the ``\emph{reduced} strategic form'', since the relevant primitive object in the corresponding definition is what we call here ``strategy'' and is often called in the literature---as pointed out in \Sref{foot:strategy}---''reduced strategy''.} which is essentially the result of bypassing the sequential nature of a dynamic game. Hence,  given a dynamic game $\Game$, its \emph{reduced strategic form} (henceforth, strategic form) representation is given by the tuple
\begin{equation*}
\Game^r := \Angles { I , Z, (S_j)_{j \in I}, \zeta, (\succsim_j)_{j \in I} },
\end{equation*} 
whose elements are defined as the corresponding ones described above.

%%%%%%%%%%%%%%%%%%%%%%%%%%%%%%%%%%%%%%%%%%
%%%%%%%%%%%%%%%%%%%%%%%%%%%%%%%%%%%%%%%%%%
%%%%%%%%%%%%%%%%%%%%%%%%%%%%%%%%%%%%%%%%%%
%%%%%%%%% SUB-Section %%%%%%%%%%%%%%%%%%%%%%%%%%%%%%%%%%%%%%%%%%
%%%%%%%%%%%%%%%%%%%%%%%%%%%%%%%%%%%%%%%%%%
%%%%%%%%%%%%%%%%%%%%%%%%%%%%%%%%%%%%%%%%%%
\subsection{Dominance Notions for the Strategic Form Representation} 
\label{subsec:dominance_strategic}

In the following, we fix a dynamic game $\Game$ and we focus on its strategic form. Thus, in particular, given a restriction $R \subseteq S$ and a player $i \in I$, a strategy $s_i \in R_i$ is \emph{strictly dominated relative to $R$ by strategy $s^{*}_i \in R_i$} if $\zeta (s^{*}_i, s_{-i}) \succ_i \zeta (s_i, s_{-i})$ for every $s_{-i} \in R_{-i}$, whereas a strategy $s_i \in R_i$ is \emph{weakly dominated relative to $R$ by strategy $s^{*}_i \in R_i$} if $\zeta (s^{*}_i, s_{-i}) \succsim_i \zeta (s_i, s_{-i})$ for every $s_{-i} \in R_{-i}$ and there exists a $s^{*}_{-i} \in R_{-i}$ such that  $\zeta (s^{*}_i, s^{*}_{-i}) \succ_i \zeta (s_i, s^{*}_{-i})$.  In the previous cases we write that $s^{*}_i$ strictly (resp., weakly) dominates $s_i$ relative to $R$ and that a strategy $s_i$ is strictly (resp., weakly) dominated relative to $R$ if there exists a strategy $s^{*}_i \in R_i$ that strictly (resp., weakly) dominates $s_i$ relative to $R$. A strategy $s^{*}_i \in R_i$ that is not weakly dominated relative to $R$ for player $i$ by any strategy in $R_i$ is \emph{admissible with respect to $R$}.\footnote{This terminology goes back to \citet[Chapter 4.11, p.79]{Luce_Raiffa_1957}.} It should be observed that, given a restriction $R \subseteq S$ and a player $i \in I$, if $R_{-i}$ is a singleton and strategy $s_i \in R_i$ is weakly dominated relative to $R$, then $s_i$ is \emph{strictly} dominated relative to $R$. 

\begin{notation}[Set of Admissible Strategies with respect to a Restriction]
Given a restriction $R \subseteq S$ and a player $i \in I$, we let $\Ad_i (R)$ denote the set of strategies of player $i$ that are admissible with respect to $R$.
\end{notation}

\citet[Definition 4, p.426]{Borgers_1993} introduced an additional dominance notion:\footnote{Actually the author refers to this dominance notion as ``Pure Strategy Dominance''  in the title, while in the main body of the paper it is simply called ``Dominance''.} given a restriction $R \subseteq S$ and a player $i \in I$, strategy $s_i \in R_i$ is \emph{B\"orgers dominated} (henceforth, B-dominated) \emph{with respect to $R$}  if $s_i \notin \Ad_i (R_i \times Q_{-i})$, for every nonempty (product) subset $Q_{-i} \subseteq R_{-i}$.

\begin{remark}[Relation Between the Dominance Notions]
\label{rem:relation_dominance}
Given a restriction $R \subseteq S$ and a player $i \in I$: 
\begin{itemize}[leftmargin=*]
\item if a strategy $s_i \in R_i$ is strictly dominated relative to $R$, then it is B-dominated with respect to $R$; 

\item if a strategy $s'_i \in R_i$ is B-dominated with respect to $R$, then $s'_i$ is weakly dominated relative to $R$.
\end{itemize}
\end{remark}

With respect to \Mref{rem:relation_dominance}, it is immediate to establish that the opposite directions do not hold.

\begin{remark}[Singleton Sets]
\label{rem:singleton}
Given a restriction $R \subseteq S$, a player $i \in I$, and a strategy $s_i \in R_i$, if $s_i \in R_i$ is B-dominated with respect to $R$, then for every singleton $Q_{-i} \subseteq R_{-i}$ there exists a strategy $s^{*}_i \in R_i$ that strictly dominates $s_i$ relative to $R_i \times Q_{-i}$.
\end{remark}

\begin{notation}[Set of B-Dominated Strategies with respect to a Restriction]
\label{not:B-dominance}
Given a restriction $R \subseteq S$ and a player $i \in I$, we let $\bd_i (R)$ denote the set of strategies of player $i$ that are B-dominated with respect to $R$.
\end{notation}

Regarding all the dominance notions just introduced, it is understood that in the following the omission of the restriction with respect to which the dominance notion is defined stands for that restriction being equal to $S$.

\begin{example}[Dynamic Game with Simultaneous Moves with an Outside Option]
\label{ex:main_example}
To see the dominance notions we have introduced at work, we take the dynamic game in \Sref{fig:example_ordinal_basic}, with two players, again Ann (viz., $\Ann$) and Bob (viz., $\Bob$).

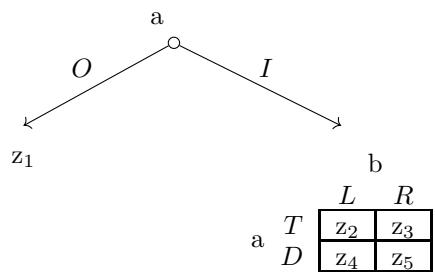
\begin{figure}[H]
\centering
\begin{tikzpicture}
%%% NODE STYLE
[info/.style={circle, draw, inner sep=1.5, fill=black},
scale=1.1] 
%%% WORKING GRID
% Basic
%\draw[step=.2cm, gray, very thin] (-6,-8) grid (6,8);
%\draw[red] (-6,0) -- (6,0);
%\draw[red] (0,-8) -- (0,8);
% x axis
%\foreach \x in {-5,-4,-3,-2,-1,1,2,3,4,5} \draw[green] (\x,-8) -- (\x,8);
% y axis
%\foreach \y in {-8,-7,-6,-5,-4,-3,-2,-1,1,2,3,4,5,6,7,8} \draw[green] (-6,\y) -- (6,\y);
%%% NODES
\node[info, fill=white] (n_1) at (0,4.5) {};
%%% PAYOFFS ARROWS
\draw[->] (n_1) -- (-1.8,3.5);
\draw[->] (n_1) -- (2,3.5);
%%% PLAYERS
% Player 1
\draw (-0.2,4.8) node {$\Ann$};
%%% PAYOFFS
\draw (-1.8,3.1) node {$\mathrm{z}_1$};
\draw (2,2.5) node {
\begin{game}{2}{2}[$\Ann$][$\Bob$]
		& $L$ 	& $R$ 	\\
$T$ 	& $\mathrm{z}_2$ 	& $\mathrm{z}_3$ 	\\
$D$ 	& $\mathrm{z}_4$ 	& $\mathrm{z}_5$ 	\\
\end{game}};
%%% ACTIONS
% Player 1
\draw (-1.1, 4.2) node {$O$};
\draw (1.1, 4.2) node {$I$};
\end{tikzpicture}
\caption{A dynamic game with simultaneous moves with an outside option.}\label{fig:example_ordinal_basic}
\end{figure}

\noindent We also provide its strategic form in \Sref{fig:example_ordinal_strategic}. We have that $S_\Ann = \{ O, IT, ID \}$ and $S_\Bob = \{ L, R \}$. Also, we posit that Ann has the following preferences over the terminal histories: $\mathrm{z}_2 \succ_{\Ann} \mathrm{z}_1 \succ_{\Ann} \mathrm{z}_5 \succ_{\Ann} \mathrm{z}_3 \sim_\Ann \mathrm{z}_4$.

\begin{figure}[H]
\centering
\begin{game}{3}{2}[$\Ann$][$\Bob$]
		& $L$ 	& $R$	\\
$O$ 	& $\mathrm{z}_1$ 	& $\mathrm{z}_1$ \\
$IT$ 	& $\mathrm{z}_2$ 	& $\mathrm{z}_3$ 	\\
$ID$ 	& $\mathrm{z}_4$ 	& $\mathrm{z}_5$ \\
\end{game}
\caption{The game in \Sref{fig:example_ordinal_basic} in its strategic form representation.}
\label{fig:example_ordinal_strategic}
\end{figure}

\noindent Thus, focusing on Ann only, with the preferences over terminal histories given above, we have that strategy $ID$ is strictly dominated by $O$: as a result, in light of what is written above concerning the relation between the various dominance notions just introduced, strategy $ID$ is B-dominated. Alternatively, we can establish the fact that this strategy is B-dominated in a more explicit way by employing the actual definition of B-dominance. Thus, given Bob's strategy $L$, we have that $ID$ is strictly dominated by both $O$ and $ID$; given Bob's strategy $R$, strategy $ID$ is strictly dominated by $O$; finally, given $S_\Bob$, we have that $ID$ is strictly dominated by $O$.
\end{example}

%%%%%%%%%%%%%%%%%%%%%%%%%%%%%%%%%%%%%%%%%
%%%%%%%%%%%%%%%%%%%%%%%%%%%%%%%%%%%%%%%%%
%%%%%%%%%%%%%%%%%%%%%%%%%%%%%%%%%%%%%%%%%
%%%%%%%%%%%%%%%%%%%%%%%%%%%%%%%%%%%%%%%%%
%%%%%%%%%%%%%%%%%%%%%%%%%%%%%%%%%%%%%%%%%
%%%%%%%% SECTION %%%%%%%%%%%%%%%%%%%%%%%%%
%%%%%%%%%%%%%%%%%%%%%%%%%%%%%%%%%%%%%%%%%
%%%%%%%%%%%%%%%%%%%%%%%%%%%%%%%%%%%%%%%%%
%%%%%%%%%%%%%%%%%%%%%%%%%%%%%%%%%%%%%%%%%
%%%%%%%%%%%%%%%%%%%%%%%%%%%%%%%%%%%%%%%%%
%%%%%%%%%%%%%%%%%%%%%%%%%%%%%%%%%%%%%%%%%
\section{Rationality \& Dominance}
\label{sec:rationality_dominance}

%%%%%%%%%%%%%%%%%%%%%%%%%%%%%%%%%%%%%%%%%%
%%%%%%%%%%%%%%%%%%%%%%%%%%%%%%%%%%%%%%%%%%
%%%%%%%%%%%%%%%%%%%%%%%%%%%%%%%%%%%%%%%%%%
%%%%%%%%% SUB-Section %%%%%%%%%%%%%%%%%%%%%%%%%%%%%%%%%%%%%%%%%%
%%%%%%%%%%%%%%%%%%%%%%%%%%%%%%%%%%%%%%%%%%
%%%%%%%%%%%%%%%%%%%%%%%%%%%%%%%%%%%%%%%%%%
\subsection{Rationality} 
\label{subsec:rationality}

Given a dynamic game $\Game$ and a player $i \in I$, we let $u_i \in \Re^Z$ denote a unique up to positive affine transformation \emph{Bernoulli utility function}\footnote{\label{foot:utility}Here, given that there is no consensus on the terminology used, following the one set forth in  \citet[Chapter 6.B, p.184]{Mas-Colell_et_al_1995}, we distinguish between the function $u_i$ (as stated above, the Bernoulli utility function) and the related function that captures expected utility, with the latter the actual von Neumann-Morgenstern utility function (and, with respect to this point, see \Sref{subsec:relation_ICD}). It should be noted that, on the contrary, \cite{Borgers_1993} follows a rather common practice in the field that does not distinguish between the two functions, according to which the term ``von Neumann-Morgenstern utility function'' is used for \emph{both} the utility function over outcomes $u_i$ and the utility function over lotteries. See in particular \citet[Footnote 12, p.184]{Mas-Colell_et_al_1995}.} of player $i$ (henceforth, utility function) such that $u_i (z) \geq u_i (z')$ if and only if $z \succsim_i z'$, for every $z, z' \in Z$, where we write that such a utility function $u_i$ is \emph{originated from the rational preference relation} $\succsim_i$, with $\Utilities_i$ denoting the set of utility functions that originate from the rational preference relation $\succsim_i$.  Also, for every $i \in I$, a \emph{conditional probability system} \footnote{\label{foot:CPS}This  corresponds to one of the primitive elements of a  \emph{conditional probability space} of \citet[Section 1.2--1.4]{Renyi_1955}. However, the name---that eventually stuck in the game-theoretical literature---actually comes from a related definition from \citet[Section 5]{Myerson_1986}. See \citet[Section 3.2]{Hammond_1994} for a discussion of the relation between these two notions.}  (henceforth, CPS) $\mu_i$ on $(2^{S_{-i}}, \mcal{H}_i)$ is a mapping
\begin{equation*}
\mu_i \Rounds { \cdot | \cdot } : 2^{S_{-i}} \times \mcal{H}_i \to [0,1]
\end{equation*}
that satisfies the following axioms:
\begin{enumerate}[label=A\arabic*), itemsep=0.5ex]
\item for every $h \in H_i$, $\mu_i \Rounds { S_{-i} (h) | S_{-i} (h) } =1$;
\item for every $h \in H_i$, $\mu_i (\cdot | S_{-i} (h))$ is a probability measure on $(2^{S_{-i}}, \mcal{H}_i)$;
\item (Chain Rule) for every $\widehat{S}_{-i} \subseteq S_{-i}$ and for every $S_{-i} (h), S_{-i} (h') \in \mcal{H}_i$, if $\widehat{S}_{-i} \subseteq S_{-i} (h') \subseteq S_{-i} (h) $, then
\begin{equation*}
\mu_i \Rounds { \widehat{S}_{-i}|S_{-i} (h) } = %
\mu_i \Rounds { \widehat{S}_{-i} | S_{-i} (h') } \cdot %
\mu_i \Rounds { S_{-i} (h')|S_{-i} (h) } .
\end{equation*} 
\end{enumerate}
Given that we let $\Delta (Y)$ denote the set of all probability measures over an arbitrary space $Y$, $[\Delta(S_{-i})]^{\mcal{H}_i}$ denotes the set of all mappings from $\mcal{H}_{i}$ to $\Delta (S_{-i})$, while we let $\Delta^{\mcal{H}_i} (S_{-i}) \subseteq [\Delta(S_{-i})]^{\mcal{H}_i}$ denote the set of CPSs on $(2^{S_{-i}}, \mcal{H}_i)$, i.e., the set of all those mappings from $\mcal{H}_i$ to $\Delta (S_{-i})$ that satisfy the axioms above. Finally, given a restriction $R \subseteq S$, we let $u_{i}|_{R} \in \Re^{Z(R)}$ denote the utility function restricted to $Z (R)$ whose extension\footnote{See \citet[Chapter 1.3, p.4]{Aliprantis_Border_2006} concerning this---standard---terminology.} is $u_i \in \Re^Z$ and
\begin{equation*}
\label{eq:CPS_strong}
\Delta^{\mcal{H}_i} (R_{-i}) := \Set %
{ \mu_i \in \Delta^{\mcal{H}_i} (S_{-i}) | %
\forall h \in H_i \ %
\big(%
R^i (h) \neq \emptyset \Lra %
\mu_i (R_{-i} (h) | S_{-i} (h)) = 1 %
\big) }.
\end{equation*}
We can now introduce the following definition, which is the main building block of this section.

\begin{definition}[Sequentially Rational Strategy Given a Restriction]
\label{def:rational_given}
Given a restriction $R \subseteq S$ and a player $i \in I$, strategy $s^{*}_i \in R_i$ is \emph{sequentially rational given} $R$ if there exist a utility function $u_{i}|_R \in \Re^{Z (R)}$ and a CPS $\mu_i \in \Delta^{\mcal{H}_i} (R_{-i})$  such that for every $h \in H_i (s^{*}_i)$, if $R^i (h)$ is nonempty, then
\begin{equation}
\label{eq:rational_given_gen} 
\sum_{s_{-i} \in R_{-i}} %
u_i (\zeta(s^{*}_i , s_{-i})) \cdot %
\mu_i (\{ s_{-i} \} | S_{-i} (h)) \geq %
\sum_{s_{-i} \in R_{-i}} %
u_i (\zeta(s_i , s_{-i})) \cdot %
\mu_i (\{ s_{-i} \} | S_{-i} (h))
\end{equation}
for every $s_i \in R_i (h)$.
\end{definition}

\begin{example}[$4$-legged Centipede]
\label{ex:centipede}
The game in \Sref{fig:centipede} is an example of a dynamic game with \emph{perfect information} with two players, namely, Ann (viz., $\Ann$) and Bob (viz., $\Bob$).

\begin{figure}[H]
\centering
\begin{tikzpicture}
%%% NODE STYLE
[info/.style={circle, draw, inner sep=1.5, fill=black},
scale=1.1] 
%%% WORKING GRID
% Basic
%\draw[step=.2cm, gray, very thin] (-6,-4) grid (6,4);
%\draw[red] (-6,0) -- (6,0);
%\draw[red] (0,-4) -- (0,4);
% x axis
%\foreach \x in {-5,-4,-3,-2,-1,1,2,3,4,5} \draw[green] (\x,-4) -- (\x,4);
% y axis
%\foreach \y in {-4,-3,-2,-1,1,2,3,4} \draw[green] (-6,\y) -- (6,\y);
%%% NODES
\node[info, fill=white] (n_1) at (-4,2.5) {};
\node[info] (n_2) at (-2,2.5) {};
\node[info] (n_3) at (0,2.5) {};
\node[info] (n_4) at (2,2.5) {};
%%% TREE ARROWS
\draw [->] (n_1) -- (n_2);
\draw [->] (n_2) -- (n_3);
\draw [->] (n_3) -- (n_4);
%%% PAYOFFS ARROWS
\draw[->]  (n_1) -- (-4,0.5);
\draw[->]  (n_2) -- (-2,0.5);
\draw[->]  (n_3) -- (0,0.5);
\draw[->]  (n_4) -- (2,0.5);
\draw[->]  (n_4) -- (3.8,2.5);
%%% PLAYERS
% Player 1
\draw (-4.15, 2.8) node {$\Ann$};
\draw (-0.2, 2.8) node {$\Ann$};
% Player 2
\draw (-2, 2.8) node {$\Bob$};
\draw (2, 2.8) node {$\Bob$};
%%% PAYOFFS
\draw (-4,0.2) node {$\mathrm{z}_1$};
\draw (-2,0.2) node {$\mathrm{z}_2$};
\draw (0,0.2) node {$\mathrm{z}_3$};
\draw (2,0.2) node {$\mathrm{z}_4$};
\draw (4.3,2.5) node {$\mathrm{z}_5$};
%%% ACTIONS
% Player 1
\draw (-3, 2.7) node {$B$};
\draw (-4.2, 1.5) node {$A$};
\draw (1, 2.7) node {$F$};
\draw (-0.2, 1.5) node {$E$};
% Player 2
\draw (-1, 2.7) node {$D$};
\draw (-2.2, 1.5) node {$C$};
\draw (3, 2.7) node {$H$};
\draw (1.8, 1.5) node {$G$};
\end{tikzpicture}
\caption{A $4$-legged Centipede Game.}
\label{fig:centipede}
\end{figure}
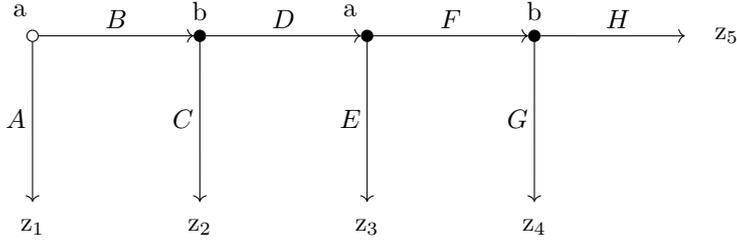

\noindent To see the intuition behind the definitions above, let Bob's preferences over terminal histories be captured by the following ordering: $\mathrm{z}_4 \succ_\Bob \mathrm{z}_2 \succ_\Bob \mathrm{z}_3 \succ_\Bob \mathrm{z}_1 \sim_\Bob \mathrm{z}_5$. Also, let $s^{*}_\Bob := DG$ and let $R := \{ A \} \times S_\Bob$ be a restriction. Clearly, for every $h \in H_\Bob (s^{*}_\Bob)$, $R_\Ann (h)$ is empty. Thus, our definition asks for a utility function and an \emph{arbitrary} CPS  such that $DG$ satisfies \Mref{eq:rational_given_gen}. On the contrary, if the restriction $R$ is defined as $R := \{ A , BE\} \times S_\Bob$, since  at $\la B \ra$ we  have that $R_\Ann (\la B \ra) = \{ BE \}$, the definition asks for a utility function and a CPS $\mu_\Bob$ such that $\mu_\Bob (R_\Ann (\la B \ra) | S_\Ann (\la B \ra)) = 1$, which---in this specific case---would correspond to a CPS $\mu_\Bob$ such that $\mu_\Bob (BE | S_\Ann (\la B \ra)) = 1$. Of course, in such an instance, $s^{*}_\Bob$ (i.e., $DG$) is not sequentially rational given $R$, because it does not satisfy (subjected) expected utility maximization with respect to the CPS $\mu_\Bob$ defined above.
\end{example}

%%%%%%%%%%%%%%%%%%%%%%%%%%%%%%%%%%%%%%%%%%
%%%%%%%%%%%%%%%%%%%%%%%%%%%%%%%%%%%%%%%%%%
%%%%%%%%%%%%%%%%%%%%%%%%%%%%%%%%%%%%%%%%%%
%%%%%%%%% SUB-Section %%%%%%%%%%%%%%%%%%%%%%%%%%%%%%%%%%%%%%%%%%
%%%%%%%%%%%%%%%%%%%%%%%%%%%%%%%%%%%%%%%%%%
%%%%%%%%%%%%%%%%%%%%%%%%%%%%%%%%%%%%%%%%%%
\subsection{Dominance Notions for the Extensive Form Representation} 
\label{subsec:dominance_extensive}

The dominance notions we introduced in \Sref{subsec:dominance_strategic} work at the level of strategies without taking into account the sequential nature of a dynamic game, i.e., they do not take into consideration the extensive form representation of a dynamic game. The notion we provide next is---on the contrary---phrased specifically to address the presence of information sets.

\begin{definition}[Conditional B-Dominance]
\label{def:conditional_B-dominance}
Given a restriction $R \subseteq S$ and a player $i \in I$, strategy $s_i \in R_i$ is \emph{conditionally B-dominated with respect to $R$} if there exists an information set $h \in H_i (s_i)$ such that $R^i (h)$ is nonempty and $s_i \in \bd_i (R^i (h))$.
\end{definition}

\begin{notation}[Set of Strategies Not Conditionally B-Dominated]
\label{not:conditional_B-dominance}
Given a restriction $R \subseteq S$ and a player $i \in I$, we let $\Und_i (R)$ denote the set of strategies of player $i \in I$ that are not conditionally B-dominated with respect to $R$.
\end{notation}

As for the dominance notions introduced in \Sref{subsec:dominance_strategic}, the omission of the restriction with respect to which conditional B-dominance is defined stands for that restriction being equal to $S$.

\begin{example}[continues=ex:main_example, name=Conditional B-Dominance at Work]
To see the definition of Conditional B-Dominance at work, we take the dynamic game in \Sref{fig:example_ordinal_basic} and---as before---we focus only on Ann's preferences over terminal histories. From \Mref{def:conditional_B-dominance}, we have to focus on the information sets where Ann is active, which---incidentally---in this specific case are all the information sets present in the game. Hence, we take Ann's conditional problems $S^\Ann (\la \varnothing \ra)$ and  $S^\Ann (\la I \ra)$, with  \Sref{fig:example_ordinal_decision} providing the corresponding graphical representation. 

\begin{figure}[H]
\centering
\begin{tikzpicture}
%%% NODE STYLE
[info/.style={circle, draw, inner sep=1.5, fill=black},
scale=1.1] 
%%% WORKING GRID
% Basic
%\draw[step=.2cm, gray, very thin] (-6,-8) grid (6,8);
%\draw[red] (-6,0) -- (6,0);
%\draw[red] (0,-8) -- (0,8);
% x axis
%\foreach \x in {-5,-4,-3,-2,-1,1,2,3,4,5} \draw[green] (\x,-8) -- (\x,8);
% y axis
%\foreach \y in {-8,-7,-6,-5,-4,-3,-2,-1,1,2,3,4,5,6,7,8} \draw[green] (-6,\y) -- (6,\y);
%%% NODES
%\node[info] (n_1) at (0,4.5) {};
%%%% PAYOFFS ARROWS
%\draw[->] (n_1) -- (-1.8,3.5);
%\draw[->] (n_1) -- (2,3.5);
%%%% PLAYERS
%% Player 1
%\draw (-0.2,4.8) node {$a$};
%%%% PAYOFFS
%\draw (-1.8,3.1) node {$2$};
%
\draw (-2,0) node {
\begin{game}{3}{2}[$\Ann$][$\Bob$]
		& $L$ 	& $R$ 	\\
$O$ 	& $\mathrm{z}_1$ 	& $\mathrm{z}_1$ 	\\
$IT$ 	& $\mathrm{z}_2$ 	& $\mathrm{z}_3$ 	\\
$ID$ 	& $\mathrm{z}_4$ 	& $\mathrm{z}_5$ 	\\
\end{game}};
\draw (2,0.2) node {
\begin{game}{2}{2}[$\Ann$][$b$]
		& $L$ 	& $R$ 	\\
$IT$ 	& $\mathrm{z}_2$ 	& $\mathrm{z}_3$ 	\\
$ID$ 	& $\mathrm{z}_4$ 	& $\mathrm{z}_5$ 	\\
\end{game}};
%%% ACTIONS
% Player 1
\draw (-2.7, 1.3) node {$S^\Ann (\la \varnothing \ra)$};
\draw (1.3, 1.3) node {$S^\Ann (\la I \ra)$};
\end{tikzpicture}
\caption{A representation of Ann's conditioning events in the game in \Sref{fig:example_ordinal_basic}.}
\label{fig:example_ordinal_decision}
\end{figure}

\noindent Observe that only Ann is active at $\la \varnothing \ra$, hence the matrix that corresponds to $S (\la \varnothing \ra)$ is relevant \emph{only} for Ann (i.e., we have that $S^\Ann (\la \varnothing \ra)$ is the conditional problem of Ann at $\la \varnothing \ra$), while both players are active at $\la I \ra$ and the matrix that corresponds to $S (\la I \ra)$ is relevant for both (i.e., we have that $S^\Ann (\la I \ra)$ and $S^\Bob (\la I \ra)$ are the conditional problems of the players at $\la I \ra$). From the analysis we performed of this game (in its strategic form) in \Sref{subsec:dominance_strategic}, we know that strategy $ID$ is B-dominated. This can be rephrased by stating that strategy $ID \in \bd_\Ann (S^\Ann (\la \varnothing \ra)$. Hence, strategy $ID \notin \Und_\Ann (S)$.
\end{example}

The following---crucial in this work---lemma establishes a link between Conditional B-Dominance and Weak Dominance and, in doing so, a connection between the extensive form and the strategic form representation of a dynamic game.

\begin{lemma}[Conditional B-Dominance \& Admissibility]
\label{lem:fundamental_CBD_WD}
Given a dynamic game $\Game$, its strategic form representation $\Game^r$, a restriction $R \subseteq S$, and a player $i \in I$, if $s_i \in \Ad_i (R)$ in $\Game^r$, then $s_i \in \Und_i (R)$ in $\Game$.
\end{lemma}

%%%%%%%%%%%%%%%%%%%%%%%%%%%%%%%%%%%%%%%%%%
%%%%%%%%%%%%%%%%%%%%%%%%%%%%%%%%%%%%%%%%%%
%%%%%%%%%%%%%%%%%%%%%%%%%%%%%%%%%%%%%%%%%%
%%%%%%%%% SUB-Section %%%%%%%%%%%%%%%%%%%%%%%%%%%%%%%%%%%%%%%%%%
%%%%%%%%%%%%%%%%%%%%%%%%%%%%%%%%%%%%%%%%%%
%%%%%%%%%%%%%%%%%%%%%%%%%%%%%%%%%%%%%%%%%%
\subsection{Characterization} 
\label{subsec:characterization}

In this section we answer the following question: given that---as outside observers on a dynamic interaction---we assume on our side only knowledge of players' ordinal preferences over outcomes and we actually observe the outcome of a dynamic game, when can we say if the players involved in the interaction acted according to sequential rationality?

With respect to this question we provide an answer  that we decline in two fashions that are intimately linked: indeed, where the first one is game-theoretical and phrased in the game-theoretical language set forth in the previous sections, the second one takes a decision-theoretic stance in the spirit of the revealed preference methodology and shows the `inner' decision-theoretic nature of previous answer.

%%%%%%%%%%%%%%%%%%%%%%%%%%%%%%%%%%%%%%%%%%
%%%%%%%%% SUB-Sub-Section %%%%%%%%%%%%%%%%%%%%%%%%%%%%%%%%%%%%%%%%%%
\subsubsection{Game-Theoretic Take} 
\label{subsubsec:GT_take}

The game-theoretical answer to our question builds on the newly introduced notions of sequential rationality given a restriction and conditional B-dominance given a restriction. Indeed, given these two notions, we can now proceed with a characterization of the former via the latter.

\begin{lemma}
\label{lem:given_rationality_conditional}
Given a restriction $R \subseteq S$ and a player $i \in I$, a strategy $s^{*}_i \in R_i$ is sequentially rational given $R$ if and only if it is not conditionally B-dominated with respect to  $R$.
\end{lemma}

We are now in position to introduce the notion of sequential rationality that lies at the core of this work, which further demonstrates that the building block of the present endeavour is indeed \Mref{def:rational_given} in light of the fact that it obtains immediately from that definition by setting the restriction $R := S$.

\begin{definition}[Sequentially Rational Strategy]
\label{def:rational}
Given a player $i \in I$, a strategy $s^{*}_i \in S_i$ is \emph{sequentially rational} if there exist a utility function $u_i \in \Re^Z$ and a CPS $\mu_i \in \Delta^{\mcal{H}_i} (S_{-i})$ such that 
\begin{equation*}
%\label{eq:rational} 
\sum_{s_{-i} \in S_{-i}} u_i (\zeta(s^{*}_i , s_{-i})) \cdot \mu_i (\{ s_{-i} \} | S_{-i} (h) ) \geq %
\sum_{s_{-i} \in S_{-i}} u_i (\zeta(s_i , s_{-i})) \cdot \mu_i (\{ s_{-i} \} | S_{-i} (h) )
\end{equation*}
for every $h \in H_i (s^{*}_i)$ and for every $s_i \in S_i (h)$.
\end{definition}

In the result that follows we show that a strategy that is sequentially rational according to \Mref{def:rational} cannot be conditionally B-dominated. Thus, this result links our notion of sequential rationality to a dominance notion based solely on \emph{pure} strategies that is employed in the extensive form representation of a dynamic game. 

\begin{theorem}
\label{th:rationality}
Given a player $i \in I$, a strategy $s^{*}_i \in S_i$ is sequentially rational if and only if it is not conditionally B-dominated.
\end{theorem}

%%%%%%%%%%%%%%%%%%%%%%%%%%%%%%%%%%%%%%%%%%
%%%%%%%%% SUB-Sub-Section %%%%%%%%%%%%%%%%%%%%%%%%%%%%%%%%%%%%%%%%%%
\subsubsection{Decision-Theoretic Take} 
\label{subsubsec:DT_take}

It is possible to provide an interpretation of our dominance characterization in \Mref{th:rationality} as a complete class theorem in the spirit of the analysis made in \citet[Chapter 8.3]{Chambers_Echenique_2016} of \citet[Proposition, p.427]{Borgers_1993} along the lines of the seminal \citet[Theorem 2.2, p.183]{Wald_1949} from the statistics literature.\footnote{See also \citet[Chapter 8.4.1]{Berger_1985} and
\citet[Chapter 8.4]{Parmigiani_Inoue_2009}.} In particular, we identify those properties that need to be satisfied by a preference relation so that, assuming that we have access to one---and only---one observation, we say that the observed outcome is compatible with the decision makers acting according to sequential rationality.

Given a player $i \in I$, a restriction $R \subseteq S$, and a strategy $s_i \in R_i$, we let $\la s_i , R \ra$ denote a \emph{preference  pair}, where the intuition behind this notion is that $s_i \in R_i$ is the strategy preferred by player $i \in I$ given the restriction $R$. Thus, given a rational preference relation $\succsim_i$:
\begin{itemize}[leftmargin=*]
\item a preference pair $\la s_i , R \ra$ is \emph{rationalized by} $\succsim_i$ if $\zeta (s_i , s_{-i}) \succsim_i \zeta (\underline{s}_i , s_{-i})$, for every $\underline{s}_i \in R_i$;

\item a preference pair $\la s_i , R \ra$ is \emph{rationalized by} $\succ_i$ if $\zeta (s_i , s_{-i}) \succ_i \zeta (\underline{s}_i , s_{-i})$, for every $\underline{s}_i \in R_i \setminus \{s_i\}$.
\end{itemize}

To introduce the crucial condition needed to establish the result we are after, for every player $i \in I$, we introduce a rational preference relation $\CDTeq_i$ to take into account the conditional problems of player $i$. Thus, we let $\CDTeq_i \subseteq S_i (h) \times S_i (h)$ denote the rational preference relation of player $i$ over her own strategies at her conditional problem $S^i (h)$, for every $h \in H_i$, with $\prefeq_i := (\CDTeq_i)_{h \in H_i}$, where the symmetric and asymmetric parts are defined as usual and $\prefeq^{h}_i\!\!\!|_{R}$ denotes the corresponding preference relation restricted to a restriction $R \subseteq S$.\footnote{See \Sref{subsubsec:DT_interpretation} for an analysis of this condition along with its relation to the literature and \Sref{parapp:DT_take} for a  presentation of the decision-theoretic framework behind this definition.}

\begin{definition}[Conditional Monotonicity Given a Restriction]
\label{def:conditional_monotonicity}
Given a restriction $R \subseteq S$, a player $i \in I$, a nonempty conditional problem $R^i (h)$, and strategies $\overline{s}_i,  \underline{s}_i \in R_i$:
\begin{itemize}[leftmargin=*]
\item (Weak Condition) if $\zeta (\overline{s}_i , s_{-i}) \succsim_i \zeta (\underline{s}_i , s_{-i})$ for every $s_{-i} \in R_{-i} (h)$, then $\overline{s}_i \prefeqhR \underline{s}_i$;

\item (Strong Condition) if $\zeta (\overline{s}_i , s_{-i}) \succsim_i \zeta (\underline{s}_i , s_{-i})$ for every $s_{-i} \in R_{-i} (h)$ and there exists an $\overline{s}_{-i} \in R_{-i} (h)$ such that $\zeta (\overline{s}_i , \overline{s}_{-i}) \succ_i \zeta (\underline{s}_i , \overline{s}_{-i})$, then $\overline{s}_i \prefeqhR \underline{s}_i$.
\end{itemize}
\end{definition}

Building on the notion just introduced, a rational preference relation $\prefeqR := \big(\!\! \prefeq^{h}_i\!\!\!|_{R} \big)_{h \in H_i}$ \emph{satisfies conditional monotonicity}, denoted $\CMsim|_{R}$, if it satisfies \Mref{def:conditional_monotonicity}, whereas it satisfies \emph{conditional subjective expected utility}, denoted $\CSEUsimR$, if it satisfies \Mref{def:rational_given}. Thus, we can now state the main result of this section, which is essentially in the spirit of the revealed preference methodology.

\begin{proposition}
\label{prop:DT_take}
Given a restriction $R \subseteq S$, a player $i \in I$, and a strategy $s^{*}_i \in R_i$, the preference pair $\la s^{*}_i , R \ra$ is rationalized by a rational preference relation $\CMsim|_{R}$ if and only if it is rationalized by a rational preference relation $\CSEUsimR$. 
\end{proposition}

%%%%%%%%%%%%%%%%%%%%%%%%%%%%%%%%%%%%%%%%%
%%%%%%%%%%%%%%%%%%%%%%%%%%%%%%%%%%%%%%%%%
%%%%%%%%%%%%%%%%%%%%%%%%%%%%%%%%%%%%%%%%%
%%%%%%%%%%%%%%%%%%%%%%%%%%%%%%%%%%%%%%%%%
%%%%%%%%%%%%%%%%%%%%%%%%%%%%%%%%%%%%%%%%%
%%%%%%%% SECTION %%%%%%%%%%%%%%%%%%%%%%%%%
%%%%%%%%%%%%%%%%%%%%%%%%%%%%%%%%%%%%%%%%%
%%%%%%%%%%%%%%%%%%%%%%%%%%%%%%%%%%%%%%%%%
%%%%%%%%%%%%%%%%%%%%%%%%%%%%%%%%%%%%%%%%%
%%%%%%%%%%%%%%%%%%%%%%%%%%%%%%%%%%%%%%%%%
%%%%%%%%%%%%%%%%%%%%%%%%%%%%%%%%%%%%%%%%%
\section{Revealed Forward Induction}
\label{sec:revealed_FI}

In this section we investigate what are the game-theoretical implications of  having players reasoning about each other's behavior in a context where only ordinal preferences over terminal histories are assumed to be transparent. 

To accomplish the task set forth above, we start by providing an abstract definition of the notion of reduction operator. Hence,  a \emph{reduction operator} is a mapping $\mbb{O}$ such that
\begin{equation*}
R \mapsto \mbb{O} (R) \subseteq R
\end{equation*}
for every restriction $R \subseteq S$. Given a reduction operator $\mbb{O}$ and two restrictions $Q$ and  $R$, $Q$ is a \emph{partial reduction} of $R$ if $\mbb{O} (R) \subseteq Q \subseteq R$, while $Q$ is a \emph{full reduction} of $R$ if $Q = \mbb{O} (R)$. An operator $\mbb{O}$ that gives as output a full reduction of $R$, for every restriction $R \subseteq S$, is called a \emph{full reduction} operator.

%%%%%%%%%%%%%%%%%%%%%%%%%%%%%%%%%%%%%%%%%%
%%%%%%%%%%%%%%%%%%%%%%%%%%%%%%%%%%%%%%%%%%
%%%%%%%%%%%%%%%%%%%%%%%%%%%%%%%%%%%%%%%%%%
%%%%%%%%% SUB-Section %%%%%%%%%%%%%%%%%%%%%%%%%%%%%%%%%%%%%%%%%%
%%%%%%%%%%%%%%%%%%%%%%%%%%%%%%%%%%%%%%%%%%
%%%%%%%%%%%%%%%%%%%%%%%%%%%%%%%%%%%%%%%%%%
\subsection{Iterating Conditional B-Dominance}
\label{subsec:solution_concepts}

%%%%%%%%%%%%%%%%%%%%%%%%%%%%%%%%%%%%%%%%%%
%%%%%%%%% SUB-Sub-Section %%%%%%%%%%%%%%%%%%%%%%%%%%%%%%%%%%%%%%%%%%
\subsubsection{Iterative Conditional B-Dominance} 
\label{subsubsec:ICBD}

We now introduce a recursive procedure that works by iteratively eliminating strategies that are conditionally B-dominated.  Thus, focusing on Conditional B-Dominance and building on \Mref{not:conditional_B-dominance}, for every restriction $R \subseteq S$, we let $\Und_{i} (R)$ denote the full reduction operator applied to $R$ that captures those player $i$'s strategies $s_i \in R_i$ that are not conditionally B-dominated with respect to $R$, i.e.,
\begin{equation}
\label{eq:U_operator}
\Und_{i} (R) := \Set { s^{*}_i \in R_i |  \forall h \in H_i (s^{*}_i) \ %
\big(  R^i (h) \neq \emptyset \Lra s^{*}_i \notin \bd_i (R^i (h))  \big) },
\end{equation}
with $\Und (R) := \prod_{j \in I} \Und_{j} (R)$.

\begin{remark}[Operator Nestedness]
\label{rem:operator_monotonicity}
Given a restriction $R \subseteq S$, by definition $\Und (R) \subseteq R$.
\end{remark}

The operator $\Und$ satisfies a property which is crucial for the proofs of the results that follow, namely, it does satisfy nonemptiness: it never gives the empty set as an output.

\begin{lemma}[Operator Nonemptiness]
\label{lem:basic_nonemptiness}
Given a restriction $R \subseteq S$, $\Und (R) \neq \emptyset$.
\end{lemma}

We can now introduce the iterative elimination procedure under scrutiny in this paper, which can be seen as an analog of Iterated Conditional Dominance of \cite{Shimoji_Watson_1998} for our class of dynamic games (with ordinal preferences).

\begin{algorithm}[Iterative Conditional B-Dominance (ICBD)]
\label{def:ICBD}
Given a dynamic game $\Game$ and a restriction $R \subseteq S$, consider the following procedure, with $n \in \bN$: 
\begin{itemize}[leftmargin=*]
\item (Step $n=0$) Let $\Und^{0} (R) := R$;
\item (Step $n \geq 1$) assuming that $\Und^{n-1} (R)$  has been defined, let 
\begin{equation*}
\Und^{n} (R) := \Und ( \Und^{n-1} (R)).
\end{equation*}
\end{itemize}
Thus, for every $k \in \bN$, $\Und^{k} (S)$ denotes the set of strategy profiles that survive the $k$-th iteration of the Iterative Conditional B-Dominance (henceforth, ICBD). Finally, 
\begin{equation*}
\Und^{\infty} (S) := \bigcap_{\ell \geq 0} \Und^{\ell} (S)
\end{equation*}
is the \emph{set of strategy profiles that survive the ICBD algorithm}.
\end{algorithm}

To see \Mref{def:ICBD} at work, we take the dynamic game in \Sref{fig:example_ordinal_basic} and we redraw it by representing the players' preferences over terminal histories via ordinal utility functions.  The resulting dynamic game is a workhorse of the literature on dynamic games (in general) and on its part which focuses on the study of forward induction (in particular).

\begin{example}[continues=ex:main_example, name=ICBD at Work]
We take the dynamic game in \Sref{fig:example_ordinal_basic} with two players  and we set numbers to represent the ordinal preferences of both players to ease the analysis. The resulting dynamic game is the so-called Battle of the Sexes with an Outside Option.

\begin{figure}[H]
\centering
\begin{tikzpicture}
%%% NODE STYLE
[info/.style={circle, draw, inner sep=1.5, fill=black},
scale=1.1] 
%%% WORKING GRID
% Basic
%\draw[step=.2cm, gray, very thin] (-6,-8) grid (6,8);
%\draw[red] (-6,0) -- (6,0);
%\draw[red] (0,-8) -- (0,8);
% x axis
%\foreach \x in {-5,-4,-3,-2,-1,1,2,3,4,5} \draw[green] (\x,-8) -- (\x,8);
% y axis
%\foreach \y in {-8,-7,-6,-5,-4,-3,-2,-1,1,2,3,4,5,6,7,8} \draw[green] (-6,\y) -- (6,\y);
%%% NODES
\node[info, fill=white] (n_1) at (0,4.5) {};
%%% PAYOFFS ARROWS
\draw[->] (n_1) -- (-1.8,3.5);
\draw[->] (n_1) -- (2,3.5);
%%% PLAYERS
% Player 1
\draw (-0.2,4.8) node {$\Ann$};
%%% PAYOFFS
\draw (-1.8,3.1) node {$2, 2$};
\draw (2,2.5) node {
\begin{game}{2}{2}[$\Ann$][$\Bob$]
		& $L$ 	& $R$ 	\\
$T$ 	& $3 ,1$ 	& $0, 0$ 	\\
$D$ 	& $0, 0$ 	& $1, 3$ 	\\
\end{game}};
%%% ACTIONS
% Player 1
\draw (-1.1, 4.2) node {$O$};
\draw (1.1, 4.2) node {$I$};
\end{tikzpicture}
\caption{Battle of the Sexes with an Outside Option.}
\label{fig:BoS_outside}
\end{figure}
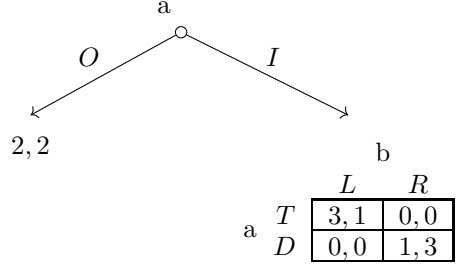

\noindent We proceed step by step to obtain $\Und^\infty (S)$ for the game in \Sref{fig:BoS_outside}.
\begin{itemize}[leftmargin=*]
\item ($n = 0$)  By definition we have that  $\Und^{0}_\Ann  (S) := S_a$ and  $\Und^{0}_\Bob (S) := S_\Bob$.
\item ($n = 1$) Strategy $ID$ is strictly dominated---hence, B-dominated---with respect to Ann's conditional problem $S^\Ann (\la \varnothing \ra)$. Thus, $\Und^{1}_\Ann (S) = \{ O, IT \}$, while $\Und^{1}_\Bob (S) = \Und^{0}_\Bob (S)$.
\item ($n = 2$) Strategy $R$ is strictly dominated---hence, B-dominated---with respect to Bob's (only) conditional problem $S^\Bob (\la I \ra)$ given $\Und^{1} (S)$. Hence, $\Und^{2}_\Ann (S) = \Und^{1}_\Ann (S)$, while $\Und^{2}_\Bob (S) = \{ L \}$.
\item ($n = 3$) Strategy $O$ is strictly dominated---hence, B-dominated---with respect to Ann's conditional problem $S^\Ann (\la \varnothing \ra)$ given $\Und^{2} (S)$. Hence, $\Und^{3}_\Ann (S) = \{ IT \}$, while $\Und^{3}_\Bob (S) =\Und^{2}_\Bob (S)$.
\item ($n \geq 4$) Nothing changes and the procedure ends. 
\end{itemize} 
Thus, we have that $\Und^{\infty}_\Ann (S) = \{ IT \}$ and $\Und^{\infty}_\Bob (S) = \{ L \}$ and---obviously---we have that $\Und^{\infty} (S) = \{ (IT, L ) \}$.
\end{example}

The next result establishes the nonemptiness of ICBD,  a rather basic property that any solution concept should satisfy. 

\begin{proposition}[Nonemptiness]
\label{prop:nonemptiness}
Given a dynamic game $\Game$:
\begin{itemize}
\item[i)] for every $k \in \bN$, $\Und^{k} (S) \neq \emptyset$, i.e., $\Und^{\infty} (S) \neq \emptyset$;

\item[ii)] there exists a $K \in \bN$ such that $\Und^{K} (S) = \Und^{K+1} (S) = \Und^{\infty} (S) \neq \emptyset$.
\end{itemize}
\end{proposition}

We have that an operator satisfies \emph{monotonicity} whenever, given restrictions $Q$ and $R$, if $Q \subseteq R$, then $\Und (Q) \subseteq \Und (R)$. It turns out that ICBD does not satisfy monotonicity, a point related to the fact that this algorithm captures forward induction reasoning that is established next.\footnote{See \cite{Perea_2018} and \cite{Catonini_2020} regarding the lack of monotonicity of solution concepts linked to forward induction reasoning.}

\begin{example}[continues=ex:main_example, name=Non-Monotonicity]
We consider the game in \Sref{fig:example_ordinal_basic} and we let $Q := (S_\Ann \setminus \{O\}) \times S_\Bob$ and $R := S_\Ann \times S_\Bob$. It is immediate to establish that $\Und (Q) = Q$, whereas we already showed that $\Und (R) = (S_\Ann \setminus \{ID\}) \times S_\Bob$. Hence, 
\begin{equation*}
\Und (Q) = (S_\Ann \setminus \{O\}) \times S_\Bob \not\subseteq (S_\Ann \setminus \{ID\}) \times S_\Bob = \Und (R),
\end{equation*}
since $ID \in \Und_\Ann (Q)$ and $ID \notin \Und_\Ann (R)$.  
\end{example}

The example that follows shows that ICBD does not satisfy \emph{order independence}, i.e., our algorithm is not robust to the order in which strategies are eliminated. In other words, different elimination orders can deliver different outcomes.

\begin{example}[Order Dependence]
\label{ex:order_dependence}
To see that ICBD does not satisfy order independence, we take the dynamic game in \Sref{fig:example_Perea} from \citet[Figure 4, p.247]{Perea_2014} and \citet[Example 4, p.138]{Chen_Micali_2013} with two players, again Ann (viz., $\Ann$) and Bob (viz., $\Bob$). 

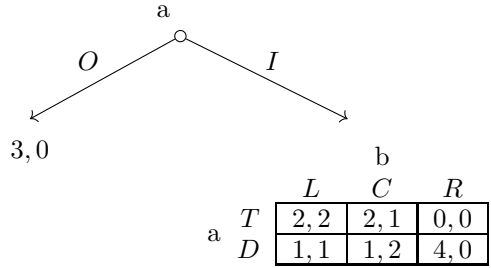
\begin{figure}[H]
\centering
\begin{tikzpicture}
%%% NODE STYLE
[info/.style={circle, draw, inner sep=1.5, fill=black},
scale=1.1] 
%%% WORKING GRID
% Basic
%\draw[step=.2cm, gray, very thin] (-6,-8) grid (6,8);
%\draw[red] (-6,0) -- (6,0);
%\draw[red] (0,-8) -- (0,8);
% x axis
%\foreach \x in {-5,-4,-3,-2,-1,1,2,3,4,5} \draw[green] (\x,-8) -- (\x,8);
% y axis
%\foreach \y in {-8,-7,-6,-5,-4,-3,-2,-1,1,2,3,4,5,6,7,8} \draw[green] (-6,\y) -- (6,\y);
%%% NODES
\node[info, fill=white] (n_1) at (0,4.5) {};
%%% PAYOFFS ARROWS
\draw[->] (n_1) -- (-1.8,3.5);
\draw[->] (n_1) -- (2,3.5);
%%% PLAYERS
% Player 1
\draw (-0.2,4.8) node {$\Ann$};
%%% PAYOFFS
\draw (-1.8,3.1) node {$3, 0$};
\draw (2,2.5) node {
\begin{game}{2}{3}[$\Ann$][$\Bob$]
		& $L$ 		& $C$ 		& $R$ 	\\
$T$ 	& $2, 2$ 	& $2, 1$ 	& $0, 0$\\
$D$ 	& $1, 1$ 	& $1, 2$ 	& $4, 0$\\
\end{game}};
%%% ACTIONS
% Player 1
\draw (-1.1, 4.2) node {$O$};
\draw (1.1, 4.2) node {$I$};
\end{tikzpicture}
\caption{A dynamic game showing order dependence.}
\label{fig:example_Perea}
\end{figure}

\noindent We have that $\Und^\infty (S) = \{ (O, C) \}$. On the contrary, if we construct an ancillary operator $\mbb{L}$ that on the first iteration deletes strategies only at the information set to which they belong and then for the remaining iterations works like $\Und$, we would have that such an operator would delete $IT$ from $S^\Ann (\la \varnothing \ra)$ \emph{only}, without deleting it \emph{also} from $S^\Ann (\la I \ra)$ on its first iteration and---as a result---we would obtain $\mbb{L}^{\infty} (S) = \{ (O, L) \}$.
\end{example}

%%%%%%%%%%%%%%%%%%%%%%%%%%%%%%%%%%%%%%%%%%
%%%%%%%%% SUB-Sub-Section %%%%%%%%%%%%%%%%%%%%%%%%%%%%%%%%%%%%%%%%%%
\subsubsection{Ordinal Strong Rationalizability} 
\label{subsubsec:strong_ordinal_rationalizability}

We now introduce a new full reduction operator, which is going to be the focus of this section. Thus, for every restriction $R \subseteq S$, we let $\OSR_{i} (R)$ denote the full reduction operator applied to $R$ that captures those player $i$'s strategies $s_i \in R_i$ that are sequentially rational given $R$ as in \Mref{def:rational_given}, with $\OSR (R) := \prod_{j \in I} \OSR_{j} (R)$. 

\begin{algorithm}[Ordinal Strong Rationalizability]
\label{def:ordinal_strong_rationalizability}
Given a dynamic game $\Game$ and a restriction $R \subseteq S$, consider the following procedure, with $n \in \bN$: 
\begin{itemize}[leftmargin=*]
\item (Step $n=0$) let $\OSR^{0} (R) := R$;
\item (Step $n \geq 1$) assuming that $\OSR^{n-1} (R)$  has been defined, let 
\begin{equation*}
\OSR^{n} (R) := \OSR ( \OSR^{n-1} (R)).
\end{equation*}
\end{itemize}
Thus, for every $k \in \bN$, $\OSR^{k} (S)$ denotes the set of strategy profiles that survive the $k$-th iteration of Ordinal Strong Rationalizability. Finally, 
\begin{equation*}
\OSR^{\infty} (S) := \bigcap_{\ell \geq 0} \OSR^{\ell} (S)
\end{equation*}
is the \emph{set of strategy profiles that survive Ordinal Strong Rationalizability}.
\end{algorithm}

Thus, Ordinal Strong Rationalizability as defined in \Mref{def:ordinal_strong_rationalizability}, as the name explicitly points out, is the ordinal version of Strong Rationalizability  where only ordinal preferences over outcomes are transparent to the players.\footnote{See also \citet[Footnote 9, p.50]{Battigalli_1997}.} Now, concerning its nonemptiness, we  answer this question in the next section.

%%%%%%%%%%%%%%%%%%%%%%%%%%%%%%%%%%%%%%%%%%
%%%%%%%%% SUB-Sub-Section %%%%%%%%%%%%%%%%%%%%%%%%%%%%%%%%%%%%%%%%%%
\subsubsection{Relation between the Procedures} 
\label{subsubsec:relation}

We now address the relation between ICBD and Ordinal Strong Rationalizability. In particular, the following two results show that the well-known relation between Iterated Conditional Dominance as an algorithmic counterpart of Strong Rationalizability applies \emph{mutatis mutandis} to ICBD and Ordinal Strong Rationalizability.

\begin{lemma}
\label{lem:ICBD_SR}
Given a dynamic game $\Game$ and a restriction $R \subseteq S$, $\Und (R) = \OSR (R)$.
\end{lemma}

\begin{proposition}
\label{prop:ICBD_SR}
Given a dynamic game $\Game$: 
\begin{itemize}
\item[i)] $\Und^{k} (S) = \OSR^{k} (S)$, for every $k \in \bN$;

\item[ii)] $\Und^{\infty} (S) = \OSR^{\infty} (S)$.
\end{itemize}
\end{proposition}

Thus, we have the nonemptiness of Ordinal Strong Rationalizability as an immediate corollary of \Mref{prop:ICBD_SR}, stated next.

\begin{corollary}[Nonemptiness of Ordinal Strong Rationalizability]
\label{cor:nonemptiness}
For every dynamic game $\Game$, for every $k \in \bN$, $\OSR^{k} (S) \neq \emptyset$, i.e., $\OSR^{\infty} (S) \neq \emptyset$. In particular, for every dynamic game there exists a $K \in \bN$ such that $\OSR^{K} (S) = \OSR^{K+1} (S) = \OSR^{\infty} (S) \neq \emptyset$.
\end{corollary}

%%%%%%%%%%%%%%%%%%%%%%%%%%%%%%%%%%%%%%%%%%
%%%%%%%%%%%%%%%%%%%%%%%%%%%%%%%%%%%%%%%%%%
%%%%%%%%%%%%%%%%%%%%%%%%%%%%%%%%%%%%%%%%%%
%%%%%%%%% SUB-Section %%%%%%%%%%%%%%%%%%%%%%%%%%%%%%%%%%%%%%%%%%
%%%%%%%%%%%%%%%%%%%%%%%%%%%%%%%%%%%%%%%%%%
%%%%%%%%%%%%%%%%%%%%%%%%%%%%%%%%%%%%%%%%%%
\subsection{Intermezzo: On Iterated Conditional Dominance \& Strong Rationalizability}
\label{subsec:relation_ICD}

In order to capture the revealed preference implications of forward induction reasoning via ICBD (or Ordinal Strong Rationalizability---for this purpose, \Mref{prop:ICBD_SR} shows the two are equivalent) we need a solution concept that is based on the existence of a profile of Bernoulli utility functions that actually captures forward induction. For this purpose, we now introduce two---tightly linked---solution concepts, which are the `cardinal' counterparts of ICBD and Ordinal Strong Rationalizability, namely, Iterated Conditional Dominance of \cite{Shimoji_Watson_1998} and Strong Rationalizability of \cite{Pearce_1984} and \cite{Battigalli_1997}.

To define these two solution concepts, we need additional notation. Thus, given a dynamic game $\Game$, we let $\Game^u$ denote a  \emph{dynamic cardinal game derived from} $\Game$, where $u := (u_j)_{j \in I}$ is a profile of Bernoulli utility functions and $u_i$ is originated from player $i$'s ordinal preferences, for every $i \in I$.\footnote{Thus, $\Game^u$ represents a dynamic game where us---as modelers---have `more' information than in \Mref{eq:dynamic_ordinal_game} concerning the players' preferences, since we know also their \emph{risk} preferences, that we are \emph{fixing} exogenously.} From $\Game^u$, we let $\overline{\Game}^u$ denote the \emph{(correlated) mixed extension} of the dynamic cardinal game $\Game^u$, where 
\begin{equation*}
U^{h}_i (\sigma_i , \mu_{i} ) := %
\sum_{s_{i} \in S_{i}} %
\sum_{s_{-i} \in S_{-i}} %
u_i (\zeta (s_i, s_{-i}) ) \cdot %
\sigma_i (\{s_{i}\}) \cdot %
\mu_i \Rounds { \{s_{-i} | S_{-i} (\la h \ra) },
\end{equation*}
is player $i$'s \emph{von Neumann--Morgenstern utility function at $h \in H_i$}, with $\sigma_i \in \Delta (S_i)$ being an arbitrary  \emph{mixed strategy} of player $i$ and $\mu_i \in \Delta^{\mcal{H}_i} (S_{-i})$ an arbitrary CPS on $S_{-i}$ (and  $U_j$ denoting the von Neumann--Morgenstern utility function at $\la \varnothing \ra \in H_j$ of a player $j \in I_{\la \varnothing \ra}$), where the function $U^{h}_{i}|_R$ corresponding to $U^{h}_{i}$ restricted to $R \subseteq S$ is naturally defined.

Given  a $\overline{\Game}^u$, a player $i \in I$,and a restriction $R \subseteq S$, a strategy $s_i \in R_i$ is \emph{strictly dominated relative to $R$ by a mixed strategy} $\sigma^{*}_i \in \Delta (R_i)$  if $U_i|_{R} (\sigma^{*}_i, s_{-i}) > U_{i}|_R (s_i, s_{-i})$, for every $s_{-i} \in R_{-i}$. Given a restriction $R \subseteq S$, we let $\md_i (R)$ denote the set of strategies of player $i \in I$ belonging to $R_i$ that are strictly dominated relative to $R$ by a mixed strategy. Also, we say that strategy $s^{*}_i \in R_i$ is \emph{sequentially rational given a restriction $R$ and a utility function $u_{i}|_R$} if there exists a CPS $\mu_i \in \Delta^{\mcal{H}_i} (R_{-i})$ such that
\begin{equation*}
s^{*}_i \in \arg \max_{s_i \in R_i} U^{h}_{i}|_R (s_i , \mu_i) .
\end{equation*}
for every $h \in H_i (s^{*}_i)$.

Now, building on the notation and terminology set forth above, we introduce two new full reduction operators. Thus, for every restriction $R \subseteq S$, we let $\ICD_{i} (R)$ denote the full reduction operator applied to $R$ that captures those player $i$'s strategies $s_i \in R_i$ that are not strictly dominated relative to $R$ by a mixed strategy at an information set $h \in H_i (s_i)$, i.e.,
\begin{equation*}
\label{eq:U_ICD_operator}
\ICD_{i} (R) := \Set { s^{*}_i \in R_i |  \forall h \in H_i (s^{*}_i) \ %
( R^i (h) \neq \emptyset \Lra s^{*}_i \notin \md_i (R^i (h)) )},
\end{equation*}
and we let $\ICD (R) := \prod_{j \in I} \ICD_{j} (R)$. As a result, \emph{Iterated Conditional Dominance} of \citet[Section 3, pp.170--171]{Shimoji_Watson_1998} is the algorithm defined as ICBD in \Mref{def:ICBD} with the operator $\Und$ substituted by the newly introduced operator $\ICD$. Also, for every restriction $R \subseteq S$, we let $\SR_{i} (R)$ denote the set of strategies $s_i \in R_i$ such that there exists a CPS $\mu_i \in \Delta^{\mcal{H}_i} (R_{-i})$ such that, for every $h \in H_i (s_i)$, the nonemptiness of $R^{i} (h)$ implies that $s^{*}_i$ is sequentially rational given $R$ and $u_i$, i.e., $\SR_{i} (R)$ denotes the full reduction operator defined as 
\begin{equation*}
\label{eq:SR_operator}
\SR_{i} (R) := %
\Set { s^{*}_i \in R_i | \exists \mu_i \in \Delta^{\mcal{H}_i} (R_{-i}) : \forall h \in H_i (s^{*}_i) \ %
\Rounds { %
R^i (h) \neq \emptyset \Lra s^{*}_i \in \arg \max_{s_i \in S_i} U^{h}_{i} \vert_R (s_i, \mu_i) %
} },
\end{equation*}
with $\SR (R) := \prod_{j \in I} \SR_{j} (R)$. As a result, \emph{Strong Rationalizability} of \citet[Definition 9, p.1042]{Pearce_1984} and \citet[Definition 2, p.46]{Battigalli_1997} is the algorithm defined as Ordinal Strong Rationalizability in \Mref{def:ordinal_strong_rationalizability} with the operator $\OSR$ substituted by the newly introduced operator $\SR$.

As pointed out above, Iterated Conditional Dominance is tightly linked to Strong Rationalizability as the following---well-known result in the literature---establishes.

\begin{remark}[{\citet[Theorem 3, p.177]{Shimoji_Watson_1998}}]
\label{rem:SH98}
Given a dynamic game $\Game$ with corresponding dynamic cardinal game in its mixed extension $\overline{\Game}^u$, $\ICD^k (S) = \SR^k (S)$, for every $k \in \bN$.
\end{remark}

Before actually tackling the problem at the core of this section, namely, if ICBD and Ordinal Strong Rationalizability can capture the revealed preference implications of forward induction reasoning by establishing a link with Iterated Conditional Dominance and Strong Rationalizability, here, we compare the predictions obtained via ICBD and Iterated Conditional Dominance\footnote{Of course, we could have picked also Ordinal Strong Rationalizability instead of ICBD and Strong Rationalizability instead of Iterated Conditional Dominance.} when a profile of utility functions $u := (u_j)_{j \in I}$ is fixed, with the caveat that in order to work with ICBD we treat $u$ as a profile of \emph{ordinal} utility functions, whereas to deal with Iterated Conditional Dominance we treat it as a profile of  \emph{cardinal} utility functions.\footnote{See \citet[Section 6]{Guarino_Ziegler_2022} for a similar take on the issue in a different context.}

Thus, given an arbitrary restriction $R \subseteq S$, the first question we want to ask is if $\Und (R) \subseteq \ICD (R)$. The answer is negative, as shown in the example that comes next.

\begin{example}[$\Und (R) \not\subseteq \ICD (R)$]
\label{ex:ICD}
To see that Iterated Conditional Dominance gives sharper predictions than ICBD, consider the dynamic game in \Sref{fig:dynamic_outside} with two players, again Ann (viz., $\Ann$) and Bob (viz., $\Bob$). 

\begin{figure}[H]
\centering
\begin{tikzpicture}
%%% NODE STYLE
[info/.style={circle, draw, inner sep=1.5, fill=black},
scale=1.1] 
%%% WORKING GRID
% Basic
%\draw[step=.2cm, gray, very thin] (-6,-8) grid (6,8);
%\draw[red] (-6,0) -- (6,0);
%\draw[red] (0,-8) -- (0,8);
% x axis
%\foreach \x in {-5,-4,-3,-2,-1,1,2,3,4,5} \draw[green] (\x,-8) -- (\x,8);
% y axis
%\foreach \y in {-8,-7,-6,-5,-4,-3,-2,-1,1,2,3,4,5,6,7,8} \draw[green] (-6,\y) -- (6,\y);
%%% NODES
\node[info, fill=white] (n_1) at (0,4.5) {};
%%% PAYOFFS ARROWS
\draw[->] (n_1) -- (-1.8,3.5);
\draw[->] (n_1) -- (2,3.5);
%%% PLAYERS
% Player 1
\draw (-0.2,4.8) node {$\Ann$};
%%% PAYOFFS
\draw (-1.8,3.1) node {$0, 0$};
\draw (2,2.5) node {
\begin{game}{3}{2}[$\Ann$][$\Bob$]
		& $L$ 	& $R$ 	\\
$T$ 	& $3 ,2$ 	& $0, 1$ 	\\
$M$	& $0, 1$	& $3, 0$	\\
$D$ 	& $1, 0$ 	& $1, 1$ 	\\
\end{game}};
%%% ACTIONS
% Player 1
\draw (-1.1, 4.2) node {$O$};
\draw (1.1, 4.2) node {$I$};
\end{tikzpicture}
\caption{Dynamic Game with an Outside Option with a fixed profile of utility functions.}
\label{fig:dynamic_outside}
\end{figure}
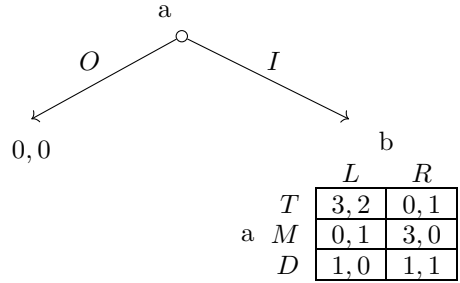

\noindent First of all, we have that $O \notin \mbb{U}^{1}_\Ann (S)$ and $O \notin \ICD^{1}_\Ann (S)$. Also, and rather crucially, we have that $D \in \Und^{1}_\Ann (S)$, but $D \notin \ICD^{1}_\Ann (S)$, since $D$ is strictly dominated relative to $S^\Ann (\la I \ra)$ by any mixed strategy $\sigma_\Ann ( \alpha(T), (1-\alpha)(M))$, with $\alpha \in (\frac{1}{3}, \frac{2}{3})$. As a result, we have $\Und^{\infty} (S) = (S_\Ann \setminus \{O\} ) \times S_\Bob$, while $\ICD^{\infty} (S) = \{ (T, L ) \}$.
\end{example}

Regarding the opposite direction, on the contrary, we can establish that $\ICD (R) \subseteq \Und (R)$ as stated in the following proposition.

\begin{proposition}[$\ICD (R) \subseteq \Und (R)$]
\label{prop:ICD_ICBD}
Given a dynamic game $\Game$, a dynamic cardinal game $\overline{\Game}^u$ in its mixed extension derived from it, and a restriction $R \subseteq S$, $\ICD (R) \subseteq \Und (R)$.
\end{proposition}

As a matter of fact, \Mref{prop:ICD_ICBD} is hardly surprising, since it essentially captures the natural idea that taking into account \emph{also} mixed strategies for (strict) dominance purposes is going to shrink the set of strategies that survive an elimination step.

%%%%%%%%%%%%%%%%%%%%%%%%%%%%%%%%%%%%%%%%%%
%%%%%%%%%%%%%%%%%%%%%%%%%%%%%%%%%%%%%%%%%%
%%%%%%%%%%%%%%%%%%%%%%%%%%%%%%%%%%%%%%%%%%
%%%%%%%%% SUB-Section %%%%%%%%%%%%%%%%%%%%%%%%%%%%%%%%%%%%%%%%%%
%%%%%%%%%%%%%%%%%%%%%%%%%%%%%%%%%%%%%%%%%%
%%%%%%%%%%%%%%%%%%%%%%%%%%%%%%%%%%%%%%%%%%
\subsection{Rationaliz(abilit)ing Forward Induction}
\label{subsec:rationalizing_FI}

The question we answer in this section is the following: given that we assume on our side as outside observers only knowledge of players' ordinal preferences over outcomes and we actually observe the players' choices in a dynamic game, when can we say that there exists a profile of utility functions $(u_j)_{j \in I}$ and a profile of CPSs $(\mu_j)_{j \in I}$ such that the choices are compatible with the players acting in a sequentially rational fashion given the profile of utilities $(u_j)_{j \in I}$ and according to forward induction reasoning?

Thus, with respect to the question stated above, we provide an answer that---as in \Sref{subsec:characterization}---we decline in two fashions: the first one  game-theoretical, with the second decision-theoretic in nature with a revealed preference flavour.

%%%%%%%%%%%%%%%%%%%%%%%%%%%%%%%%%%%%%%%%%%
%%%%%%%%% SUB-Sub-Section %%%%%%%%%%%%%%%%%%%%%%%%%%%%%%%%%%%%%%%%%%
\subsubsection{Game-Theoretic Take} 
\label{subsubsec:ICBD_GT_take}

To answer the question stated above, first of all, we ask ourselves  whether for every strategy $s_i$ of a player $i$ that is not conditionally B-dominated given a restriction $R \subseteq S$ there exists a utility function $u_i$ that agrees with the assumed ordinal preferences such that $s_i$ is not conditionally dominated (i.e., it does not survive an application $\ICD$ operator) given $R$.\footnote{We are grateful to Andr\'{e}s Perea for having originally raised this issue.} Thus, in the analysis that follows, we let $\Und_i (R) [\succsim_i]$ denote the $\Und$ operator applied on a restriction $R \subseteq S$ for a player $i \in I$ given a dynamic game where player $i$'s preferences are given by $\succsim_i$, whereas we let $\ICD_i (R) [u_i]$ (resp., $\SR_i (R) [u_i]$) denote the $\ICD_i$ operator (resp., the $\SR_i$ operator) applied on $R \subseteq S$ for $i \in I$ given a dynamic cardinal game where player $i$'s preferences are given by a Bernoulli utility function $u_i$ originated from $\succsim_i$.

As a matter of fact, given this setting, we can establish a relation between ICBD and Iterated Conditional Dominance.

\begin{lemma}
\label{lem:ICD_general}
Given a dynamic game $\Game$, a restriction $R \subseteq S$, and a player $i \in I$,
\begin{equation*}
\Und_i (R) [\succsim_i] = %
\bigcup_{u_i \in \Utilities_i} \ICD_i (R) [u_i].
\end{equation*}
\end{lemma}

\Mref{lem:ICD_general} is crucial for our purposes, as the result that follows shows.

\begin{proposition}
\label{prop:ICD_general}
Given a dynamic game $\Game$, for every $k \in \bN$ there exists a profile of utility functions $u := (u_j)_{j \in I}$ such that
\begin{equation*}
\Und^{k} (S) [\succsim] = %
\SR^{k} (S) [u^{k}] .
\end{equation*}
In particular, there exists a profile of utility functions $u^{\infty}$ such that
\begin{equation*}
\Und^{\infty} (S) [\succsim] = %
\SR^{\infty} (S) [u^{\infty}] .
\end{equation*}
\end{proposition}

Now, in light of \Mref{rem:SH98}, we obtain immediately the following result,  strictly speaking a corollary of \Mref{prop:ICD_general} that we state as a theorem since it answers exactly the question stated above and it captures one of the two---the other being the related one concerning sequential rationality---essential issues addressed in this paper, namely, the possibility of a choice-theoretic foundation of forward induction.

\begin{theorem}
\label{th:ICD_general}
Given a dynamic game $\Game$, for every $i \in I$ and $k \in \bN$ there exists a utility function $u_i \in \Utilities_i$ such that
\begin{equation*}
\Und^{k}_i (S) [\succsim_i] = %
\SR^{k}_i (S) [u^{k}_i] .
\end{equation*}
In particular, there exists a utility function $u^{\infty}_i \in \Utilities$ such that
\begin{equation*}
\Und^{\infty}_i (S) [\succsim_i] = %
\SR^{\infty}_i (S) [u^{\infty}_i] .
\end{equation*}
\end{theorem}

It turns out that, always having in mind the question set forth above, we can actually strengthen \Mref{th:ICD_general} by establishing a result which would read as  \Mref{th:ICD_general} with the caveat that \emph{one} utility function works for every iteration,\footnote{See \cite{Fishburn_1978} for a similar endeavour where the focus is on Nash equilibrium.} with the result being a  corollary of  \citet[Corollary 1, Section 5.1]{Gafarov_Salcedo_2015}.\footnote{See also \citet[Proposition 3, p1888]{Weinstein_2016} and Footnote 8 therein.} 

\begin{theorem}
\label{th:ICD_general_utility}
Given a dynamic game $\Game$, for every $i \in I$ there exists a utility function $u^{*}_i \in \Utilities_i$ such that 
\begin{equation*}
\Und^{k}_i (S) [\succsim_i] = %
\SR^{k}_i (S) [u^{*}_i] 
\end{equation*}
for every $k \in \bN$
\end{theorem}

%%%%%%%%%%%%%%%%%%%%%%%%%%%%%%%%%%%%%%%%%%
%%%%%%%%% SUB-Sub-Section %%%%%%%%%%%%%%%%%%%%%%%%%%%%%%%%%%%%%%%%%%
\subsubsection{Decision-Theoretic Take} 
\label{subsubsec:ICBD_DT_take}

We now perform an analog of the analysis in \Sref{subsubsec:GT_take} by employing the same notion used there of preference pair and, for this purpose, we identify a given solution concept with the operator it is built upon. Hence, given a solution concept $\mathbb{C} \in \Sets {\Und, \OSR, \ICD, \SR }$, a player $i \in I$, and a restriction $R \subseteq S$, a preference pair $\la s_i , R \ra$ is \emph{rationalized by} $\mathbb{C}$ if $s_i \in \mathbb{C}^{\infty}_i (R) \subseteq R_i$.

Given the notions set forth above along with the results established in \Sref{subsubsec:ICBD_GT_take}, the following result is immediate.

\begin{proposition}
\label{prop:ICBD_DT_take}
Given a player $i \in I$, a restriction $R \subseteq S$, and a strategy profile $s^{*} \in R$, the preference pair $\la s^{*}_i, R \ra$ is rationalized by $\Und$ if and only if it is rationalized by $\SR$. 
\end{proposition}

The result should be interpreted in the spirit of the revealed preference methodology: the fact that a strategy profile is preferred given a certain restriction assuming transparency of ordinal preferences and---in an informal sense---forward induction reasoning ensures us that there exists a profile of utility functions and CPSs such that the strategy profile belongs to the set of strongly rationalizable strategy profiles.\footnote{It should be observed that, much in the same spirit in which \citet[Corollary, p.732]{Lo_2000} and the preceding discussion relate \citet[Theorem, p.730]{Lo_2000} when put in a game-theoretical context to the results in \cite{Epstein_1997a}, the same could be done for \Mref{prop:ICBD_DT_take} modulo details.}

%%%%%%%%%%%%%%%%%%%%%%%%%%%%%%%%%%%%%%%%%
%%%%%%%%%%%%%%%%%%%%%%%%%%%%%%%%%%%%%%%%%
%%%%%%%%%%%%%%%%%%%%%%%%%%%%%%%%%%%%%%%%%
%%%%%%%%%%%%%%%%%%%%%%%%%%%%%%%%%%%%%%%%%
%%%%%%%%%%%%%%%%%%%%%%%%%%%%%%%%%%%%%%%%%
%%%%%%%% SECTION %%%%%%%%%%%%%%%%%%%%%%%%%
%%%%%%%%%%%%%%%%%%%%%%%%%%%%%%%%%%%%%%%%%
%%%%%%%%%%%%%%%%%%%%%%%%%%%%%%%%%%%%%%%%%
%%%%%%%%%%%%%%%%%%%%%%%%%%%%%%%%%%%%%%%%%
%%%%%%%%%%%%%%%%%%%%%%%%%%%%%%%%%%%%%%%%%
%%%%%%%%%%%%%%%%%%%%%%%%%%%%%%%%%%%%%%%%%
\section{The Backward Induction Outcome}
\label{sec:relevant_ties}

\Sref{subsec:relation_ICD} showed us that, when we fix a profile of utility functions $u$ and we treat it alternatively as ordinal or cardinal depending on the solution concept we want to use, ICBD cannot really be compared with its `cardinal' counterpart Iterated Conditional Dominance, since the two are based on different assumptions. However, there is an important class of games where---due to their nature---the behavioral predictions of these two solution concepts should really be the same, since risk preferences should not play any role in the analysis given the context provided by the class. Thus, the focus of this section is exactly on this class of games, i.e., generic\footnote{\label{foot:genericity}See \citet[Chapter 4.7, p.186]{Myerson_1991}.} dynamic games with perfect information, where the notion of genericity we employ is the following, which goes back to \citet[Section 4, p.48]{Battigalli_1997}:  a dynamic game with perfect information $\Game$ satisfies the \emph{``No Relevant Ties'' condition} (henceforth, NRT condition)  if, for every $z, z' \in Z$ and for every $i \in I$, if $z \neq z'$ and $i \in I_h$, where the information set $h \in H$ is the last common predecessor of $z$ and $z'$, then it is not the case that $z \sim_i z'$. In the rest of this section---and related proofs---a dynamic game that satisfies the NRT condition is called an \emph{NRT dynamic game} (where it has to be observed that such a construct is---\emph{a fortiori}---a dynamic game of perfect information).

\begin{example}[continues=ex:centipede, name=ICBD at Work in an NRT Dynamic Game]
We now take the game in \Sref{fig:centipede} and, as in \Sref{fig:BoS_outside}, we set numbers to represent the players' ordinal preferences. The result is the game in \Sref{fig:reny_centipede}, i.e., Reny's Centipede Game, introduced for the first time in \cite{Reny_1992}.\footnote{We depict this game by following the representation in \citet[Figure 2, p.271]{Battigalli_Siniscalchi_2002}, while the original version of the game can be found  in \citet[Figure 3, p.637]{Reny_1992}.} 

\begin{figure}[H]
\centering
\begin{tikzpicture}
%%% NODE STYLE
[info/.style={circle, draw, inner sep=1.5, fill=black},
scale=1.1] 
%%% WORKING GRID
% Basic
%\draw[step=.2cm, gray, very thin] (-6,-4) grid (6,4);
%\draw[red] (-6,0) -- (6,0);
%\draw[red] (0,-4) -- (0,4);
% x axis
%\foreach \x in {-5,-4,-3,-2,-1,1,2,3,4,5} \draw[green] (\x,-4) -- (\x,4);
% y axis
%\foreach \y in {-4,-3,-2,-1,1,2,3,4} \draw[green] (-6,\y) -- (6,\y);
%%% NODES
\node[info, fill=white] (n_1) at (-4,2.5) {};
\node[info] (n_2) at (-2,2.5) {};
\node[info] (n_3) at (0,2.5) {};
\node[info] (n_4) at (2,2.5) {};
%%% TREE ARROWS
\draw [->] (n_1) -- (n_2);
\draw [->] (n_2) -- (n_3);
\draw [->] (n_3) -- (n_4);
%%% PAYOFFS ARROWS
\draw[->]  (n_1) -- (-4,0.5);
\draw[->]  (n_2) -- (-2,0.5);
\draw[->]  (n_3) -- (0,0.5);
\draw[->]  (n_4) -- (2,0.5);
\draw[->]  (n_4) -- (3.8,2.5);
%%% PLAYERS
% Player 1
\draw (-4.15, 2.8) node {$\Ann$};
\draw (-0.2, 2.8) node {$\Ann$};
% Player 2
\draw (-2, 2.8) node {$\Bob$};
\draw (2, 2.8) node {$\Bob$};
%%% PAYOFFS
\draw (-4,0.2) node {$3, 0$};
\draw (-2,0.2) node {$1, 2$};
\draw (0,0.2) node {$2, 1$};
\draw (2,0.2) node {$0, 3$};
\draw (4.3,2.5) node {$4, 0$};
%%% ACTIONS
% Player 1
\draw (-3, 2.7) node {$B$};
\draw (-4.2, 1.5) node {$A$};
\draw (1, 2.7) node {$F$};
\draw (-0.2, 1.5) node {$E$};
% Player 2
\draw (-1, 2.7) node {$D$};
\draw (-2.2, 1.5) node {$C$};
\draw (3, 2.7) node {$H$};
\draw (1.8, 1.5) node {$G$};
\end{tikzpicture}
\caption{The $4$-legged Centipede Game from \cite{Reny_1992}.} \label{fig:reny_centipede}
\end{figure}
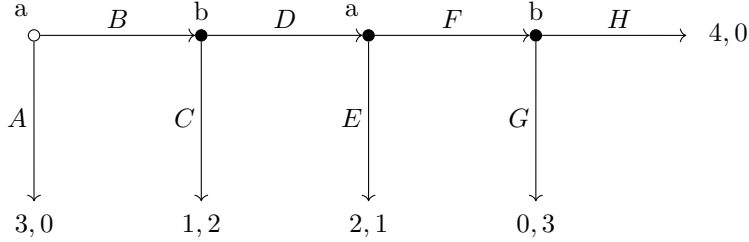

\noindent As for the Battle of the Sexes with an Outside Option, we now proceed step by step to obtain $\Und^\infty (S)$ for the game in  \Sref{fig:reny_centipede}.
\begin{itemize}[leftmargin=*]
\item ($n = 0$)  By definition we have that  $\Und^{0}_\Ann  (S) := S_\Ann$ and  $\Und^{0}_\Bob (S) := S_\Bob$.
\item ($n = 1$) Strategy $BE$ is strictly dominated---hence, B-dominated---with respect to Ann's conditional problem $S^\Ann (\la \varnothing \ra)$. Also, strategy $DH$ is strictly dominated---hence, B-dominated---with respect to Bob's conditional problem $S^\Bob (\la B \ra)$. Thus, $\Und^{1}_\Ann (S) = \{ A, BF \}$, while $\Und^{1}_\Bob (S) = \{ C, DG \}$.
\item ($n = 2$) Strategy $BF$ is strictly dominated---hence, B-dominated---with respect to Ann's conditional problem $S^\Ann (\la \varnothing \ra)$ given $\Und^{1} (S)$. Hence, $\Und^{2}_\Ann (S) = \{ A \}$, while $\Und^{2}_\Bob (S) = \Und^{1}_\Bob (S)$.
\item ($n = 3$) Strategy $C$ is strictly dominated---hence, B-dominated---with respect to Bob's conditional problem $S^\Bob (\la B \ra)$ given $\Und^{2} (S)$. Hence, $\Und^{3}_\Ann (S) = \Und^{2}_\Ann (S)$, while $\Und^{3}_\Bob (S) =\{ DG \}$.
\item ($n \geq 4$) Nothing changes and the procedure ends. 
\end{itemize} 
Thus, we have that $\Und^{\infty}_\Ann (S) = \{ A \}$ and $\Und^{\infty}_\Bob (S) = \{ DG \}$ and---obviously---we have that $\Und^{\infty} (S) = \{ (A, DG ) \}$. Hence, in particular, we have that $\zeta (\mbb{U}^{\infty} (S) )= \la A \ra$, that is, the ICBD procedure selects in this NRT dynamic game the \emph{unique} outcome $\la A \ra \in Z$.
\end{example}

In dynamic games with perfect information and in generic dynamic games with perfect information in particular, there is general agreement that the informal notion of backward induction reasoning applied to standard strategies is captured by Subgame Perfect Equilibrium of \citet[Definition, p.308]{Selten_1965} (see \citet[Section 6.2]{Osborne_Rubinstein_1994} for definitions and results). As it is well known, generic dynamic games with perfect information and NRT dynamic games\footnote{See, for example, \citet[Theorem 4.7, p.186]{Myerson_1991}.} have a unique backward induction outcome, which we denote by $z^{BI}$, that can be obtained via Subgame Perfect Equilibrium.  Interestingly, under a certain condition,\footnote{See also \Sref{subsubsec:generic_dynamic_games} for additional conceptual background.} the unique backward induction outcome $z^{BI}$ can be obtained by means of the algorithm known as Iterated Admissibility applied to the strategic form representation of an NRT dynamic game.  Hence, before identifying this condition, first we recall  some definitions apt for the study of games in their strategic form along with the definition of Iterated Admissibility. 

Thus, we employ the following  terminology  (see \citet[Definition 6, p.227]{Marx_Swinkels_1997}), which should be---not surprisingly---reminiscent of the one we used in \Sref{sec:revealed_FI} for operators. Given a game $\Game^r$ in its strategic form representation, a set $\mbf{R}$ is a \emph{reduction of} $S$ \emph{by weak dominance} if $\mbf{R}$ can be obtained from $S$ as $\mbf{R} :=  \bigcap_{\ell \geq 0} \mbf{R}^\ell$,\footnote{ \citet[Definition 6, p.227]{Marx_Swinkels_1997} employ a different notation: they write $\mbf{R} :=  S \setminus \mbf{R}^1, \dots, \mbf{R}^\ell$, where $\mbf{R}^k$ is the set of strategies \emph{deleted} at the $k$-th iteration of the reduction process, for every $k \in \{1, \dots , \ell\}$. On the contrary, in our notation, $\mbf{R}^k$ is the set of strategies that \emph{survive} the reduction process at the $k$-th iteration.} by setting $\mbf{R}^0 := S$, with $\mbf{R}^{n+1}$ derived from $\mbf{R}^{n}$ by letting $\mbf{R}^{n+1} := \mbf{R}^{n} \setminus \{s\}$, with $s := (s_j)_{j \in I}$ and there exists an $i \in I$ such that $s_i$ is weakly dominated on $\mbf{R}^{n}_i$ by an element of $\mbf{R}^{n}_i \setminus \{s_i\}$. Also, $\mbf{R}$ is a \emph{full} reduction of $S$ by weak dominance  if $\mbf{R}$ is a reduction of $S$ by weak dominance and there exist no strategies in $\mbf{R}$ that are weakly dominated. By proceeding recursively, we now introduce \emph{Pure Iterated Admissibility} (henceforth, Iterated Admissibility), which is defined as the maximal\footnote{An algorithm is \emph{maximal} with respect to a dominance notion if all strategies of all players that are  dominated according to that notion are removed at every step.} algorithm such that $\IA^{0}_i := S_i$,  $\IA^{n+1}_i := \Set { s_i \in \IA^{n}_i | s_i \in \Ad_i (\IA^{n}) }$, and $\IA^{\infty}_i  := \bigcap_{\ell \geq 0} \IA^{\ell}_i$, where $\IA^{k} := \prod_{j \in I} \IA^{k}_j$ for every $k \geq 0$. As a result, $\IA^{\infty}  := \bigcap_{\ell \geq 0} \IA^{\ell}$ is the set of strategy profiles that survive Iterated Admissibility, which can be proved via standard arguments to be nonempty.  

Whereas it is obvious that $\mbf{A}^{\infty}$ is a full reduction of $S$ by weak dominance, it has to be observed that it is well-known that Iterated Admissibility is \emph{not} order independent, i.e., it is not robust to the order in which strategies are eliminated.\footnote{See the example in \citet[Figure 63.1, p.63]{Osborne_Rubinstein_1994}.} As a result, given an arbitrary game in its strategic form $\Game^r$, it is possible to have many full reductions based on weak dominance, each giving a different set of strategy profiles as output. However, \citet[Equation 2, p.223]{Marx_Swinkels_1997} identifies a condition---which is the one we are after and we state next---that, if satisfied by a game, ensures that different full reductions based on weak dominance are equivalent outcome-wise.

\begin{definition}[Transference of Decision-Maker Indifference]
\label{def:TDI}
The strategic form representation $\Game^r$ of a dynamic game $\Game$ satisfies ``Transference of Decision-Maker Indifference'' condition (henceforth, TDI) if, for every $i  \in I$,
\begin{equation}
\label{eq:TDI}
\zeta (s_i , s_{-i}) \sim_i \zeta (s'_i , s_{-i}) \Longrightarrow %
\zeta (s_i , s_{-i}) \sim  \zeta (s'_i , s_{-i}),
\end{equation}
for every $s_i , s'_i \in S_i$ and $s_{-i} \in S_{-i}$.
\end{definition}

For the purpose of the present work, the TDI condition is particularly important in light of the remark that follows, that can easily be established by a contrapositive argument.

\begin{remark}[NRT Dynamic Games \& TDI] 
\label{rem:NRT_TDI}
The strategic form $\Game^r$ of an NRT dynamic game $\Game$ satisfies the TDI condition as in  \Mref{def:TDI}.\footnote{See also \citet[Section 8, p.77]{Battigalli_Friedenberg_2012}.}
\end{remark}

We now go back to Reny's Centipede Game in \Sref{fig:reny_centipede} to see at work what is written above.

\begin{example}[continues=ex:centipede, name=IA \& Backward Induction in NRT Dynamic Games]
We take the game in \Sref{fig:reny_centipede} and we represent in its strategic form in  \Sref{fig:strategic_centipede}.
\begin{figure}[H]
\centering
\begin{game}{3}{3}[$\Ann$][$\Bob$]
		& $C$ 		& $DG$	& $DH$\\
$A$ 	& $3, 0$ 	& $3, 0$ 	& $3, 0$ \\
$BE$ 	& $1, 2$ 	& $2, 1$ 	& $2, 1$ \\
$BF$ 	& $1 ,2$ 	& $0, 3$ 	& $4, 0$ \\
\end{game}
\caption{The strategic form representation of Reny's Centipede Game.}
\label{fig:strategic_centipede}
\end{figure}
\noindent To see that this game satisfies \Mref{def:TDI}, we reformulate \Mref{eq:TDI} in terms of a profile of \emph{ordinal} utility functions $v$ representing the players preferences, i.e., we have  
\begin{equation}
\label{eq:ex_TDI}
v_i (s_i , s_{-i}) = v_i (s'_i , s_{-i}) \Longrightarrow %
v_j (s_i , s_{-i}) = v_j (s'_i , s_{-i}).
\end{equation}
Now, it is easy to see that this game satisfies \Mref{eq:ex_TDI}. Concerning Iterated Admissibility, we have the following.
\begin{itemize}[leftmargin=*]
\item ($n = 0$)  By definition we have that  $\IA^{0}_\Ann  := S_\Ann$ and  $\IA^{0}_\Bob := S_\Bob$.
\item ($n = 1$) Strategy $BE$ is strictly dominated by $A$. Thus, $\IA^{1}_\Ann = \{ A, BF \}$, while $\IA^{1}_\Bob = \IA^{0}_\Bob$.
\item ($n = 2$) Strategy $C$ is weakly dominated by $DG$, that also strictly dominates $DH$. Hence, $\IA^{2}_\Ann = \IA^{1}_\Ann $, while $\IA^{2}_\Bob  = \{ DG \}$.
\item ($n = 3$) Strategy $BF$ is strictly dominated by $A$. Hence, $\IA^{3}_\Ann  = \{ A \}$, while $\IA^{3}_\Bob  =\IA^{2}_\Bob $.
\item ($n \geq 4$) Nothing changes and the procedure ends. 
\end{itemize}
Thus, we have that $\IA^{\infty}  = \{ (A, DG) \}$ and $\zeta (\IA^{\infty}) = \{ \la A \ra \} = \{ z^{BI} \}$.
\end{example}

What we showed above concerning the game in \Sref{fig:reny_centipede}, i.e., the fact that $\zeta(\IA^{\infty}) = \{ z^{BI} \}$, does not obtain by accident. Indeed, 
not only TDI makes Iterated Admissibility order independent with respect to outcomes (as pointed out above), but it is also  exactly the condition which ensures that Iterated Admissibility---again, no matter how performed---selects  $z^{BI}$ as the only outcome in the strategic form corresponding to an NRT dynamic game.\footnote{See \Mref{rem:corollary_MS97} with respect to this point.} This is captured by the result that we state next, that---indeed---links Iterated Admissibility and backward induction reasoning.

\begin{remark}[{\citet[Section VI, p.240]{Marx_Swinkels_1997}}]
\label{rem:Moulin}
Given a dynamic game $\Game$ whose strategic form representation $\Game^r$ satisfies the TDI condition,  $\zeta (\IA^{\infty}) = \{ z^{BI} \}$.
\end{remark}

In general, it is not possible to establish an inclusion relation between Iterated Admissibility and ICBD. Indeed, it can be shown an example of a dynamic ordinal game where $\IA^\infty \not\subseteq \Und^{\infty} (S)$, where such an example is a straightforward adaptation to our ordinal framework of the counterexample provided in \cite{Catonini_2023}  in the standard `cardinal' framework to disprove the longstanding conjecture in the field that stated that Iterated Admissibility with weak dominance by possibly mixed actions refines Strong Rationalizability.\footnote{In particular, to obtain such an example, by using the representation and the notation employed in \cite{Catonini_2023}, it is enough to change Bob's utility at $(D, L)$ from $0$  (as in \citet[Figure 2, p.6]{Catonini_2023}) to $1$ (of course, by treating this number `ordinally').}  Nonetheless, and crucially for our endeavour, when the dynamic game $\Game$ is an NRT game, we can provide a converse of \Mref{lem:fundamental_CBD_WD}, which leads to the lemma stated next.

\begin{lemma}
\label{lem:fundamental_NRT}
Given an NRT dynamic game $\Game$, its strategic form representation $\Game^r$  satisfying the TDI condition, a restriction $R \subseteq S$, and a player $i \in I$,  $s_i \in \mbb{U}_i (R)$ if and only if $s_i \in \Ad_i (R)$ in $\Game^r$.
\end{lemma}

\Mref{lem:fundamental_NRT} happens to be the missing ingredient to establish the result we are after that we state next, which is a translation in an ordinal setting of the result that goes in the literature as Battigalli's theorem established in \citet[Theorem 4, p.53]{Battigalli_1997},\footnote{See \Sref{subsubsec:generic_dynamic_games} for a discussion of the results mentioned in this section and \Sref{foot:BI_literature} therein for the related literature.} in light of the tight connections between Conditional B-Dominance and Weak Dominance established above.

\begin{proposition}
\label{prop:relevant_ties}
Given an NRT dynamic game $\Game$ whose strategic form representation $\Game^r$ satisfies the TDI condition, $\zeta (\mbb{U}^{\infty} (S)) = \{ z^{BI} \}$. 
\end{proposition}

The proof of \Mref{prop:relevant_ties} can be found in the appendix. Here, we just point out that we essentially already provided all the elements to derive it, since we mentioned above that \cite{Marx_Swinkels_1997} show that Iterated Admissibility is order independent with respect to outcomes in all games that satisfy TDI, i.e., no matter which algorithmic procedure based on weak dominance we employ, we always obtain the same outcome in this class of games. Thus, as soon as we realize (via an application of \Mref{lem:fundamental_NRT}) that---on a dynamic game $\Game$ whose  strategic form representation $\Game^r$ satisfies the TDI condition---ICBD induces the existence of an algorithmic procedure based on full reduction of weakly dominated strategies on $\Game^r$, the result is established.

%%%%%%%%%%%%%%%%%%%%%%%%%%%%%%%%%%%%%%%%%
%%%%%%%%%%%%%%%%%%%%%%%%%%%%%%%%%%%%%%%%%
%%%%%%%%%%%%%%%%%%%%%%%%%%%%%%%%%%%%%%%%%
%%%%%%%%%%%%%%%%%%%%%%%%%%%%%%%%%%%%%%%%%
%%%%%%%%%%%%%%%%%%%%%%%%%%%%%%%%%%%%%%%%%
%%%%%%%% SECTION %%%%%%%%%%%%%%%%%%%%%%%%%
%%%%%%%%%%%%%%%%%%%%%%%%%%%%%%%%%%%%%%%%%
%%%%%%%%%%%%%%%%%%%%%%%%%%%%%%%%%%%%%%%%%
%%%%%%%%%%%%%%%%%%%%%%%%%%%%%%%%%%%%%%%%%
%%%%%%%%%%%%%%%%%%%%%%%%%%%%%%%%%%%%%%%%%
%%%%%%%%%%%%%%%%%%%%%%%%%%%%%%%%%%%%%%%%%
\section{Applications}
\label{sec:applications}

%%%%%%%%%%%%%%%%%%%%%%%%%%%%%%%%%%%%%%%%%%
%%%%%%%%%%%%%%%%%%%%%%%%%%%%%%%%%%%%%%%%%%
%%%%%%%%%%%%%%%%%%%%%%%%%%%%%%%%%%%%%%%%%%
%%%%%%%%% SUB-Section %%%%%%%%%%%%%%%%%%%%%%%%%%%%%%%%%%%%%%%%%%
%%%%%%%%%%%%%%%%%%%%%%%%%%%%%%%%%%%%%%%%%%
%%%%%%%%%%%%%%%%%%%%%%%%%%%%%%%%%%%%%%%%%%
\subsection{On Sophisticated Voting}
\label{subsec:binary_agendas}

In this section, we show the importance of \Mref{prop:relevant_ties} in the context of a certain kind of binary agendas.\footnote{See \citet[Chapter 4.10]{Myerson_1991} for a definition of binary agendas similar to ours and a discussion of the model.} These are rather fundamental voting procedures that can be described informally as follows: a (finite) set of individuals votes on a (finite) set of alternatives, where the voting process is comprised itself of a sequence of voting procedures between two alternatives. Given their fundamental nature and their structure amenable to game-theoretical analysis, much work has been done on these voting procedures, starting from the seminal \cite{Farquharson_1969},\footnote{See also \cite{Niemi_1983} for a reappraisal of this book.}  which has been actually one of the first contributions (along with \cite{Dummett_Farquharson_1961}) to frame voting as a non-cooperative game. A crucial aspect of this work has been the introduction of the notion of \emph{sophisticated voting}, which corresponds to Iterated Admissibility applied to the strategic form representation of the voting process. Moving from this book, \cite{McKelvey_Niemi_1978}, \cite{Moulin_1979}, and \cite{Gretlein_1983}---among others---have further investigated the relation between the analysis presented in \cite{Farquharson_1969} in terms of \emph{sophisticated voting} and  the backward induction outcome. Thus, given this background, in this section we study a particular form of binary agendas, where, at every step of the process,  individuals have to vote \emph{sequentially} between two alternatives.\footnote{See \cite{Sloth_1993} for an analysis with a special focus of sequential voting.} As a result, this setting gives rise to a dynamic game with perfect information, whose class has been---of course---the focus of our analysis in \Sref{sec:relevant_ties}. Related to our analysis, it is important to observe that our ordinal setting, based on rational preference relations, without any reliance on Bernoulli utilities and probability measures, is particularly well-suited for the analysis of voting procedures (e.g., see  \citet[Footnote 3, p.1338]{Moulin_1979}).

To link what is written above to the result we state in this section, we now  provide a formal description of the following elements:

\begin{itemize}
\item[a)] the structure of the voting process;

\item[b)] how individuals are going to vote at each step of the voting process;

\item[c)] according to which decision rule an alternative is chosen at every step of the voting process. 
\end{itemize}
In what follows, we address each point to then provide in this context  a result that is related to \Mref{prop:relevant_ties}.

Given the list above, first of all, we address Point (a). To do so we describe the structure of the voting process by introducing the notion of binary agenda. A \emph{binary agenda} $\Psi$ is a tuple
\begin{equation*}
\Psi := \Angles { K, I, (\unrhd_j)_{j \in I}, (T, \widehat{\sqsubseteq}), \psi },
\end{equation*}
where $K$ is finite set of (social) alternatives, $I$ is a finite set of individuals having preferences over $K$ represented for every $i \in I$ by a rational preference relation $\unrhd_i \subseteq K \times K$, $(T, \widehat{\sqsubseteq})$ is a finite arborescence with root $n_0$, and $\psi$ is a correspondence  $\psi: T \cto K$ satisfying the following properties:
\begin{enumerate}[label=\arabic*)]
\item $\psi (n_0) = K$;

\item for every nonterminal $n \in T$, $\abs{T(n)} = 2$, with $T(n)$ denoting the set of \emph{immediate} successors of $n$; 

\item for every nonterminal $n \in T$ and $n' , n'' \in T (n)$, $\psi (n) = \psi (n') \cup \psi (n'')$.
\end{enumerate}

The definition of binary agenda we just provided remains silent regarding Point (b), which captures the form that the voting process takes at every step. In other words, the definition above does not describe how individuals are going to vote at every step, i.e., if simultaneously or sequentially. Thus, with respect to this point, we assume that at every step the voting process takes places \emph{sequentially}, which  is implemented by assuming that, every time players   are called to cast their vote between two alternatives, a fixed linear order $\gg$ is  exogenously  imposed on $I$, with $\overline{i}$ (resp., $\underline{i}$) denoting the first (resp., last) player to vote. As a result, a binary agenda $\Psi$ induces a dynamic game with perfect information $\Game_{\Psi}$ as in  \Mref{eq:dynamic_ordinal_game}, with $A_i := K$, $Z := K$, and $\succsim_i := \unrhd_i$, where in the following we write that $\Psi$ induces $\Game_\Psi$ or that $\Game_\Psi$ is induced by $\Psi$. Clearly, we have for every $i \in I$ that $A_i := K$ (since this represents the very act of casting a vote) and  $Z := K$ (since a terminal history in $\Game_{\Psi}$ represents an alternative $k \in K$ selected via the voting procedure). Regarding the translation of $T$ in terms of the arborescence $(X, \sqsubseteq)$ of \Mref{eq:dynamic_ordinal_game} (with related---singleton---information sets belonging to $H$), the following points are in order:
\begin{itemize}[leftmargin=*]
\item the empty history $\la \varnothing \ra$ corresponds to the root $n_0 \in T$;

\item for every $n \in T$ there exists a subarborescence $X' \subseteq X$ with the history $x$ corresponding to $n$ being the initial history of this arborescence such that $I_x := \{\overline{i}\}$;

\item for every $x \in X$, $\abs{X (x)} = 2$,  with $X(x)$ denoting the set of \emph{immediate} successors of $x$.

\item for every $x \in X$ with corresponding $h \in H$ such that $I_h := \{ \underline{i}\}$, there exist $n', n'' \in T$ such that $X (x) := \{x', x''\}$ with $x'$ corresponding to $n'$ and $x''$ corresponding to $n''$.
\end{itemize}

We now have to specify according to  which decision rule an alternative is chosen at every step of the voting process. Thus, to deal with Point (c) above, we introduce a further assumption on the players' preferences over the alternatives in $K$. In particular, we assume that the following \emph{``No Indifference'' condition} holds: for every $i \in I$, $\unrhd_i$ is a linear order  represented by $\rhd_i$. Now, armed  with the ``No Indifference'' condition along with the additional assumption that $\abs{I}$ is odd to avoid ties,\footnote{This assumption is actually unnecessary, since requiring $\abs{I}$ to be odd can be avoided---for example---by giving to a particular voter the power to break ties. Nonetheless, we adopt it, since it is standard in the literature.} we introduce the decision rule that we are going to employ to choose between alternatives at every step of the voting process. Thus, given that the \emph{majority rational preference relation} $\blacktriangleright \subseteq K \times K$ is defined as $k \blacktriangleright k'$ if there exists a $J \subseteq I$ such that $\abs{J} > \abs{I \setminus J}$ and $k \rhd_j k'$ for every $j \in J$, we define the \emph{majority rule} as the decision rule that chooses---between two alternatives---the alternative that is (strictly) preferred according to the majority rational preference relation. In the following, we use the expression ``majority voting'' to refer to a voting procedure that employs the majority rule.

With the framework introduced above at our disposal, we can now start the process of linking the study of binary agendas with sequential majority voting to our work in \Sref{sec:relevant_ties}. The following remark points out the relevancy of the TDI condition in this context as well.

\begin{remark}[{\citet[Section 3, p.269]{Hummel_2008}}]
\label{rem:TDI_voting}
A  binary agenda $\Psi$ with sequential majority voting satisfying the  ``No Indifference'' condition induces a dynamic game with perfect information $\Game_{\Psi}$ whose strategic form representation $\Game^{r}_{\Psi}$ satisfies the TDI condition.
\end{remark}

We can now state the main result of this section, that follows immediately  from \Mref{rem:TDI_voting} and \Mref{prop:relevant_ties}. In linking the backward induction outcome obtained in binary agendas with sequential majority voting to forward induction reasoning (via the ICBD procedure), it sheds a new light on the notion of \emph{sophisticated voting} of  \cite{Farquharson_1969}.

\begin{proposition}
\label{prop:sophisticated_forward}
Given a  dynamic game with perfect information $\Game_{\Psi}$ induced by a binary agenda $\Psi$ with sequential majority voting satisfying the ``No Indifference'' condition, $\zeta (\Und^{\infty} (S)) = \{ z^{BI} \}$.
\end{proposition}

%%%%%%%%%%%%%%%%%%%%%%%%%%%%%%%%%%%%%%%%%%
%%%%%%%%%%%%%%%%%%%%%%%%%%%%%%%%%%%%%%%%%%
%%%%%%%%%%%%%%%%%%%%%%%%%%%%%%%%%%%%%%%%%%
%%%%%%%%% SUB-Section %%%%%%%%%%%%%%%%%%%%%%%%%%%%%%%%%%%%%%%%%%
%%%%%%%%%%%%%%%%%%%%%%%%%%%%%%%%%%%%%%%%%%
%%%%%%%%%%%%%%%%%%%%%%%%%%%%%%%%%%%%%%%%%%
\subsection{On Farsightedness}
\label{subsec:farsightedness}

A question that has been extensively studied in the game-theoretical literature  from the birth of this field is how we can capture via game-theoretical tools a notion of \emph{``standard of behavior''} (as in the words of \citet[Chapter 4.6, pp.40--43]{vonNeumann_Morgenstern_1953}). Such standard of behavior should ideally capture both those \emph{outcomes} that cannot be rationally---in an informal sense---replaced by a coalition of  farsighted players and those \emph{coalitions} that would form given the assumptions regarding the players set forth above. It turns out that, from a technical standpoint, the question has been operationalized via the quest for a notion of \emph{stability} that possesses a list of desirable properties.\footnote{Related to this issue are \cite{Pomatto_2022} and  \cite{Luo_2009b,Luo_2009}.} In particular, and rather informally, given a set of outcomes $O$, the following are the properties that have been deemed crucial to deem that set a standard of behavior, i.e., a stable set:
\begin{itemize}[leftmargin=*]
\item \emph{Internal Stability:} no outcome $o \in O$ should be considered strictly worse than another $o' \in O$; 

\item \emph{External Stability:} given an $o \notin O$, there exists an outcome $o' \in O$  that is strictly preferred to $o$.
\end{itemize}
Whereas these two notions led to the solution concept known as \emph{von Neumann--Morgenstern Stable Sets}, \cite{Harsanyi_1974} pointed out a deficiency of this solution concept, namely, its inability to capture farsighted behavior.

Thus, in order to fix this shortcoming, starting from \citet[Pp.301--302]{Chwe_1994}, social environments have been extensively employed as the reference theoretical framework to tackle the problem delineated above. Thus, a \emph{social environment} is a tuple 
\begin{equation*}
\label{eq:social_environment}
\mscr{S} := \left\la %
I , O, (\SEs_j)_{j \in I} , \big( \effrel[J] \big)_{J \in \Coalitions} 
\right\ra ,
\end{equation*}
where $I$ is a set of \emph{players} with $\Coalitions := \wp_+ (I)$ denoting the set of \emph{coalitions},\footnote{As it is customary, we let $\wp_+ (X) := 2^X \setminus \{\emptyset\}$.} $O$ is a set of \emph{outcomes}, $\SEs_i \subseteq O \times O$ is player $i$'s rational preference relation over outcomes, with $i \in I$ arbitrary, and $\effrel[J] \subseteq O \times O$ is a binary relation called the \emph{effectiveness relation} of coalition $J$, with $J \in \Coalitions$ arbitrary. The crucial element of a social environment is the effectiveness relation $\effrel[J]$ of a coalition $J$, that allows the modeler to describe the fact that a coalition can change the status quo of the interaction: in particular, $o \effrel[J] o'$ captures the idea that $o \in O$ is the status quo and that coalition $J$ can make $o' \in O$ the new status quo, with the understanding that this does \emph{not} mean that this new status quo is automatically going to be enforced; indeed, it is possible to have that $o' \effrel[J'] o''$, which means that---moving from $o' \in O$ as a status quo---another coalition $J'$ can make $o''$ the new status quo. Given what is written above, we have that $o \effrel[\emptyset] o$ denotes the no-move at status quo $o$.

Various solution concepts specifically designed for social environments have been proposed in the literature to tackle the issue of capturing farsightedness, starting from the \emph{largest consistent set} of \citet[Section ``More Definitions'', pp.302--304]{Chwe_1994}.\footnote{See also \cite{Xue_1998}, \cite{Ray_Vohra_2015}, and \cite{Dutta_Vohra_2017}.} However, here, we focus on an alternative take on this issue, provided by \cite{Herings_et_al_2004},\footnote{See also \cite{Karos_Robles_2021}.} where the problem is addressed by opportunely translating a social environment in a dynamic game to then use Strong Rationalizability of \cite{Pearce_1984} and \cite{Battigalli_1997} to capture farsightedness. Crucially, in order to employ Strong Rationalizability, the endeavour in \cite{Herings_et_al_2004} is based on the notion of Bernoulli utility function in the definition of a social environment, which is \emph{not} in line with this literature, that typically starts---as we did above---with a profile of rational preference relations.

Thus, we now describe\footnote{See \citet[Section 3]{Herings_et_al_2004} for a proper introduction to the setting, along with relevant examples.} the essentials of the setting of \cite{Herings_et_al_2004} to show how it is possible to translate the endeavour performed there in our---weaker, since based only on ordinal preferences---framework, exploiting ICBD. For this purpose, we fix a social environment $\mscr{S}$ with $o^* \in O$ denoting the original status quo and we show how to obtain a corresponding---in a sense to be made precise---dynamic game $\Game [\mscr{S}]$, which is a dynamic ordinal game built on the social environment $\mscr{S}$ (alternatively called, the derived dynamic game). Thus, the \emph{dynamic ordinal game built on the social environment} $\mscr{S}$ is a dynamic ordinal game where $I_0 := I \cup \{0\}$ denotes the set of original players $I$ augmented with a dummy individual $0$ (which is \emph{not} a player), and 
\begin{equation*}
\mcal{M} := \Set { \Rounds{o \effrel[J] o'} | o, o' \in O , o \effrel[J] o' } \cup
\Set { o \effrel[\emptyset] o | o \in O }
\end{equation*}
denotes the set of  all possible moves and no-moves. Following the terminology set forth in \cite{Herings_et_al_2004}, every history $(a^1, \dots, a^k)$ corresponds to a \emph{stage} $k$, with $k \in \{1, \dots, n\}$. Thus, at every stage $k \in \bN$ every player $i \in I$ is active and observes the actions taken by the remaining players in $I \setminus \{i\}$, with 
\begin{equation*}
\mcal{M} \big(  (a^1, \dots, a^k)  \big) := \Set { \Rounds{o \effrel[J] o'} \in \mcal{M} | \omega ((a^1, \dots, a^k)) = o } 
\end{equation*}
denoting the set of feasible moves after history $(a^1, \dots, a^k)$ given that the status quo is $o$, with $\omega ((a^1, \dots, a^k))$ denoting the function that maps histories to corresponding outcomes considered as a---temporary---status quo. Now, whereas an action $\alpha^{k}_0$ of the dummy individual 
at history $(a^1, \dots, a^k)$ is a permutation of $\mcal{M} \big(  (a^1, \dots, a^k)  \big)$ that provides an order according to which the moves are implemented, an \emph{action} of player $i \in I$ at history $(a^1, \dots, a^k)$ is a function 
\begin{equation*}
\alpha^{k}_i : \mcal{M} \big(  (a^1, \dots, a^k)  \to \{0, 1\},
\end{equation*}
with the understanding that $\alpha^{k}_i \Rounds{ \Rounds{o \effrel[J] o'} } = 1$ means that $i \in J$ agrees to join the proposal of the coalition $J \in \Coalitions$ to move from $o$ to $o'$, whereas $\alpha^{k}_i \Rounds{ \Rounds{o \effrel[J] o'} } = 0$ means that $i \in S$ is blocking the proposal made by the coalition $J$. Finally, since in social environments negotiations on outcomes have to eventually stop to implement the agreed outcome, to capture this natural point, it is assumed that there exists an exogenously imposed $N \in \bN$ such that $(a^1, \dots , a^N)$ is a terminal history, with $X := \bigcup^{N}_{k = 0} X^k$ denoting the set of all histories.

Given the setting described above, for every player $i \in I$, strategies are naturally defined, with ICBD defined accordingly.\footnote{With the caveat that  we need to adapt our definition of ICBD, originally phrased for finite games, to the present---infinite---setting, where it should be observed that in this case the extension is straightforward.} In particular,  we let $\Und^{\infty} (S) [k]$ denote the set of strategies that survive ICBD when $\Game [\mscr{S}]$ consists of at most $k + 1$ stages, with $Z^{\infty} [k] := \zeta (\Und^{\infty} (S) [k])$. As a result, we let $Z^{\infty} := \lim \sup_{k \to \infty} Z^{\infty} [k]$ denote the set of  \emph{ordinally socially rationalizable outcomes}, which allows us to state the next proposition, which is an ordinal version \citet[Theorem 2, p.148]{Herings_et_al_2004} in our setting.\footnote{The proof is omitted, since it is an immediate adaptation of \citet[Proof of Theorem 2, p.155]{Herings_et_al_2004}.}

\begin{proposition}
\label{prop:social_environments}
Given a social environment $\mscr{S}$ and the dynamic ordinal game $\Game [\mscr{S}]$ built on $\mscr{S}$, $Z^\infty \neq \emptyset$. 
\end{proposition}

One clarification is in order, namely that, as a matter of fact, \citet[Definition 4, p.146]{Herings_et_al_2004} implements a `cautious' form of Strong Rationalizability that is based on the notion of \emph{Cautious Rationalizability} set forth in \citet[Definition 11, pp.1045--1046]{Pearce_1984}. However, it is immediate to establish that the definition of ICBD can opportunely modified in this direction.\footnote{And, for this purpose, it is relevant the definition of cautious sequential rationality we set forth in \Sref{subsubapp:characterization}.}

%%%%%%%%%%%%%%%%%%%%%%%%%%%%%%%%%%%%%%%%%%
%%%%%%%%%%%%%%%%%%%%%%%%%%%%%%%%%%%%%%%%%%
%%%%%%%%%%%%%%%%%%%%%%%%%%%%%%%%%%%%%%%%%%
%%%%%%%%% SUB-Section %%%%%%%%%%%%%%%%%%%%%%%%%%%%%%%%%%%%%%%%%%
%%%%%%%%%%%%%%%%%%%%%%%%%%%%%%%%%%%%%%%%%%
%%%%%%%%%%%%%%%%%%%%%%%%%%%%%%%%%%%%%%%%%%
\subsection{On Money-Burning Games}
\label{subsec:on_money-burning_games}

As pointed out in \Sref{subsec:motivation_results}, after its introduction in \cite{Kohlberg_1981} and \cite{Kohlberg_Mertens_1986}, forward induction has gained interest in the game-theoretical literature not only for its conceptual appeal, but also in light of its potential applications. Indeed, starting from the Intuitive Criterion of \cite{Cho_Kreps_1987}, forward induction has been considered an important idea to refine the potential outcomes compatible with a given (in that literature, typically  equilibrium-based) solution concept. Nonetheless, whereas forward induction reasoning has cutting power, it is not always the case that employing this form of reasoning singles out a specific outcome in an interaction. Thus, it is important to identify classes of games where forward induction reasoning performs best by actually identifying a single outcome as the result of an interaction. In the spirit of this quest, \cite{Ben-Porath_Dekel_1992} identify a class of games with this property, namely, money-burning games. Thus, we now formalize this class of games in their ordinal version to then address how ICBD performs in this context.

In the following, we let $I := \{\Ann, \Bob\}$ and, without loss of generality in light of standard representation results, we represent the---ordinal---preferences of the players via (unique up to monotone transformations) ordinal utility functions $v_\Ann$ and $v_\Bob$. Thus, the \emph{base game} of the money-burning game is a tuple $\Game^\mathsf{b} := \la I , (A_j , v_j)_{j \in I} \ra$ where there exist $a^{*}_\Ann \in A_\Ann$ and $a^{*}_\Bob \in A_\Bob$ such that: 
\begin{itemize}[leftmargin=*]
\item $v_\Ann (a^{*}_\Ann, a^{*}_\Bob) > v_\Ann (a_\Ann, a_\Bob)$, for every $(a_\Ann , a_\Bob) \in A_\Ann \times A_\Bob$; 

\item $v_\Bob (a^{*}_\Ann, a^{*}_\Bob) > v_\Bob (a^{*}_\Ann, a_\Bob)$, for every $a_\Bob \in A_\Bob$.
\end{itemize}
Thus, given that we let $\Burnt := \bN$ and we let $\varep \in \Re_{>0}$ be arbitrary, the \emph{money-burning $\varep$-game} $\Game^{\mathsf{mb}} [\varep]$ is the dynamic ordinal game where $H_\Ann := \{ \la \varnothing \ra \} \cup \Set { n \varep | n \in \Burnt }$ and $H_\Bob :=  \Set { n \varep | n \in \Burnt }$, $S_\Ann := \Burnt \times A_\Ann$ and $S_\Bob := [A_\Bob]^\Burnt$, and
\begin{align*}  
v_\Ann \big( (n, a_\Ann) , s_\Bob (n) \big) & := v_\Ann ( a_\Ann , a_\Bob ) - \varep n , \\
v_\Bob \big( (n, a_\Ann) , s_\Bob (n) \big) & := v_\Bob ( a_\Ann , a_\Bob ) ,
\end{align*}
with $s_\Bob (n) := a_\Bob$, for every $n \in \Burnt$. As a result, the class of money-burning games is comprised of all money-burning $\varep$-game, with $\varep \in \Re_{>0}$.

The next result, upon which we build in \Sref{subsubsec:abstract_GT_choice_theory}, shows that ICBD captures the \emph{unique} outcome compatible with forward induction reasoning in ordinal money-burning games.\footnote{See \cite{Shimoji_2002} for a similar result linking Iterated Conditional Dominance and money-burning games.}

\begin{proposition}
\label{prop:money-burning}
Given a base game $\Game^{\mathsf{b}}$, there exists a $\delta \in \Re_{>0}$ such that, for every $\varep \in (0, \delta)$,  $\zeta (\Und^{\infty} (S)) = \big\{ \big( (0, a^{*}_\Ann) , a^{*}_\Bob \big)\big\}$ in the money-burning $\varep$-game $\Game^{\mathsf{mb}} [\varep]$.
\end{proposition}

One point is in order, namely that---strictly speaking---ICBD as set forth in \Sref{subsubsec:ICBD} applies \emph{only} to finite dynamic ordinal games (as we defined them in \Sref{subsec:dynamic_ordinal_games}). However, the extension of ICBD to the class of---obviously not finite---money-burning games, given the particularly `simple' structure of these dynamic games, is actually straightforward.

%%%%%%%%%%%%%%%%%%%%%%%%%%%%%%%%%%%%%%%%%
%%%%%%%%%%%%%%%%%%%%%%%%%%%%%%%%%%%%%%%%%
%%%%%%%%%%%%%%%%%%%%%%%%%%%%%%%%%%%%%%%%%
%%%%%%%%%%%%%%%%%%%%%%%%%%%%%%%%%%%%%%%%%
%%%%%%%%%%%%%%%%%%%%%%%%%%%%%%%%%%%%%%%%%
%%%%%%%% SECTION %%%%%%%%%%%%%%%%%%%%%%%%%
%%%%%%%%%%%%%%%%%%%%%%%%%%%%%%%%%%%%%%%%%
%%%%%%%%%%%%%%%%%%%%%%%%%%%%%%%%%%%%%%%%%
%%%%%%%%%%%%%%%%%%%%%%%%%%%%%%%%%%%%%%%%%
%%%%%%%%%%%%%%%%%%%%%%%%%%%%%%%%%%%%%%%%%
%%%%%%%%%%%%%%%%%%%%%%%%%%%%%%%%%%%%%%%%%
\section{Discussion}
\label{sec:discussion}

%%%%%%%%%%%%%%%%%%%%%%%%%%%%%%%%%%%%%%%%%%
%%%%%%%%%%%%%%%%%%%%%%%%%%%%%%%%%%%%%%%%%%
%%%%%%%%%%%%%%%%%%%%%%%%%%%%%%%%%%%%%%%%%%
%%%%%%%%% SUB-Section %%%%%%%%%%%%%%%%%%%%%%%%%%%%%%%%%%%%%%%%%%
%%%%%%%%%%%%%%%%%%%%%%%%%%%%%%%%%%%%%%%%%%
%%%%%%%%%%%%%%%%%%%%%%%%%%%%%%%%%%%%%%%%%%
\subsection{On Decision-Theoretic Issues}
\label{subsec:on_decision-theoretic_issues}

%%%%%%%%%%%%%%%%%%%%%%%%%%%%%%%%%%%%%%%%%%
%%%%%%%%% SUB-Sub-Section %%%%%%%%%%%%%%%%%%%%%%%%%%%%%%%%%%%%%%%%%%
\subsubsection{On the Decision-Theoretic Interpretation of Conditional B-Dominance} 
\label{subsubsec:DT_interpretation}

\cite{Lo_2000} links B-Dominance to a notion of monotonicity that is the natural `static' counterpart of our \Mref{def:conditional_monotonicity}. Rather crucially, \citet[Lemma 2, p.729]{Lo_2000} shows that this notion of monotonicity characterizes Axiom P3 ``Eventwise Monotonicity'' as in \citet[P3, p.26]{Savage_1972}\footnote{The axiom is stated in \citet[P3, p.26]{Savage_1972} in a slightly different---but equivalent---form in comparison to \citet[Section 2.1, p.750]{Machina_Schmeidler_1992}, where it has been called for the first time ``Eventwise Monotonicity''. It is important to point out that it has been shown recently in \cite{Hartmann_2020} that, whereas this axiom is necessary for the derivation of Savage's subjective expected utility representation theorem in presence of a finite set of outcomes, it is actually redundant when the set of outcomes is infinite.} in a finite setting. This is a crucial point, since \citet[Theorem, p.730]{Lo_2000} shows that a decision maker's \emph{ordinal} preferences that abide \emph{simply} Axiom P3 are observationally indistinguishable to those \emph{cardinal} preferences that admit the existence of a utility function and a probability measures. This is a particularly interesting point, since \citet[Theorem, p.730]{Lo_2000} builds \emph{only} on Axiom P3, with \citet[Section 2]{Lo_2000} being explicit in dropping the so-called Savage's Axiom P2 ``Sure-Thing Principle'' as in \citet[P2, p.23]{Savage_1972}.

Interestingly, our result is only superficially similar to  \citet[Theorem, p.730]{Lo_2000}. Indeed, whereas our \Mref{def:conditional_monotonicity} looks---as it really should be---like a dynamic counterpart of the monotonicity notion set forth in \cite{Lo_2000}, we cannot dispense with  Savage's Axiom P2 ``Sure-Thing Principle''. As a matter of fact, this is a well-known consequence of the relation between dynamic choice and this axiom, that we uncover formally in  \Sref{parapp:DT_take} by paying special attention to how this all relates to \Mref{def:conditional_monotonicity}.\footnote{See \cite{Al-Najjar_Weinstein_2009} and \cite{Siniscalchi_2009} for an in-depth discussion of the issues related to dynamic choice with a special emphasis on ambiguity.}

%%%%%%%%%%%%%%%%%%%%%%%%%%%%%%%%%%%%%%%%%%
%%%%%%%%% SUB-Sub-Section %%%%%%%%%%%%%%%%%%%%%%%%%%%%%%%%%%%%%%%%%%
\subsubsection{Beliefs in Dynamic Games \& Alternative Notions of Sequential Rationality} 
\label{subsubsec:structural_rationality}

\cite{Siniscalchi_2022} is complementary to the present work with respect to the problem of assessing a player's beliefs in the course of a dynamic game. Incidentally, this has immediate implications on the embraced notion of sequential rationality.\footnote{See also \cite{Siniscalchi_2020}, which provides an axiomatic foundation to structural preferences, and \cite{Siniscalchi_Forthcoming}, where structural rationality is employed in a game-theoretical context (see also \citet[Section 6.5]{Siniscalchi_2022} with respect to this point).}

In particular,  \cite{Siniscalchi_2022} moves from the following two observations: (1) according to the conceptual stance presented in \cite{Savage_1972}, beliefs should be elicitable (i.e., it should be in principle possible to relate them to observable behavior); (2) dynamic games present a specific challenge in providing a notion of sequential rationality that takes into account that a player's beliefs cannot necessarily be observed at every information set (e.g., when such an information set is off the predicted path of play). Thus, to address the two points described above, the author introduces in \citet[Definition 3, p.2441]{Siniscalchi_2022} a novel notion of sequential rationality that he calls ``Structural Rationality'' according to which a strategy is structurally rational given a belief in the form of a consistent\footnote{See \citet[Section 3]{Siniscalchi_2022} for the details behind the notion of consistency of a CPS.} CPS if there exists no other strategy of that player that gives her a higher \emph{ex ante} expected payoff against all feasible perturbations---\emph{\`a la} \cite{Selten_1975} and \cite{Bewley_2002}---of that player's belief. Given this framework, \citet[Theorem 2, p.2448]{Siniscalchi_2022}, shows that offering side-bets at the beginning of the game under structural rationality  allows to elicit beliefs in an incentive-compatible way at every information set and, as such, it provides a theoretical foundation to the usage of the strategy method as in \cite{Selten_1967}.

%%%%%%%%%%%%%%%%%%%%%%%%%%%%%%%%%%%%%%%%%%
%%%%%%%%% SUB-Sub-Section %%%%%%%%%%%%%%%%%%%%%%%%%%%%%%%%%%%%%%%%%%
\subsubsection{On Decision-Theoretic Foundations of Expected Utility in Games} 
\label{subsubsec:DT_foundation_EU}

Our endeavour starts from assuming that every player has preferences over terminal histories of a dynamic game represented via a rational preference relation. Moving from this, we introduce in \Mref{def:rational} a notion of sequential rationality according to which a strategy of a player is sequentially rational if we can produce \emph{both} a utility function which agrees with the payer's rational preference relation  \emph{and} a conditional subjective probability---in the form of a CPS---according to which the strategy is a subjective expected utility maximizer. 

Two works, i.e., \cite{Gilboa_Schmeidler_2003} and \cite{Perea_2021}, deal with similar issues  and can be viewed as complementary to our approach in answering specifically the question of what is that utilities actually represent in a game-theoretical context, with the important caveat that they both work with state-dependent preferences.\footnote{With respect to this point, see in particular \citet[Section 1.2, p.187]{Gilboa_Schmeidler_2003} and \citet[Section 7(d), p.27]{Perea_2021}.} Thus, to answer this question, both papers work in the expected utility framework, with beliefs---as probability measures---assumed as a primitive notion. Moving from this very primitive notion, these works derive the main building block of their endeavours, which are \emph{conditional preference relations}, i.e., preference relations over acts and states of nature conditional on beliefs (as probability measures) on states of nature, with the understanding that these conditional preference relations are different from those described in \Sref{subsubsec:DT_take}, since in that case beliefs---as CPSs---are \emph{not} a primitive and the conditional element boils down to the family of information sets. The main result of both papers is an axiomatization of expected utility maximization in a game-theoretical context (i.e., \citet[Theorem, p.189]{Gilboa_Schmeidler_2003} and \citet[Theorem 5.1, p.19]{Perea_2021}), thus providing an expected utility representation to conditional preference relations.

%%%%%%%%%%%%%%%%%%%%%%%%%%%%%%%%%%%%%%%%%%
%%%%%%%%%%%%%%%%%%%%%%%%%%%%%%%%%%%%%%%%%%
%%%%%%%%%%%%%%%%%%%%%%%%%%%%%%%%%%%%%%%%%%
%%%%%%%%% SUB-Section %%%%%%%%%%%%%%%%%%%%%%%%%%%%%%%%%%%%%%%%%%
%%%%%%%%%%%%%%%%%%%%%%%%%%%%%%%%%%%%%%%%%%
%%%%%%%%%%%%%%%%%%%%%%%%%%%%%%%%%%%%%%%%%%
\subsection{On Technical Aspects of the Results}
\label{subsec:on_technical_aspects_of_the_results}

%%%%%%%%%%%%%%%%%%%%%%%%%%%%%%%%%%%%%%%%%%
%%%%%%%%% SUB-Sub-Section %%%%%%%%%%%%%%%%%%%%%%%%%%%%%%%%%%%%%%%%%%
\subsubsection{Restrictions} 
\label{subsubsec:restrictions}

In \Sref{subsec:strategies_reduced_strategic_form}, we define a restriction as a nonempty subset $R \subseteq S$ with a product structure, i.e., of the form $R := \prod_{j \in I} R_j$. However, in general, concerning the term ``restriction'', the usage in the literature varies. 

Here, we compare our terminology in particular to two seminal papers of the literature on game--theoretical algorithmic procedures, namely, \cite{Marx_Swinkels_1997} and \cite{Shimoji_Watson_1998}. Thus, as a matter of fact, our terminology follows \citet[Section IV, p.226]{Marx_Swinkels_1997} and, related to this, the fact that the definition of Iterated Admissibility in \Sref{sec:relevant_ties} is based on product restrictions is an established convention, as mentioned also in \citet[p.131]{Chen_Micali_2013}. On the contrary, it is important to point out that what are called ``restrictions'' in \citet[Section 2, p.164]{Shimoji_Watson_1998}  are dispensed  from having a product structure, because the term is essentially used there to capture a given player $i$'s conditioning events, that---focusing on the typical example of conditioning events in dynamic games---are of the form $S_{i} (h) \times S_{-i} (h)$, which (as pointed out in \Sref{subsec:strategies_reduced_strategic_form} and \Sref{foot:Mailath_et_al_1993}) do not necessarily have a product structure.

%%%%%%%%%%%%%%%%%%%%%%%%%%%%%%%%%%%%%%%%%%
%%%%%%%%% SUB-Sub-Section %%%%%%%%%%%%%%%%%%%%%%%%%%%%%%%%%%%%%%%%%%
\subsubsection{Genericity \& The Backward Induction Outcome} 
\label{subsubsec:generic_dynamic_games}

Since ICBD is the algorithmic counterpart of Ordinal Strong Rationalizability and captures forward induction reasoning in dynamic games with ordinal preferences, as already pointed out, our \Mref{prop:relevant_ties} provides an `ordinal' foundation to the so-called Battigalli's theorem (see \citet[Theorem 4, p.53]{Battigalli_1997}).\footnote{\label{foot:BI_literature}In the standard setting with Bernoulli utility functions, this result can be proved in various ways: \citet[Theorem 2, Section 3]{Heifetz_Perea_2015}, \citet[Theorem 4.1, p.126]{Perea_2018}, and \citet[Corollary 3, p.216]{Catonini_2020} obtain it constructively. Also, it is an immediate corollary of \citet[Theorem 4, p.28]{Chen_Micali_2012}, which is the working paper version of \cite{Chen_Micali_2013}. As mentioned in \citet[Section 1]{Heifetz_Perea_2015}, it is also possible to derive it by means of \citet[Theorem 1, p.230]{Marx_Swinkels_1997}, as we do with \Mref{prop:relevant_ties} (see \Sref{subapp:proofs_relevant_ties}). The original proof in \citet[pp.59-60]{Battigalli_1997} borrowed arguments from \cite{Reny_1992}, that---in turn---were based on properties of Fully Stable Sets of \citet[Section 3.4]{Kohlberg_Mertens_1986}.}  This result in the theory of dynamic games states that, given an NRT dynamic game (with cardinal preferences), the (unique) outcome $z^{BI}$ can be induced by both Subgame Perfect Equilibrium \emph{and} Strong Rationalizability. The importance of this result lies in the fact that Subgame Perfect Equilibrium in generic dynamic games is considered the prototypical application of \emph{backward} induction reasoning, while Strong Rationalizability is considered a solution concept that captures \emph{forward} induction reasoning. Thus, our `ordinal' foundation of this theorem is particularly appealing in light of the fact that---as pointed out in \citet[Section 4, p.52]{Battigalli_1997}---knowledge of players' risk attitudes should not play any role in generic dynamic games with perfect information.

Closing on a bibliographical note, the \Mref{rem:Moulin} in  \Sref{sec:relevant_ties}
has an ancestor in a similar result which relies on assuming that the dynamic game of perfect information is \emph{generic}\footnote{As in \Sref{foot:genericity}.} (see \citet[Theorem 1, p.85]{Moulin_1986}). This result can be found for the first time in \citet[Proposition 2]{Moulin_1979}, even if the idea of backward induction is not explicitly mentioned. However, \cite{Gretlein_1982} shows the presence of two problems in the proof of \citet[Proposition 2]{Moulin_1979}, amended in \cite{Gretlein_1983}. Thus, as pointed out in \citet[p.113]{Gretlein_1983}, the first paper where the result is stated---by putting a special emphasis on the strict ordering over terminal histories of a given dynamic game with perfect information---and correctly proved is \cite{Rochet_1980},\footnote{While  writing these lines, we still had not have access to a copy of this paper: unfortunately, this work is presently unavailable, as confirmed by the author, who kindly tried to retrieve it without success (private communication). Hence, our decision of adding  \cite{Moulin_1986} above (with full bibliographical details) as another  reference for the result.} where analogous bibliographical points can be found in \citet[Section II, pp.221-222, \& Section VI, p.240]{Marx_Swinkels_1997}. It has to be observed that \cite{Moulin_1979}, \cite{Rochet_1980}, \cite{Gretlein_1983}, and \cite{Moulin_1986} work with a condition actually stronger than the TDI condition, called the ``One-to-One'' Condition (see \citet[Section 1, p.1340]{Moulin_1979}),\footnote{See also Equation (2) in \citet[Section 2, p.1393]{Armbruster_Boge_1983}.} which---as emphasized in \citet[Section III, p.222]{Marx_Swinkels_1997}---is always satisfied by the strategic form of a generic dynamic game.

%%%%%%%%%%%%%%%%%%%%%%%%%%%%%%%%%%%%%%%%%%
%%%%%%%%%%%%%%%%%%%%%%%%%%%%%%%%%%%%%%%%%%
%%%%%%%%%%%%%%%%%%%%%%%%%%%%%%%%%%%%%%%%%%
%%%%%%%%% SUB-Section %%%%%%%%%%%%%%%%%%%%%%%%%%%%%%%%%%%%%%%%%%
%%%%%%%%%%%%%%%%%%%%%%%%%%%%%%%%%%%%%%%%%%
%%%%%%%%%%%%%%%%%%%%%%%%%%%%%%%%%%%%%%%%%%
\subsection{On ICBD \& Different Approaches to Game Theory}
\label{subsec:ICBD_GT}

%%%%%%%%%%%%%%%%%%%%%%%%%%%%%%%%%%%%%%%%%%
%%%%%%%%% SUB-Sub-Section %%%%%%%%%%%%%%%%%%%%%%%%%%%%%%%%%%%%%%%%%%
\subsubsection{Epistemic Foundation of ICBD} 
\label{subsubsec:epistemic_foundation}

As pointed out in \citet[Section 5.1]{Borgers_1993}, the epistemic  foundation of the iterative procedure based on B-dominance is essentially the same as the one of Correlated Rationalizability of \cite{Bernheim_1984} and \cite{Pearce_1984} made by \cite{Boge_Eisele_1979}, \cite{Brandenburger_Dekel_1987}, and \cite{Tan_Werlang_1988}. The only difference lies in the definition of the event in the type structure that captures the rationality of a player: indeed, this has to be opportunely modified according to the notion of rationality introduced in \cite{Borgers_1993}.

The same can be written concerning the epistemic foundation of ICBD, in the sense that the epistemic foundation of ICBD is captured as in \citet[Sections 3-4]{Battigalli_Siniscalchi_2002} by the event \emph{Rationality and Common Strong Belief in Rationality} in a belief-complete\footnote{See \citet[Section 3]{Siniscalchi_2008} by bearing in mind that the author calls this notion ``completeness''.} type structure such as the Canonical Hierarchical Structure constructed in \citet[Section 2]{Battigalli_Siniscalchi_1999}, with the definition of sequential rationality changed accordingly.

%%%%%%%%%%%%%%%%%%%%%%%%%%%%%%%%%%%%%%%%%%
%%%%%%%%% SUB-Sub-Section %%%%%%%%%%%%%%%%%%%%%%%%%%%%%%%%%%%%%%%%%%
\subsubsection{On Abstract Game-Theoretical Choice Theory} 
\label{subsubsec:abstract_GT_choice_theory}

Starting from \cite{Sprumont_2000},\footnote{See also \cite{Yanovskaya_1980}, \cite{Galambos_2005}, \cite{Demuynck_Lauwers_2009}, with \cite{Yanovskaya_1980} being the actual first contribution---in Russian---on the topic.} a growing stream of literature has focused on approaching game-theoretical solution concepts with the tools proper of abstract choice theory in the spirit of the revealed preference methodology applied to game theory.\footnote{See \citet[Chapter 10]{Chambers_Echenique_2016} for a textbook presentation of this topic.} Endeavours belonging to this literature typically move from the notion of \emph{joint choice function}, an exact translation in a game-theoretical context of the notion of choice function\footnote{See, for example,  \citet[Chapter 2.1, p.16]{Chambers_Echenique_2016}.} proper of abstract choice theory, whose objective is to capture those outcomes in a game that an outside observer sees as the result of the actual interaction taking place represented as a game. Building on this notion, this literature answers the question of what are the properties that a joint choice function has to satisfy in order to be rationalizable (in the sense of abstract choice theory) by a given game-theoretical solution concept, where---and rather crucially---such a notion of rationalizability begs for the existence of a profile of rational preference relations capturing the players' preferences with respect to which the observed outcome has to be a maximal element.

With respect to this kind of endeavours, \cite{Ray_Zhou_2001}, \cite{Ray_Snyder_2013}, and \cite{Schenone_2020}  focused on dynamic games and on subgame perfect equilibria and backward induction in particular.\footnote{See also \cite{Xu_Zhou_2007}, \cite{Bossert_Suzumura_2010}, \cite{Rehbeck_2014}, and \cite{Li_Tang_2017} for a different take on game-theoretical revealed preference efforts with a focus on dynamic games.} Thus, regarding this literature, there are no contributions on forward induction in general and---more specifically---on the different properties that a joint choice function should satisfy to capture the outcomes obtained by players reasoning according to forward induction in comparison to backward induction. Now, given that one obvious problem to tackle those issues is the lack of a definition of an `ordinal' solution concept capable of capturing forward induction,  ICBD can provide a `benchmark' definition of forward induction reasoning for endeavours belonging to this literature, since ICBD provides outcomes compatible with forward induction reasoning in presence of ordinal preferences. 

However, it is important to stress that, even if ICBD can be a step towards addressing these issues, it seems that the problem of capturing \emph{in general} the differences in the properties that joint choice functions have to satisfy when capturing forward versus backward induction for \emph{all} classes of dynamic games is unmanageable. Nonetheless, it is possible to single out two classes of dynamic games where this objective should be achievable, in particular in light of the fact that forward induction has a strong bite there, namely: NRT dynamic games and money-burning games.\footnote{The analysis should also cover dynamic games with an outside option, in the spirit of the dynamic game represented in \Sref{fig:example_ordinal_basic} that has been the main character of \Mref{ex:main_example} throughout this work. However, it should be underlined with respect to this point that the class of generic $2$-person dynamic games with an outside option identified in \citet[Section 3, p.485]{vanDamme_1989} is too `large' for the stated purposes, since the main condition identified there is that the profile of actions (in the subgame) which arises when the first player decides to opt in is an \emph{equilibrium}. Of course, this is stronger than what we could achieve via ICBD as an iteration process.}

Thus, as mentioned above, in these two classes, forward induction reasoning via ICBD has  a strong bite, by actually giving a unique outcome. And now, having set forth the context, one point is in order. In addressing the different properties that joint choice functions should satisfy to capture forward versus backward induction in these classes of games, those properties should lead---no matter what---to the \emph{same} outcome (obtained via a different `logic') in NRT dynamic games, in light of \Mref{prop:relevant_ties}.

One---closing---point is in order: this literature has focused on equilibrium-based solution concepts, e.g., Nash Equilibrium (as in the seminal \cite{Sprumont_2000}) or Subgame Perfect Equilibrium (as in \cite{Ray_Zhou_2001} and \cite{Schenone_2020}), which seems at odds with the idea advocated here of employing a Rationalizability-type solution concept such as ICBD as a benchmark definition for forward induction reasoning for the purpose of endeavours in the spirit of abstract game-theoretical choice theory. However, two aspects of ICBD have to be emphasized:
\begin{enumerate}[label=\roman*)]
\item ICBD is extremely powerful in the classes of games singled out above, thus, not having the multitude of predictions that is typically considered a shortcoming of rationalizability-type solution concepts;

\item ICBD has a clear decision-theoretic foundation (as set forth in \Sref{sec:rationality_dominance}). 
\end{enumerate}
In particular, regarding point (ii), it is important to stress that it should put ICBD on par with equilibrium-based solution concepts for these kind of endeavours, since the properties that a joint choice function has to satisfy should implicitly capture a decision-theoretic stance in the (outcome) selection process and equilibrium-based solution concepts should---at least in principle---be based on axiomatic (and decision-theoretic in nature) derivations in the spirit of  \cite{Govindan_Wilson_2012} along the lines set forth in \citet[Section E, p.1036]{Kohlberg_Mertens_1986}.\footnote{See in particular \citet[Section 6.2, p.1671]{Govindan_Wilson_2012}. See also \cite{Brandenburger_Friedenberg_2009}.}

%%%%%%%%%%%%%%%%%%%%%%%%%%%%%%%%%%%%%%%%%%
%%%%%%%%%%%%%%%%%%%%%%%%%%%%%%%%%%%%%%%%%%
%%%%%%%%%%%%%%%%%%%%%%%%%%%%%%%%%%%%%%%%%%
%%%%%%%%%%%%%%%%%%%%%%%%%%%%%%%%%%%%%%%%%%
%%%%%%%%%%%%%%%%%%%%%%%%%%%%%%%%%%%%%%%%%%
%%%%%%%%%%%%%%%%%%%%%%%%%%%%%%%%%%%%%%%%%%
%%%%%%%%%%%%%%%%%%%%%%%%%%%%%%%%%%%%%%%%%
%%% APPENDIX
\appendix

\section*{Appendix}
%\section*{Appendices}

%%%%%%%%%%%%%%%%%%%%%%%%%%%%%%%%%%%%%%%%%
%%%%%%%%%%%%%%%%%%%%%%%%%%%%%%%%%%%%%%%%%
%%%%%%%%%%%%%%%%%%%%%%%%%%%%%%%%%%%%%%%%%
%%%%%%%%%%%%%%%%%%%%%%%%%%%%%%%%%%%%%%%%%
%%%%%%%%%%%%%%%%%%%%%%%%%%%%%%%%%%%%%%%%%
%%%%%%%% SECTION %%%%%%%%%%%%%%%%%%%%%%%%%
%%%%%%%%%%%%%%%%%%%%%%%%%%%%%%%%%%%%%%%%%
%%%%%%%%%%%%%%%%%%%%%%%%%%%%%%%%%%%%%%%%%
%%%%%%%%%%%%%%%%%%%%%%%%%%%%%%%%%%%%%%%%%
%%%%%%%%%%%%%%%%%%%%%%%%%%%%%%%%%%%%%%%%%
%%%%%%%%%%%%%%%%%%%%%%%%%%%%%%%%%%%%%%%%%
\section{Proofs}
\label{app:proofs}

%%%%%%%%%%%%%%%%%%%%%%%%%%%%%%%%%%%%%%%%%%
%%%%%%%%%%%%%%%%%%%%%%%%%%%%%%%%%%%%%%%%%%
%%%%%%%%%%%%%%%%%%%%%%%%%%%%%%%%%%%%%%%%%%
%%%%%%%%% SUB-Section %%%%%%%%%%%%%%%%%%%%%%%%%%%%%%%%%%%%%%%%%%
%%%%%%%%%%%%%%%%%%%%%%%%%%%%%%%%%%%%%%%%%%
%%%%%%%%%%%%%%%%%%%%%%%%%%%%%%%%%%%%%%%%%%
\subsection{\texorpdfstring{Proofs of \Sref{sec:rationality_dominance}}{Proofs of Section 3}}
\label{subapp:rationality_dominance}

%%%%%%%%%%%%%%%%%%%%%%%%%%%%%%%%%%%%%%%%%%
%%%%%%%%% SUB-Section %%%%%%%%%%%%%%%%%%%%%%%%%%%%%%%%%%%%%%%%%%
\subsubsection{\texorpdfstring{Proofs of \Sref{subsec:dominance_extensive}}{Proofs of Section 3.2}}
\label{subsubapp:rationality_dominance}

We recall some definitions from \cite{Mailath_et_al_1993} and \cite{Shimoji_Watson_1998}, by opportunely modifying them to take into account the presence of restrictions.\footnote{\label{foot:general_info}The original definitions are phrased in terms of \emph{generic} information sets, where here we use the term ``generic'' with the meaning that the objects under scrutiny can be extensive form as well as \emph{normal form} information sets (see \citet[Definition 3, p.279]{Mailath_et_al_1993}). Here, we state these definitions by referring directly to their extensive form representation (of course, by employing our notation).} Given two restrictions $Q$ and $R$ such that $Q \subseteq R \subseteq S$, and a player $i \in I$, strategies $s_i, s'_i \in Q_i$ \emph{agree on $Q_{-i}$} if 
\begin{equation*}
%\label{eq:agree}
\zeta (s_i, s_{-i}) \sim \zeta (s'_i ,s_{-i})
\end{equation*}
for every $s_{-i} \in Q_{-i}$, whereas $s_i , s'_i \in Q_i$ are $Q$\emph{-replacements}  if $s_i$ and $s'_i$ agree on $R_{-i} \setminus Q_{-i}$.

Now, given a restriction $R \subseteq S$, a player $i \in I$, and an information set $h \in H_i$ such that  $R^i (h) \neq \emptyset$, strategies $s_i, s'_i \in R_i (h)$ are \emph{strong $R^i (h)$-replacements} if, in addition of being $R^i (h)$-replacements, for every $h' \in H_i$ with corresponding $R^i (h')$, if $R^i (h) \cap R^i (h') \neq \emptyset$, then $s_i \in R_i (h')$ if and only if $s'_i \in R_i (h')$. \citet[Section 2, p.168]{Shimoji_Watson_1998} prove for both standard strategies and strategies that the collection of conditional problems (i.e., with $S$ as the reference restriction in our definitions) of a  player satisfies the following property, that they call the ``Strong Replacement Property'': for every $i \in I$, for every conditional problem $S^i (h)$, and for every $s_i, s'_i \in S_i (h)$ there exists a strategy $s^{*}_i \in S_i (h)$ such that:
\begin{enumerate}[label=\roman*)]
\item $s_i$ and $s^{*}_i$ are strong $S^i (h)$-replacements,
\item  $s'_i$  and $s^{*}_i$ agree on $S_{-i} (h)$.
\end{enumerate}
In the following lemma, we restate the Strong Replacement Property by taking into account the presence of an arbitrary restriction $R \subseteq S$.

\begin{lemma}[Strong Replacement Property Given a Restriction]
\label{lem:strong_replacement_property}
Given a restriction $R \subseteq S$, a player $i \in I$, an information set $h \in H_i$ such that $R^i (h)$ is nonempty, and  strategies $s_i, s'_i \in R_i (h)$, there exists a strategy $s^{*}_i \in R_i (h)$ such that:
\begin{enumerate}[label=\roman*)]
\item $s_i$ and $s^{*}_i$ are strong $R^i (h)$-replacements,
\item  $s'_i$  and $s^{*}_i$ agree on $R_{-i} (h)$.
\end{enumerate}
\end{lemma}

\begin{proof}[Proof of \Mref{lem:strong_replacement_property}]
Let a restriction $R \subseteq S$, a player $i \in I$, an information set $\overline{h} \in H_i$ with $R^i (\overline{h})$ nonempty, and strategies $s_i, s'_i \in R_i (\overline{h})$ (with \emph{a fortiori} $s_i, s'_i \in R_i$) be arbitrary. Now, for every $h \in H_i (s_i) \cap H_i (s'_i)$, define strategy $s^{*}_i$ as follows:
\begin{equation*}
s^{*}_i (h) :=
\begin{cases}
s'_i (h), & \text{ if } h = \overline{h}, \\[7pt]
s_i (h), & \text{otherwise}.
\end{cases}
\end{equation*}
Thus, we have by construction that $s^{*}_i \in R_i (\overline{h})$ is well-defined and:
\begin{enumerate}[label=\roman*)]
\item $s_i$ and $s^{*}_i$ are strong $R^i (\overline{h})$-replacements,
\item  while $s'_i$  and $s^{*}_i$ agree on $R_{-i} (\overline{h})$.
\end{enumerate}
As a result, what is written above establishes the lemma, with the understanding that---of course, given our notational conventions introduced in \Sref{sec:game-theoretical_framework}---we have \emph{a fortiori} that $s^{*}_i \in R_i$. 
\end{proof}

The result that follows links the Strong Replacement Property given an arbitrary restriction $R \subseteq S$ as in \Mref{lem:strong_replacement_property} to weak dominance relative to $R$. 

\begin{lemma}[Strong Replacement Property \& Weak Dominance]
\label{lem:strong_replacement}
Given a restriction $R \subseteq S$, a player $i \in I$, and a strategy $\underline{s}_i \in R_i$, if there exists an information set $h \in H_i (\underline{s}_i)$ such that $R^i (h)$ is nonempty and there exists a strategy $\overline{s}_i \in R_i (h)$ such that $\underline{s}_i$ is weakly dominated relative to $R^i (h)$ by $\overline{s}_i$, then there exists a strategy $s^{*}_i \in R_i$ that agrees with $\overline{s}_i$ on $R_{-i} (h)$ and that agrees with $\underline{s}_i$ on $R_{-i} \setminus R_{-i} (h)$ that weakly dominates $\underline{s}_i$ relative to $R$. 
\end{lemma}

\begin{proof}[Proof of \Mref{lem:strong_replacement}]
Let a restriction $R \subseteq S$, a player $i \in I$, and a strategy $\underline{s}_i \in R_i$ be arbitrary and assume that there exists an information set $h \in H_i (\underline{s}_i)$ such that $R^{i} (h)$ is nonempty and there exists a strategy $\overline{s}_i \in R_i (h)$ such that $\underline{s}_i$ is weakly dominated relative to $R (h)$ by $\overline{s}_i$. Thus, from \Mref{lem:strong_replacement_property}, there exists a strategy $s^{*}_i \in R_i (h)$ that agrees with $\overline{s}_i$ on $R_{-i} (h)$ and such that $\underline{s}_i$ and $s^{*}_i$ are strong $R^i (h)$-replacements, which, in particular, implies that $\underline{s}_i$ and $s^{*}_i$ agree on $R_{-i} \setminus R_{-i} (h)$. That $s^{*}_i$ weakly dominates $\underline{s}_i$ relative to $R$ is an immediate consequence of the fact that---by construction---we have that:
\begin{itemize}[leftmargin=*]
\item $\zeta (s^{*}_i , \overline{s}_{-i}) \sim_i \zeta (\underline{s}_i , \overline{s}_{-i})$ for every $\overline{s}_{-i} \in R_{-i} \setminus R_{-i} (h)$, since $s^{*}_i$ and $\underline{s}_i$ agree on $R_{i} \setminus R_{i} (h)$;

\item $\zeta (s^{*}_i , s_{-i}) \succsim_i \zeta (\underline{s}_i , s_{-i})$ for every $s_{-i} \in R_{-i} (h)$ and there exists a $s^{*}_{-i} \in R_{-i} (h)$ such that $\zeta (s^{*}_i , s^{*}_{-i}) \succ_i \zeta (\underline{s}_i , s^{*}_{-i})$, since $s^{*}_i$ agrees with $\overline{s}_i$ on $R_{-i} (h)$ and $\overline{s}_i$ weakly dominates $\underline{s}_i$ relative to $R (h)$.
\end{itemize}
Hence, what is written above establishes the result.
\end{proof}

\begin{proof}[Proof of \Mref{lem:fundamental_CBD_WD}]
Fix a dynamic game $\Game$ with its corresponding strategic form $\Game^r$ and let $R \subseteq S$ be an arbitrary restriction. We proceed by proving the contrapositive letting $\underline{s}_i \in R_i$ be arbitrary and assuming that $\underline{s}_i \notin \Und_i (R)$. Thus, there exists an information set $h \in H_i (\underline{s}_i)$ such that $R^{i} (h)$ is nonempty and $\underline{s}_i \in \bd_i (R^i (h))$, i.e., for every nonempty $Q_{-i} (h) \subseteq R_{-i} (h)$,  $\underline{s}_i \notin \Ad_i (R_{i} (h) \times Q_{-i} (h))$. Since the statement involves a universal quantifier, we have that it is---\emph{a fortiori}---true for $R_{-i} (h)$, i.e., $s_i \notin \Ad_i (R_i (h) \times R_{-i} (h))$. Hence, there exists a strategy $\overline{s}_i \in R_i (h)$ that weakly dominates $\underline{s}_i$ relative to $R^i (h)$. Now, we can invoke \Mref{lem:strong_replacement} in light of the fact that all its conditions are met: thus, there exists a strategy $s^{*}_i \in R_i$ that agrees with $\overline{s}_i$ on $R_{-i} (h)$ and that agrees with $\underline{s}_i$ on $R_{-i} \setminus R_{-i} (h)$ that weakly dominates $\underline{s}_i$ relative to $R$.  Hence, $\underline{s}_i \notin \Ad_i (R)$, which establishes the result.
\end{proof}

%%%%%%%%%%%%%%%%%%%%%%%%%%%%%%%%%%%%%%%%%%
%%%%%%%%% SUB-Sub-Section %%%%%%%%%%%%%%%%%%%%%%%%%%%%%%%%%%%%%%%%%%
\subsubsection{\texorpdfstring{Proofs \& Technical Aspects of \Sref{subsec:characterization}}{Proofs \& Technical Aspects of Section 3.3}}
\label{subsubapp:characterization}

%%%%%%%%%%%%%%%%%%%%%%%%%%%%%%%%%%%%
%%% Paragraph
%
\paragraph{\texorpdfstring{Proofs of \Sref{subsubsec:GT_take}}{Proofs of Section 3.3.1}}
\label{par:GT_take}

For the purpose of the proofs belonging to this section, we formalize a new---ancillary---notion. To introduce it, in the following, for every CPS $\mu_i \in \Delta^{\mcal{H}_i} (S_{-i})$, we let $\supp \mu_i (\cdot | S_{-i} (h))$ denote its support  given $S_{-i} (h)$. Thus, given  a player $i \in I$ and  a restriction $R$, a strategy $s^{*}_i \in R_i$ is \emph{cautiously sequentially rational given $R$} if there exist a utility function $u_{i}|_R \in \Re^{Z (R)}$ and a CPS $\mu_i \in \Delta^{\mcal{H}_i} (S_{-i})$, with $\supp \mu_i (\cdot | S_{-i} (h)) =  R_{-i} (h)$,\footnote{\label{foot:cautious}The usage of the term ``cautious'' in this definition is related to this full-support condition, as in \citet[Appendix B, p.1049]{Pearce_1984} (see also  \citet[Definition 3, p.426]{Borgers_1993} and discussion thereafter).} for every $h \in H_i (s^{*}_i)$ for which  $R^{i} (h)$ is nonempty, such that 
\begin{equation}
\label{eq:rational_given} 
\sum_{s_{-i} \in R_{-i}} u_i (\zeta(s^{*}_i , s_{-i})) \cdot \mu_i (\{ s_{-i} \} | S_{-i} (h)) \geq %
\sum_{s_{-i} \in R_{-i}} u_i (\zeta(s_i , s_{-i})) \cdot \mu_i (\{ s_{-i} \} | S_{-i} (h)) 
\end{equation}
for every $s_i \in R_i (h)$.

\begin{remark}
\label{rem:given_rationality}
Given a restriction $R \subseteq S$ and a player $i \in I$, if there exists a $Q_{-i} \subseteq R_{-i}$ such that strategy $s^{*}_i \in R_i$ is cautiously sequentially rational given $R_i \times Q_{-i}$, then $s^{*}_i$ is sequentially rational given $R$.
\end{remark}

We can now proceed with the steps that lead to the proof of \Mref{th:rationality}.

%%%%%%%%%%%%%%%%%%%%%%%%%%%%%%%%%%%%%%%%
% CLAIM
%
\begin{lemma}
\label{lem:claim}
Fix a restriction $R \subseteq S$, a player $i \in I$, a strategy $s^{*}_i \in R_i$, and assume that, for every information set $h \in H_i (s^{*}_i)$, if $R^{i} (h)$ is nonempty, then $s^{*}_i$ is not weakly dominated relative to $R^i (h)$. Then there exist a utility function $u_{i}|_{R} \in \Re^{Z (R)}$ and a CPS $\mu_i \in \Delta^{\mcal{H}_i} (S_{-i})$ with $\supp \mu_i (\cdot | S_{-i} (\overline{h})) =  R_{-i} (\overline{h})$, for every $\overline{h} \in H_i (s^{*}_i)$ for which $R^{i} (\overline{h}) \neq \emptyset$, such that 
\begin{equation*}
\sum_{s_{-i} \in R_{-i}} u_i (\zeta(s^{*}_i , s_{-i})) \cdot \mu_i (\{ s_{-i} \} | S_{-i} (h)) \geq %
\sum_{s_{-i} \in R_{-i}} u_i (\zeta(s_i , s_{-i})) \cdot \mu_i (\{ s_{-i} \} | S_{-i} (h)) 
\end{equation*}
for every $s_i \in R_i (h)$.
\end{lemma}
%
%
%
% PROOF CLAIM
%
%
%
\begin{proof}
We fix a restriction $R \subseteq S$, a player $i \in I$, a strategy $s^{*}_i \in R_i$, and we assume that, for every information set $h \in H_i (s^{*}_i)$, if $R^{i} (h)$ is nonempty, then $s^{*}_i$ is not weakly dominated relative to $R^i (h)$. We let $\zeta (s^{*}_i ,\cdot) \in Z^{S_{-i}}$ denote the section of the outcome function $\zeta \in Z^S$ given $s^{*}_i \in R_i$. Also, we let $\underline{\alpha}$ denote the outcome for player $i$ that is minimal with respect to the rational preference relation $\succsim_i$ restricted to the outcomes that can be obtained given  $\zeta (s^{*}_i ,\cdot)$. Finally, we define 
\begin{equation*}
\underline{R}_{-i} := \Set { s_{-i} \in R_{-i} |  \zeta (s^{*}_i , s_{-i} ) = \underline{\alpha} }
\end{equation*}
and we let 
\begin{equation*}
\underline{R}_{i} := \Set { s_{i} \in R_i | \exists s_{-i} \in R_{-i} : \underline{\alpha} \succ_i \zeta (s_{i} , s_{-i}) }.
\end{equation*}
Given the notation set forth above, to establish the result we proceed by strong induction on the cardinality of $R_{-i}$.
%
%
%
%%% INDUCTION
%
% n = 0
%
\begin{itemize}[leftmargin=*]
\item $|R_{-i}| = 1$. Let $\widetilde{s}_{-i}$ be the unique element of $R_{-i}$ and assume that, for every information set $h \in H_i (s^{*}_i)$, if $R^{i} (h)$ is nonempty, then $s^{*}_i$ is not weakly dominated relative to $R^{i} (h)$. Trivially, since  $\underline{R}_{-i} = \{ \widetilde{s}_{-i} \}$, we have that  $\zeta (s^{*}_i , \widetilde{s}_{-i}) = \underline{\alpha}$. Now, define the utility function $\overline{u}_i$ as
\begin{equation*}
\overline{u}_i (z) :=%
\begin{cases}
1, & \text{ if } z \succsim_i \underline{\alpha} ,\\[7pt]
0, & \text{ if } \underline{\alpha} \succ_i z,
\end{cases}
\end{equation*}
with $z \in Z (R)$, and, for every $h \in H_i (s^{*}_i)$ such that $R^{i} (h) \neq \emptyset$, let $\mu_i \Rounds { \cdot | R_{-i} (h) } := \delta_{\{ \widetilde{s}_{-i} \}}$, where we let $\delta_{\{ \widetilde{s}_{-i} \}}$ denote the Dirac measure over $\widetilde{s}_{-i}$. Clearly, $\mu_i$ trivially satisfies the required condition that $\supp \mu_i \Rounds { \cdot | S_{-i} (\overline{h}) } =  R_{-i} (\overline{h})$, for every $\overline{h} \in H_i (s^{*}_i)$ for which $R_{-i} (\overline{h}) \neq \emptyset$. From the fact that $s^{*}_i$ is by assumption not weakly dominated relative to $R_i \times \Sets { \widetilde{s}_{-i} }$ and \Mref{rem:singleton}, $s^{*}_i$ is not strictly dominated relative to  $R_i \times \Sets { \widetilde{s}_{-i} }$. Hence, for every $s_i \in R_i$, $\zeta (s^{*}_i, \widetilde{s}_{-i}) \succsim_i \zeta (s_i, \widetilde{s}_{-i})$, i.e., $\underline{\alpha} \succsim_i \zeta (s_i, \widetilde{s}_{-i})$. Thus, opportunely rephrasing \Mref{eq:rational_given}, we have that
\begin{align*} 
\sum_{s_{-i} \in R_{-i}} u_i (\zeta(s^{*}_i , s_{-i})) \cdot \mu_i \Rounds {\{ s_{-i} \} | R_{-i} (h) } & = %
\overline{u}_i (\zeta (s^{*}_i , \widetilde{s}_{-i})) \cdot \mu_i \Rounds { \{ \widetilde{s}_{-i} \} | R_{-i} (h) } =\\%
& = \overline{u}_i (\zeta (s^{*}_i , \widetilde{s}_{-i})) \cdot \delta_{\{ \widetilde{s}_{-i} \}} =  \\%
& = \overline{u}_i (\underline{\alpha}) \cdot \delta_{\{ \widetilde{s}_{-i} \}} = 1 \geq %
\overline{u}_i (\zeta (s_i , \widetilde{s}_{-i})) \cdot \delta_{\{ \widetilde{s}_{-i} \}} ,
\end{align*}
for every $h \in H_i (s^{*}_i)$ and for every $s_i \in R_i (h)$, establishing the result.
%
%
% n > 0
%
\item $|R_{-i}| = n$, with $n > 1$. Assume that the statement is true for every $k \in \bN$ such that $1 < k \leq n$ and  that there exists no information set $h \in H_i (s^{*}_i)$ such that $s^{*}_i$ is weakly dominated relative to $R^{i} (h)$ by a strategy in $R_i (h)$. We split the proof of the strong induction step in two cases.

\begin{case}[$R_{-i} \setminus \underline{R}_{-i} \neq \emptyset$]
We have that $|R_{-i} \setminus \underline{R}_{-i}| < |R_{-i}|$ by assumption and that $s^{*}_i \in R_{i} \setminus \underline{R}_{i}$ by construction. By means of the strong induction hypothesis, this allows us to assume that the claim holds for $R_{i} \setminus \underline{R}_{i} \times R_{-i} \setminus \underline{R}_{-i}$. Thus, as a result of what is written above, we have in particular that there exist a utility function $\overline{u}_{i}|_{R} \in \Re^{Z (R)}$ and a CPS $\overline{\mu}_i \in \Delta^{\mcal{H}_i} (S_{-i})$ with $\supp \overline{\mu}_i \Rounds { \cdot | S_{-i} (\overline{h}) } =  (R_{-i} \setminus \underline{R}_{-i}) \cap S_{-i} (\overline{h})$, for every $\overline{h} \in H_i (s^{*}_i)$ for which $(R_{-i} \setminus \underline{R}_{-i}) \cap S_{-i} (\overline{h}) \neq \emptyset$ such that $s^{*}_i$ is a maximizer of $\overline{u}_i$ in $R_{-i} \setminus \underline{R}_{-i}$ with respect to $\overline{\mu}_i$. We now define a utility function $u^{\delta}_i (z) \in \Re^{Z (R)}$, with $\delta \in \Re_{>0}$, as follows:
\begin{equation}
\label{eq:case_1_utility}
u^{\delta}_i (z) :=%
\begin{cases}
\overline{u}_i  (z), & \text{ if } z \succsim_i \underline{\alpha} ,\\[7pt]
\overline{u}_i (z) - \delta, & \text{ if } \underline{\alpha} \succ_i z.
\end{cases}
\end{equation}
Also, for every $\varep \in (0, 1)$, we define a CPS $\mu^{\varep}_i \in \Delta^{\mcal{H}_i} (S_{-i})$ as 
\begin{equation*}
\mu^{\varep}_i \Rounds { \{ s_{-i} \} | S_{-i} (\overline{h}) } :=%
\begin{cases}
(1- \varep) \cdot \overline{\mu}_i \Rounds { \{ s_{-i} \} | S_{-i} (\overline{h}) } , & \text{ if } s_{-i} \in R_{-i} \setminus \underline{R}_{-i} ,\\%[10pt]
\\
\displaystyle \frac{\varep}{|\underline{R}_{-i}|}, & \text{ if } s_{-i} \in \underline{R}_{-i}, 
\end{cases}
\end{equation*}
for every $\overline{h} \in H_i (s^{*}_i)$ such that $R^{i} (\overline{h}) \neq \emptyset$, and notice that $\supp \mu^{\varep}_i \Rounds { \cdot | S_{-i} (\overline{h}) } =  R_{-i} (\overline{h})$. To see that this is actually a CPS, let  $\widehat{R}_{-i} \subseteq R_{-i}$ and observe that we have
\begin{equation}
\label{eq:CPS_proof}
\mu^{\varep}_i \Rounds { \widehat{R}_{-i} | S_{-i} (\overline{h})} = %
(1- \varep) \cdot \overline{\mu}_i \Rounds { \widehat{R}_{-i} \cap (R_{-i} \setminus \underline{R}_{-i}) | S_{-i} (\overline{h}) } + \varep
\end{equation}
for every $\overline{h} \in H_i (s^{*}_i)$ such that $R^{i} (\overline{h}) \neq \emptyset$.  We now show that the definition above satisfies Axioms (A1)-(A3) in the definition of CPS. The fact that Axiom (A2) is satisfied is immediate from \Mref{eq:CPS_proof}. It is immediate also that Axiom (A1) is satisfied since, by  letting $h \in H_i (s^{*}_i)$ be arbitrary, given $S_{-i} (h)$ as the conditioning event and $R^{i} (h)$ nonempty, we have  
\begin{align*}
\mu^{\varep}_i \Rounds {S_{-i} (h) | S_{-i} (h)} & = %
\mu^{\varep}_i \Rounds {R_{-i} (h) | S_{-i} (h)}  = \\%
& = (1- \varep) \cdot \overline{\mu}_i \Rounds { R_{-i} (h) \cap (R_{-i} \setminus \underline{R}_{-i}) | S_{-i} (h) } + \varep = 1,
\end{align*}
since $\overline{\mu}_i \Rounds { R_{-i} (h) \cap (R_{-i} \setminus \underline{R}_{-i}) | S_{-i} (h) } = 1$. Focusing on Axiom (A3), let $\varep \in (0, 1)$ and the conditioning events $S_{-i} (h), S_{-i} (h') \in \mcal{H}_i$ be arbitrary, assume that $\widehat{R}_{-i} \subseteq R_i$ as before and---additionally---that $\widehat{R}_{-i} \subseteq S_{-i} (h') \subseteq S_{-i} (h)$, and notice that we have that 
%
%
%%%%%%%%
\begin{align*}
\frac{\mu^{\varep}_i \Rounds{ \widehat{R}_{-i}|S_{-i} (h)} - \varep}%
{\mu^{\varep}_i \Rounds { S_{-i} (h') | S_{-i} (h)} - \varep} & =%
\frac% NOM
{(1- \varep) \cdot \overline{\mu}_i \Rounds { \widehat{R}_{-i} \cap (R_{-i} \setminus \underline{R}_{-i}) | S_{-i} (h) }}%
%DEN
{(1- \varep) \cdot \overline{\mu}_i \Rounds {S_{-i} (h') \cap (R_{-i} \setminus \underline{R}_{-i})| S_{-i} (h) }} = \\%
& = (1- \varep) \cdot \frac% NOM
{\overline{\mu}_i \Rounds { \widehat{R}_{-i} \cap (R_{-i} \setminus \underline{R}_{-i}) | S_{-i} (h) }}%
%DEN
{\overline{\mu}_i \Rounds {S_{-i} (h') \cap (R_{-i} \setminus \underline{R}_{-i})| S_{-i} (h) }} = \\%
& = (1- \varep) \cdot \overline{\mu}_i \Rounds { \widehat{R}_{-i} \cap (R_{-i} \setminus \underline{R}_{-i}) | S_{-i} (h') }  = \\
& = \mu^{\varep}_i \Rounds { \widehat{R}_{-i} | S_{-i} (h')} - \varep,
\end{align*}
%%%%%%
%
%
where the second to last equality comes from the fact that $\overline{\mu}_i$ is a CPS. Thus, notice that if we choose $\varep$ small enough, $s^{*}_i$ continues to be an expected utility maximizer of $\overline{u}_i$ with respect to $R_i \setminus \underline{R}_i$ according to $\mu^{\varep}_i$, since utility is continuous with respect to CPSs at every $S_{-i} (h)$. We let $\overline{\varep}$ denote such  value, we fix it, and we let $\delta$ vary. Since increasing $\delta$ affects only the utility of the strategies in $\underline{R}_i$, which can be made in this way arbitrarily small, as a result, for a $\delta$ large enough, which we denote by $\overline{\delta}$, $s^{*}_i$ is an expected utility maximizer in $R_i$ of $u^{\overline{\delta}}_i$ with respect to  the CPS $\mu^{\overline{\varep}}_i$.
\end{case}

\begin{case}[$R_{-i} \setminus \underline{R}_{-i} = \emptyset$]
Let $\mu_i \in \Delta^{\mcal{H}_i} (S_{-i})$ be an arbitrary CPS with support on $R_{-i}$. Let $\overline{u}_{i}|_{R} \in \Re^{Z (R)}$ be an arbitrary utility function and let $u^{\delta}_i$ be defined as in \Mref{eq:case_1_utility} from Case (1). Thus, with $\delta$ large enough, the utility function $u^{\delta}_i$ and the CPS $\mu_i$ satisfy the desired properties.
\end{case}
\end{itemize}
This completes the proof of the lemma. \qedhere
\end{proof}

\begin{lemma}
\label{lem:main_lemma}
Given a restriction $R \subseteq S$ and a player $i \in I$,  a strategy $s^{*}_i \in R_i$ is cautiously sequentially rational given $R$ if and only if, for every information set $h \in H_i (s^{*}_i)$, if $R^{i} (h)$ is nonempty, then $s^{*}_i$ is not weakly dominated relative to $R^{i} (h)$.
\end{lemma}

\begin{proof}
$[\Rightarrow]$ We assume that $s^{*}_i \in R_i$ is cautiously sequentially rational given $R$ and---proceeding by contradiction---that there exists an information set $\overline{h} \in H_i (s^{*}_i)$ such that $s^{*}_i$ is weakly dominated relative to $R^{i} (\overline{h})$, with $R^{i} (\overline{h})$ nonempty. Hence, there exists a strategy $\overline{s}_i \in R_i (\overline{h})$ such that  $\zeta (\overline{s}_i, s_{-i}) \succsim_i \zeta (s^{*}_i, s_{-i})$ for every $s_{-i} \in R_{-i} (\overline{h})$ and there exists a $s^{*}_{-i} \in R_{-i} (\overline{h})$ such that  $\zeta (\overline{s}_i, s^{*}_{-i}) \succ_i \zeta (s^{*}_i, s^{*}_{-i})$. From the assumption that $s^{*}_i$ is cautiously sequentially rational given $R$, there exist a utility function $u_{i}|_{R} \in \Re^{Z (R)}$ and a CPS $\mu_i \in \Delta^{\mcal{H}_i} (S_{-i})$, with $\supp \mu_i \Rounds { \cdot | S_{-i} (h) } =  R_{-i} (\overline{h})$ for every $h \in H_i (s^{*}_i)$ for which $R_{-i} (h) \neq \emptyset$,  such that \Mref{eq:rational_given} is satisfied by $s^{*}_i$. But, since $\overline{s}_i$ weakly dominates $s^{*}_i$ at $\overline{h}$ and $\supp \mu_i \Rounds { \cdot | S_{-i} (\overline{h}) } =  R_{-i} (\overline{h})$, we have that 
\begin{equation*} 
\sum_{s_{-i} \in R_{-i}} u_i (\zeta (\overline{s}_i , s_{-i})) \cdot %
\mu_i \Rounds {\{ s_{-i} \} | S_{-i} (\overline{h})} > %
\sum_{s_{-i} \in R_{-i}} u_i (\zeta (s^{*}_i , s_{-i})) \cdot %
\mu_i \Rounds {\{ s_{-i} \}| S_{-i} (\overline{h}) \}}, 
\end{equation*}
contradicting the fact that \Mref{eq:rational_given} is satisfied by $s^{*}_i$ at every $h \in H_i (s^{*}_i)$.\\
\noindent $[\Leftarrow]$ Assume that for every information set $h \in H_i (s^{*}_i)$, if $R^{i} (h)$ is nonempty, then $s^{*}_i$ is not weakly dominated relative to $R^{i} (h)$ by a strategy in $R_i (h)$. Then the result follows immediately from \Mref{lem:claim}.
\end{proof}

\begin{proof}[Proof of \Mref{lem:given_rationality_conditional}]
$[\Rightarrow]$ Let $R \subseteq S$ be an arbitrary restriction and assume that $s^{*}_i$ is sequentially rational given $R$. Hence, there exist a utility function $u_{i}|_{R} \in \Re^{Z (R)}$ and a CPS $\mu_i \in \Delta^{\mcal{H}_i} (R_{-i})$ such that for every $h \in H_i (s^{*}_i)$ if $R^{i} (h)$ is nonempty, then \Mref{eq:rational_given_gen} is satisfied by $s^{*}_i$ for every $s_i \in R_i (h)$. We proceed by contradiction and we assume that $s^{*}_i$ is conditionally B-dominated with respect to $R$. Hence, there exists an information set $\overline{h} \in H_i (s^{*}_i)$ such that $R^i (\overline{h})$ is nonempty and $s^{*}_i \in \bd_i (R^i (\overline{h}))$. Thus, for every $Q_{-i} (\overline{h}) \subseteq R_{-i} (\overline{h})$, we have that $s^{*}_i$ is weakly dominated relative to $R_i (\overline{h}) \times Q_{-i} (\overline{h})$ by a strategy $s_i \in R_i (\overline{h})$. Let $Q^{\mu_i}_{-i} (\overline{h}):= \supp \mu_i \Rounds { \cdot | S_{-i} (\overline{h})}$, with $Q^{\mu_i}_{-i} (\overline{h}) \subseteq R_{-i} (\overline{h})$ by definition, and  let $\overline{s}_i \in R_i (\overline{h})$ denote the strategy that weakly dominates $s^{*}_i$ relative to $R_i (\overline{h}) \times Q^{\mu_i}_{-i} (\overline{h})$. Thus,  $\zeta (\overline{s}_i, s_{-i}) \succsim_i \zeta (s^{*}_i, s_{-i})$ for every $s_{-i} \in Q^{\mu_i}_{-i} (\overline{h})$ and there exists a $s^{*}_{-i} \in Q^{\mu_i}_{-i} (\overline{h})$ such that  $\zeta (\overline{s}_i, s^{*}_{-i}) \succ_i \zeta (s^{*}_i, s^{*}_{-i})$. But then we have that
\begin{equation*}
\sum_{s_{-i} \in R_{-i}} u_i (\zeta(\overline{s}_i , s_{-i})) \cdot %
\mu_i \Rounds {\{ s_{-i} \} | S_{-i} (\overline{h}) } > %
\sum_{s_{-i} \in R_{-i}} u_i (\zeta(s^{*}_i , s_{-i})) \cdot %
\mu_i \Rounds { \{ s_{-i} \} | S_{-i} (\overline{h}) },
\end{equation*}
contradicting the fact that \Mref{eq:rational_given_gen} is satisfied by $s^{*}_i$ at every $h \in H_i (s^{*}_i)$.\\
\noindent $[\Leftarrow]$ Let $R \subseteq S$ be an arbitrary restriction and assume that $s^{*}_i$ is not conditionally B-dominated with respect to $R$. Hence, for every $h \in H_i (s^{*}_i)$, if $R^i (h)$ is nonempty, then there exists a nonempty subset $Q_{-i} (h) \subseteq R_{-i} (h)$ such that $s^{*}_i \in \Ad_i (R_{i} (h) \times Q_{-i} (h))$. Let $Q_{-i} := \bigcup_{\overline{h} \in H_i (s^{*}_i)} Q_{-i} (\overline{h})$ and notice that, from \Mref{lem:main_lemma},  $s^{*}_i$ is cautiously sequentially rational given $R_i \times Q_{-i}$. Hence, from \Mref{rem:given_rationality}, $s^{*}_i$ is---\emph{a fortiori}---sequentially rational given $R$.
\end{proof}

\begin{proof}[Proof of \Mref{th:rationality}]
The result follows immediately from \Mref{lem:given_rationality_conditional} with $R := S$.
\end{proof}

%%%%%%%%%%%%%%%%%%%%%%%%%%%%%%%%%%%%
%%% Paragraph
%
\paragraph{\texorpdfstring{Decision-Theoretic Foundation of  \Sref{subsubsec:DT_take}}{Decision-Theoretic Foundation of Section 3.3.2}}
\label{parapp:DT_take}

We let $\States$ denote the set of \emph{states of nature} endowed with $\sigma$-algebra $\sAlg_\States \subseteq 2^{\States}$,  $\Outcomes$ denote the set of \emph{consequence}, and we let $\Acts := \Outcomes^\States$ denote the set of \emph{acts}. Given an event $E \in \sAlg_\States$ and acts $f, g \in \Acts$, we let $g^{f}_E$ denote the act that is equivalent to $f$ on $E$ and to $g$ on $E^c$. An $f \in \Acts$ is \emph{constant} if there exists a consequence $x \in \Outcomes$ such that $f (\state) = x$ for every $\state \in \States$, where---as it is customary---we let a constant act be denoted by the consequence it is attached to. Now, we endow the set of acts $\Acts$ with a binary relation $\prefeq \subseteq \Acts \times \Acts$ (with symmetric and asymmetric parts canonically defined), that induces a binary relation $\outprefeq \subseteq \Outcomes \times \Outcomes$, and we say that an event $E \in \sAlg_\States$ is \emph{null} (with respect to $\prefeq$) if 
\begin{equation*}
f^{h}_E \prefeq g^{h'}_E \iff f \prefeq g ,
\end{equation*}
for every $f, g, h, h' \in \Acts$, where we let $\NonNull$ denote the set of non-null states, with $\sAlg_\NonNull$ denoting the corresponding $\sigma$-algebra. Finally, we let $\Sets { \prefeq^E }_{E \in \sAlg_\NonNull}$ denote a family of binary relations indexed by non-null events.

We can now state the axioms introduced in \cite{Ghirardato_2002} relevant for our purposes by following the numeration used there.\footnote{Alternatively, staying truer to the tree-like structure of our setting as described in \Sref{subsec:dynamic_ordinal_games}, we could have employed a version of the setting in \citet[Section 3.1]{Siniscalchi_2011} with an opportune translation of the axioms.}

\begin{GhirardatoAxiom}[Weak Order]
\label{ax:Ghirardato_Weak_Order}
For every $E \in \sAlg_\NonNull$, the binary relation $\prefeq^E \subseteq \Acts \times \Acts$ is a weak order.
\end{GhirardatoAxiom}

\begin{GhirardatoAxiom}[Dynamic Consistency]
\label{ax:Ghirardato_Dynamic_Consistency}
For every $E \in \sAlg_\NonNull$ and $f, g \in \Acts$,
\begin{equation}
\label{eq:Ghirardato_Dynamic_Consistency}
f \prefeq^E g \iff g^{f}_E \prefeq g .
\end{equation}
\end{GhirardatoAxiom}

\begin{GhirardatoAxiom}[Ordinal Preference Consistency]
\label{ax:Ghirardato_Ordinal_Preference_Consistency}
For every $E \in \sAlg_\NonNull$ and $x, y \in \Outcomes$,
\begin{equation}
\label{eq:Ghirardato_Ordinal_Preference_Consistency}
x \outprefeq y \iff x \outprefeqE y .
\end{equation}
\end{GhirardatoAxiom}

\setcounter{GhirardatoAxiom}{4}
\begin{GhirardatoAxiom}[Non-Triviality]
\label{ax:Ghirardato_Non-Triviality}
There exist $x, y \in \Outcomes$ such that $x \outpref y$.
\end{GhirardatoAxiom}

\setcounter{GhirardatoAxiom}{6}
\begin{GhirardatoAxiom}[Consequentialism]
\label{ax:Ghirardato_Consequentialism}
For every $E \in \sAlg_\NonNull$ and $f, g \in \Acts$, if $f (\state) = g (\state)$ for every $\state \in E$, then $f \bm{\sim}^E g$.
\end{GhirardatoAxiom}

\setcounter{GhirardatoAxiom}{8}
\begin{GhirardatoAxiom}[Bayesian Updating]
\label{ax:Ghirardato_Bayesian_Updating}
There exists an $\overline{h} \in \Acts$ such that for every $E \in \sAlg_\NonNull$ and $f, g \in \Acts$
\begin{equation}
\label{eq:Ghirardato_Bayesian_updating}
f \prefeq^E g \iff \overline{h}^{f}_E \prefeq \overline{h}^{g}_E .
\end{equation}
\end{GhirardatoAxiom}

The following result links the setting introduced above and \Mref{def:conditional_monotonicity}, which is essentially this very proposition stated for a finite setting.

\begin{namedtheorem}[{Theorem 3 (\citet[Chapter 2.7, p.26]{Savage_1972})}]
\label{th:Savage_Th3}
Given an event $E \in \sAlg_\States$, a finite partition $\mscr{E}$ of $E$, and acts $f, g \in \Acts$:
\begin{itemize}[leftmargin=*]
\item (Weak Condition) if $f (\state) \outprefeq g (\state)$ for every $E_k \in \mscr{E}$ and $\state \in E_k$, then $f \prefeq^E g$;

\item (Strong Condition) if $f (\state) \outpref g (\state)$ for every $E_k \in \mscr{E}$ and $\state \in E_k$ and there exists an $E_{k^*} \in \mscr{E}$ such that $f (\state^*) \outpref g (\state^*)$ for every $\state^* \in E_{k^*}$, then $f \pref^E g$.
\end{itemize}
\end{namedtheorem}

Rather crucially, as pointed out above, this is an actual result in \cite{Savage_1972}, where it is shown that this is equivalent to Axiom P3 ``Eventwise Monotonicity'' as stated in \citet[P3, p.26]{Savage_1972}. It should be stressed that in the framework of \cite{Ghirardato_2002}, this is a result that obtains only if all the axioms stated above are assumed. This last point is particularly important, because---in particular---it implies that Axiom P2 ``Sure-Thing Principle'' of \citet[P2, p.23]{Savage_1972} has to be assumed, which is a point spelled out in \citet[Lemmata 1, 2, \& 3, pp.89--91]{Ghirardato_2002}.

%%%%%%%%%%%%%%%%%%%%%%%%%%%%%%%%%%%%%%%%%%
%%%%%%%%%%%%%%%%%%%%%%%%%%%%%%%%%%%%%%%%%%
%%%%%%%%%%%%%%%%%%%%%%%%%%%%%%%%%%%%%%%%%%
%%%%%%%%% SUB-Section %%%%%%%%%%%%%%%%%%%%%%%%%%%%%%%%%%%%%%%%%%
%%%%%%%%%%%%%%%%%%%%%%%%%%%%%%%%%%%%%%%%%%
%%%%%%%%%%%%%%%%%%%%%%%%%%%%%%%%%%%%%%%%%%
\subsection{\texorpdfstring{Proofs of \Sref{sec:revealed_FI}}{Proofs of Section 4}}
\label{subapp:proofs_section_4}

%%%%%%%%%%%%%%%%%%%%%%%%%%%%%%%%%%%%%%%%%%
%%%%%%%%% SUB-Sub-Section %%%%%%%%%%%%%%%%%%%%%%%%%%%%%%%%%%%%%%%%%%
\subsubsection{\texorpdfstring{Proofs of \Sref{subsec:solution_concepts}}{Proofs of Section 4.1}}
\label{subsubapp:proofs_section_4}

To prove \Mref{lem:basic_nonemptiness} we recall some basic definitions from order theory, which we use to point out a property of weak dominance. Given an arbitrary set $Y$, a binary relation $\gtrdot \subseteq Y \times Y$ on $Y$ is a \emph{strict partial order} if: (Irreflexivity) for every $x \in Y$, it is not the case that $x \gtrdot x$; (Transitivity) for every $x, y, z \in Y$, if $x \gtrdot y$ and $y \gtrdot z$, then $x \gtrdot z$. An arbitrary (finite) set $Y$ endowed with a strict partial order is called a \emph{(finite) strict poset}.

\begin{remark}
\label{rem:strict_poset}
Given a finite strict poset $(Y, \gtrdot)$, there exists a maximal element in $Y$. 
\end{remark}

For every $i \in I$, weak dominance as in \Sref{subsec:dominance_strategic} is a binary relation on the strategy set $S_i$, which is a strict partial order.\footnote{See \citet[Section 3]{Gilboa_et_al_1990}.}

\begin{proof}[Proof of \Mref{lem:basic_nonemptiness}]
\label{proof:basic_nonemptiness}
Let $R \subseteq S$ be an arbitrary restriction. Proceeding by contradiction, assume that $\Und (R) = \emptyset$. Hence, there exists a player $i \in I$ such that $\Und_{i} (R) = \emptyset$, i.e., for every $s_i \in R_i$ there exists an information set $h \in H_i (s_i)$ such that $R^i (h)$ is nonempty and $s_i \in \bd_i (R^i (h))$. Thus, for every $Q_{-i} (h) \subseteq R_{-i} (h)$ there exists a strategy $\overline{s}_i \in R_i (h)$ such that $\overline{s}_i$ weakly dominates $s_i$ relative to $R_i (h) \times Q_{-i} (h)$. From \Mref{lem:strong_replacement}, there exists a strategy $s^{*}_i \in R_i (h)$ that agrees with $\overline{s}_i$ on $R_{-i} (h)$ and that agrees with $s_i$ on $R_{-i} \setminus R_{-i} (h)$ that weakly dominates $s_i$ relative to $R$.  Since $s_i$ was chosen arbitrarily, the same argument can be applied to every strategy in $R_i$, i.e., every strategy $\widetilde{s}_i \in R_i$ is weakly dominated relative to $R$ by a strategy $\widetilde{s}^{*}_i \in R_i$. However, this implies that there exists no maximal element in $R_i$ with respect to weak dominance, which is a strict partial ordering. Hence, from the finiteness of the dynamic games under scrutiny and \Mref{rem:strict_poset}, we obtain the  desired contradiction.
\end{proof}

\begin{proof}[Proof of \Mref{prop:nonemptiness}]
We prove the two properties in order:
\begin{itemize}
\item[i)] From \Mref{lem:basic_nonemptiness}, for every $R \subseteq S$, $\Und (R)$ is nonempty. Also, from \Mref{rem:operator_monotonicity}, for every $R \subseteq S$, $\Und (R) \subseteq R$ implies that $\Und^{k} (R) \subseteq \Und^{k - 1} (R)$, for every $k \in \bN$. Hence, for every $\ell \in \bN$, $\Und^{\ell} (S)$ is nonempty, i.e., $\bigcap_{\ell \geq 0} \Und^{\ell} (S) \neq \emptyset$.

\item[ii)] From Property (i) established above, since $S$ is finite, there exists a $K \in \bN$ such that  $\Und^{K} (S) = \Und^{K+1} (S) = \Und^{\infty} (S)$.
\end{itemize}
Thus, what is written above establishes the result. 
\end{proof}

%%%%%%%%%%%%%%%%%%%%%%%%%%%%%%%%%%%%%%%%%%
%%%%%%%%% SUB-Sub-Section %%%%%%%%%%%%%%%%%%%%%%%%%%%%%%%%%%%%%%%%%%
\subsubsection{\texorpdfstring{Proof of \Sref{subsec:relation_ICD}}{Proof of Section 4.2}}
\label{subsubapp:proofs_section_4.2}

\begin{proof}[Proof of \Mref{prop:ICD_ICBD}]
We proceed by proving the contrapositive. Thus, let $R \subseteq S$ and $s \notin \Und (R)$ be arbitrary. Hence, there exists a player $i \in I$ and an information set $h \in H_i (s_i) $ with $R^i (h)$ nonempty such that $s_i \in \bd_i (R^i (h))$. By definition of B-dominance, this implies that for every $Q_{-i}  (h) \subseteq R_{-i} (h)$ there exists a strategy $\widetilde{s}_i \in R_i (h)$ such that $\widetilde{s}_i$ weakly dominates $s_i$ relative to $R_i (h) \times Q_{-i} (h)$. In particular, from \Mref{rem:singleton}, for every singleton subset  $\{ s_{-i}\} \subseteq R_{-i} (h)$ there exists a strategy $\overline{s}_i \in R_i (h)$ such that $\overline{s}_i$ strictly dominates $s_i$ relative to $R_i (h) \times \{ s_{-i}\}$. From the finiteness assumption and the fact that $R^i (h)$ is nonempty, there exists at least one $s^{*}_{-i}$ such that $\{ s^{*}_{-i} \} \subseteq R_{-i} (h)$. Let $s^{*}_i \in R_i (h)$ denote the corresponding strategy of player $i$ such that $s^{*}_i$ strictly dominates $s_i$ relative to $R_i (h) \times \{ s^{*}_{-i} \}$. Hence, recalling the notation of Dirac measure employed in the proof of \Mref{lem:claim}, strategy $s_i \in \md_i (R^i (h))$ via the (degenerate) mixed strategy $\delta_{\{s^{*}_i\}} \in \Delta (R_i (h))$.
\end{proof}

%%%%%%%%%%%%%%%%%%%%%%%%%%%%%%%%%%%%%%%%%%
%%%%%%%%% SUB-Sub-Section %%%%%%%%%%%%%%%%%%%%%%%%%%%%%%%%%%%%%%%%%%
\subsubsection{\texorpdfstring{Proofs of \Sref{subsec:rationalizing_FI}}{Proofs of Section 4.3}}
\label{subsubapp:proofs_section_4.3}

For the purpose of the proofs that follow, we introduce new notation. In particular,  for every $i \in I$ and $R \subseteq S$, we let $\md^{u_i}_i (R)$ denote the set of strategies of player $i$ that are strictly dominated relative to $R$ by mixed strategies given the utility function $u_i \in \mcal{U}_i$. 

Before proceeding with the proof, a clarification is in order.  Whereas in the following we repeatedly refer to  \citet[Proposition, p.427]{Borgers_1993} in presence of a given restriction $R \subseteq S$, two points need to be emphasized regarding this usage.
\begin{enumerate}
\item Strictly speaking, no reference to restrictions $R \subseteq S$ is made in \citet[Proposition, p.427]{Borgers_1993}. However, It is immediate to establish such a result from  \Mref{lem:given_rationality_conditional} opportunely modified to deal with the static case.
\item The ``given'' in the notion we use has to be read in light of the previous point and \emph{not} in the way in which \cite{Borgers_1993} uses this term, which corresponds---has already mentioned in \Sref{subsubapp:characterization} at  \Sref{foot:cautious}---to our notion of \emph{cautious} sequential rationality given a restriction.
\end{enumerate}

\begin{proof}[Proof of \Mref{lem:ICD_general}]
$[\Rightarrow]$  We let $R \subseteq S$ and $i \in I$ be arbitrary, we assume that $s_i \in \Und_i (R) [\succsim_i]$, and we proceed by contradiction. Hence, we assume that $s_i \notin \ICD_i (R) [u_i]$, for every $u_i \in \Utilities_i$ and corresponding $u_{i}|_R$. Thus, for every $u_i \in \Utilities_i$  and corresponding $u_{i}|_R$ there exists an information set $h^{u_i} \in H_i (s_i)$ such that $R^{i} (h)$ is nonempty and $s_i \in \md_i (R^i (h^{u_i}))$, where we collect all these information sets in the set  $\mcal{H}_i [s_i | \Utilities_i]$, i.e., 
\begin{equation*}
\mcal{H}_i [s_i | \Utilities_i] := \Set { h^{u_i} \in H_i (s_i) | %
\exists u_i \in \Utilities_i : \ %
R^{i} (h^{u_i}) \neq \emptyset , \ %
s_i \in \md_i (R^i (h^{u_i})) } .
\end{equation*} 
Now, we observe that every information set $\widehat{h} \in \mcal{H}_i [s_i | \Utilities_i]$ induces a $2$-player (static)  ordinal game with players $i$ and $-i$ and strategy sets $R_{i} (\widehat{h}) \times R_{-i} (\widehat{h})$  at which $s_i \notin \bd_i (R^i (\widehat{h}))$ in light of our original assumption that $s_i \in \Und_i (R) [\succsim_i]$. However, the fact that  $s_i \notin \bd_i (R^i (\widehat{h}))$ implies in the corresponding $2$-player static game by  \citet[Proposition, p.427]{Borgers_1993} and \citet[Lemma 3, p.1048]{Pearce_1984} that there exists a utility function $\widehat{u}_i \in \mcal{U}_i$ with corresponding $\widehat{u}_{i}|_{R}$ such that $s_i \notin \md^{\widehat{u}_i}_i (R^i (\widehat{h}))$, which leads to a contradiction and establishes the result.\\ 
\noindent $[\Leftarrow]$ We proceed by proving the contrapositive. Thus, we let $R \subseteq S$ and $i \in I$ be arbitrary and we assume that $s_i \notin \Und_i (R) [\succsim_i]$, that is, strategy $s_i$ is conditionally B-dominated with respect to $R$. Hence, there exists an information set $\widehat{h} \in H_i (s_i)$ such that $R^{i} (\widehat{h})$ is nonempty and $s_i \in \bd_i (R^i (\widehat{h}))$. Now, observe that the information set $\widehat{h} \in H_i (s_i)$ induces a $2$-player (static)  ordinal game with players $i$ and $-i$ and strategy sets $R_{i} (\widehat{h}) \times R_{-i} (\widehat{h})$. Thus,  from  \citet[Proposition, p.427]{Borgers_1993} taking into account the presence of the restriction $R \subseteq S$, it follows that $s_i$ is not rational given $R$ for every $\mu_i \in \Delta (R_{-i} (\widehat{h}))$ and $u_{i}|_R \in \Re^{Z (R)}$. Hence, from \citet[Lemma 3, p.1048]{Pearce_1984}, it follows that, for every utility function $u_i \in \Utilities_i$ restricted to $R$, strategy $s_i$ is strictly dominated relative to $R^i (\widehat{h})$ by a mixed strategy $\sigma^{*}_i \in \Delta (R_{i} (\widehat{h}))$, which establishes the result.
\end{proof}

\begin{proof}[Proof of \Mref{prop:ICD_general}]
We prove the result by strong induction.
\begin{itemize}[leftmargin=*]
\item ($n = 0$) It is immediate to establish that there exists a $u^0 \in \Re^{\abs{I} \cdot Z}$ such that 
\begin{equation*}
\Und^{1} (S) [\succsim] = \Und ( \Und^{0} (S) ) [\succsim] = %
\Und (S) [\succsim ] = %
\ICD (S) [u^0] = \ICD ( \ICD^{0} (S) ) [u^0] = % 
\ICD (S) [u^0] 
\end{equation*}
in light of the fact that the third equality comes from \Mref{lem:ICD_general} with $R := S$.

\item ($n \geq 1$) Assume that there exists a $u^{n-1} \in \Re^{\abs{I} \cdot Z}$ such that $\Und^{n} (S) [\succsim] = \ICD^{n} (S) [u^{n-1}]$. Now, let $R :=  \Und^{n} (S) [\succsim] = \ICD^{n} (S) [u^{n-1}]$. Thus, from \Mref{lem:ICD_general},  there exists a $u^{n} \in \Re^{\abs{I} \cdot Z}$ such that 
\begin{equation*}
\Und^{n+1} (S) [\succsim] = \Und ( \Und^{n} (S) ) [\succsim] = %
\Und (R) [\succsim ] = %
\ICD (R) [u^n] = \ICD ( \ICD^{n} (S) ) [u^n] = % 
\ICD^{n+1} (S) [u^n] .
\end{equation*}
\end{itemize}
Thus, what is written above establishes the result.
\end{proof}

%%%%%%%%%%%%%%%%%%%%%%%%%%%%%%%%%%%%%%%%%%
%%%%%%%%%%%%%%%%%%%%%%%%%%%%%%%%%%%%%%%%%%
%%%%%%%%%%%%%%%%%%%%%%%%%%%%%%%%%%%%%%%%%%
%%%%%%%%% SUB-Section %%%%%%%%%%%%%%%%%%%%%%%%%%%%%%%%%%%%%%%%%%
%%%%%%%%%%%%%%%%%%%%%%%%%%%%%%%%%%%%%%%%%%
%%%%%%%%%%%%%%%%%%%%%%%%%%%%%%%%%%%%%%%%%%
\subsection{\texorpdfstring{Proofs of \Sref{sec:relevant_ties}}{Proofs of Section 5}}
\label{subapp:proofs_relevant_ties}

Given that in this section we focus on NRT dynamic games, which are \emph{a fortiori} dynamic games with perfect information, it is understood that when we use the word ``information set'' (with related notation), that corresponds to a history.

\begin{proof}[Proof of \Mref{lem:fundamental_NRT}]
$[\Rightarrow]$  Let $\Game$ be an NRT dynamic game with $R \subseteq S$ an arbitrary restriction and $i \in I$ an arbitrary player. We prove the contrapositive and let $\underline{s}_i \notin \Ad_i (R)$ be arbitrary. If there exists an $s^{*}_i$ that strictly dominates $\underline{s}_i$ relative to $R$, then the result is established. Thus, we assume that there exists an $\widehat{s}_i \in R_i$ that weakly dominates $\underline{s}_i$ relative to $R$. Hence, we can obtain the sets
\begin{align*}
\overline{R}_{-i} & := \Set { \overline{s}_{-i} \in R_{-i} | %
\zeta (\widehat{s}_i, \overline{s}_{-i}) \sim_i \zeta (\underline{s}_i, \overline{s}_{-i}) } , \\
\widehat{R}_{-i} & := \Set { \widehat{s}_{-i} \in R_{-i} | %
\zeta (\widehat{s}_i, \widehat{s}_{-i}) \succ_i \zeta (\underline{s}_i, \widehat{s}_{-i}) } , 
\end{align*}
that partition $R_{-i}$, with $\widehat{R}_{-i}$ nonempty. Since $\widehat{s}_i \neq \underline{s}_i$, there exists an information set $h^* \in H_i (\widehat{s}_i) \cap H_i (\underline{s}_i)$ (i.e., a history, given the setting) such that $\widehat{s}_i (h^*) \neq \underline{s}_i (h^*)$.  Now, given the assumption that the dynamic game is NRT, for every $s^{*}_{-i} \in R_{-i} (h^*)$ it has to be the case that
\begin{equation*}
\zeta (\widehat{s}_i, s^{*}_{-i}) \not\sim_i \zeta (\underline{s}_i, s^{*}_{-i})  ,  
\end{equation*}
which, in light of the fact that $\widehat{s}_i \in R_i$ weakly dominates $\underline{s}_i$ relative to $R$, implies \emph{a fortiori} that
\begin{equation*}
\zeta (\widehat{s}_i, s^{*}_{-i}) \succ_i \zeta (\underline{s}_i, s^{*}_{-i})  ,  
\end{equation*}
for every $s^{*}_{-i} \in R_{-i} (h^*)$, where it has to be observed that this implies that $R_{-i} (h^*) \subseteq \widehat{R}_{-i}$. Hence, $\widehat{s}_i$ strictly dominates $\underline{s}_i$ relative to $R^i (h^*)$ and it follows from \Mref{rem:relation_dominance} that $\underline{s}_i \in \bd_i (R^i (h^*))$. Thus, it follows from the definition of the $\Und$ operator that $\underline{s}_i \notin \Und_i (R)$, which establishes the result.

%%%
\noindent $[\Leftarrow]$ The result is an immediate corollary of \Mref{lem:fundamental_CBD_WD}. 
\end{proof}

Now, we fix a dynamic game $\Game$ whose strategic form representation  $\Game^r$ satisfies the TDI condition as in \Mref{def:TDI} and we define the following algorithmic procedure on $\Game^r$:  $\mbf{U}^{0}_i := S_i$,  $\mbf{U}^{n+1}_i := \Und^{n+1}_i (S)$, and $\mbf{U}^{\infty}_i  := \bigcap_{\ell \geq 0} \mbf{U}^{\ell}_i$, where $\mbf{U}^{k} := \prod_{j \in I} \mbf{U}^{k}_j$ for every $k \geq 0$. Finally, $\mbf{U}^{\infty}  := \bigcap_{\ell \geq 0} \mbf{U}^{\ell}$ is the set of strategy profiles that survive the ICBD algorithm \emph{in} $\Game^r$. Observe that, in the specific class of games under scrutiny,  $\mbf{U}^{\infty}$ is well-defined. 

The---crucial for the purpose of our proof---fact that $\mbf{U}^{\infty}$ gives rise to a full reduction by weak dominance on $S$ is stated and proved next. To accomplish this goal, first of all, we let $\cbd_i (R)$ denote the set of player $i$'s strategy that are conditionally B-dominated with respect to $R \subseteq S$, with $\cbd (R) := \prod_{j \in I} \cbd_j (R)$. This allows us to rewrite \Mref{lem:fundamental_NRT} in the following, more compact, form:
\begin{equation}
\label{eq:fundamental_CBD_WD}
s_i \in \cbd_i (R) \Llra s_i \notin \Ad_i (R),
\end{equation}
for every $R \subseteq S$, $i \in I$,  and $s_i \in R_i$. Also, we can now rewrite $\mbb{U}^n (R)$ as
\begin{equation}
\label{eq:operator_rewriting}
\Und^{n} (R) = \Und^{n-1} (R) \setminus \cbd ( \Und^{n-1} (R)). 
\end{equation}

\begin{lemma}
\label{lem:full_red}
Given a dynamic game $\Game$ with corresponding  strategic form representation $\Game^r$ satisfying the  TDI condition, $\mbf{U}^\infty$ is a full reduction by weak dominance on $S$.
\end{lemma}

\begin{proof}
Fix a  dynamic game $\Game$ with corresponding strategic form representation $\Game^r$  satisfying the  TDI condition as in \Mref{def:TDI}. We prove by induction on $n \in \bN$ that $\Und^{\infty}$ as defined above is a reduction by weak dominance. 
\begin{itemize}[leftmargin=*]
\item ($n = 0$) By definition, we have that $\mbf{U}^0 := \Und^0 (S) = S$.
\item ($n \geq 1$) Assume the result has been established for $n \in \bN$. To establish that $\mbf{U}$ is a reduction by weak dominance, first of all, observe that we have by definition that $\mbf{U}^{n+1} = \Und^{n+1} (S)$. From \Mref{eq:operator_rewriting}, this in turn is equivalent to $\Und^{n+1} (S) = \Und^{n} (S) \setminus \cbd (\Und^n (S))$. The result follows immediately from \Mref{lem:fundamental_NRT} in the form of \Mref{eq:fundamental_CBD_WD}.
\end{itemize}
Since the dynamic games under scrutiny are finite, what is written above establishes that $\mbf{U}^\infty$ is a reduction by weak dominance. That $\mbf{U}^\infty$ is additionally a \emph{full} reduction by weak dominance is an immediate consequence of \Mref{lem:fundamental_NRT} and \Mref{prop:nonemptiness}. Hence, this establishes the result.
\end{proof}

\begin{remark}[{\citet[Corollary 1, p.230]{Marx_Swinkels_1997}}]
\label{rem:corollary_MS97}
Given a dynamic game with corresponding strategic form representation $\Game^r$  satisfying the TDI condition, if $\mbf{X}^\infty$ and $\mbf{Y}^\infty $ are two full reductions by weak dominance on $\Game^r$, then $\zeta (\mbf{X}^\infty ) = \zeta (\mbf{Y}^\infty)$. 
\end{remark}

\begin{proof}[Proof of \Mref{prop:relevant_ties}]
Fix an NRT dynamic game with corresponding strategic form representation $\Game^r$. Since the dynamic game satisfies the NRT condition, from \Mref{rem:NRT_TDI}, $\Game^r$ satisfies the TDI condition. Hence, from \Mref{rem:Moulin}, $\IA^\infty$ applied on $\Game^r$ delivers the backward induction outcome, i.e., $\zeta ( \IA^{\infty}) = \{ z^{BI} \}$. The result follows immediately from \Mref{lem:full_red} and \Mref{rem:corollary_MS97}  with $\mbf{U}^\infty$ and $\IA^{\infty}$.
\end{proof}

%%%%%%%%%%%%%%%%%%%%%%%%%%%%%%%%%%%%%%%%%%
%%%%%%%%%%%%%%%%%%%%%%%%%%%%%%%%%%%%%%%%%%
%%%%%%%%%%%%%%%%%%%%%%%%%%%%%%%%%%%%%%%%%%
%%%%%%%%% SUB-Section %%%%%%%%%%%%%%%%%%%%%%%%%%%%%%%%%%%%%%%%%%
%%%%%%%%%%%%%%%%%%%%%%%%%%%%%%%%%%%%%%%%%%
%%%%%%%%%%%%%%%%%%%%%%%%%%%%%%%%%%%%%%%%%%
\subsection{\texorpdfstring{Proofs of \Sref{sec:applications}}{Proofs of Section 6}}
\label{subapp:proofs_applications}

%%%%%%%%%%%%%%%%%%%%%%%%%%%%%%%%%%%%%%%%%%
%%%%%%%%% SUB-Sub-Section %%%%%%%%%%%%%%%%%%%%%%%%%%%%%%%%%%%%%%%%%%
\subsubsection{\texorpdfstring{Proof of \Sref{subsec:on_money-burning_games}}{Proof of Section 6.3}}
\label{subsubapp:proofs_on_money-burning_games}

To provide the proof of \Mref{prop:money-burning}, we introduce the following notation: for every $n \in \Burnt$, we let $A^{\ell}_i [n] := \Und^{\ell}_i (S) \cap S_i (n)$, with $i \in \Sets {\Ann, \Bob}$, and $A^\ell [n] := A^{\ell}_\Ann [n] \times A^{\ell}_\Bob [n]$, for $\ell \in \bN$, with $\Burnt^\ell := \Set {n \in \Burnt | A^{\ell}_\Ann [n] \neq \emptyset }$. Given the notation set forth above, the following remark is an immediate consequence of the definition of the $\Und$ operator in this context.

\begin{remark}
\label{rem:money-burning}
Given an $\ell \in \bN$, $s_\Bob \notin \Und^{\ell}_\Bob (S)$ if and only if there exists an $n \in \Burnt^\ell$ such that for every $Q^{\ell -1}_\Ann [n] \subseteq A^{\ell-1}_\Ann [n]$, $s_\Bob (n) \notin \Ad_\Bob (Q^{\ell-1}_\Ann [n] \times A^{\ell-1}_\Bob [n])$. 
\end{remark}

\begin{proof}[Proof of \Mref{prop:money-burning}]
We let 
\begin{equation*}
\delta := \min \Set {v_\Ann (a^{*}_\Ann , a^{*}_\Bob) - v_\Ann (a_\Ann , a_\Bob) | %
(a_\Ann , a_\Bob) \in A_\Ann \setminus \{a^{*}_\Ann\} \times A_\Bob } ,
\end{equation*}
with $\delta \in \Re_{>0}$ and we let $\varep \in (0 , \delta)$.  

\begin{claim}
\label{claim:mb1}
For every $\ell \in \bN$, $\Und^{\ell}_\Bob (S) = \prod_{n \in \Burnt} A^{\ell}_\Bob [n]$.
\end{claim}

\begin{proofclaim}
Immediate in light of \Mref{rem:money-burning}.
\end{proofclaim}

\begin{claim}
\label{claim:mb2}
Given an $\ell \in \bN$, if $(n, a^{*}_\Ann) \in \Und^{\ell}_\Ann (S)$, then there exists a $s_\Bob \in \Und^{\ell}_\Bob (S)$ such that $s_b (n) := a^{*}_\Bob$ with $a^{*}_\Bob \in A^{\ell}_\Bob [n]$.
\end{claim}

\begin{proofclaim}
We fix an $\ell \in \bN$, we assume that $(n, a^{*}_\Ann) \in \Und^{\ell}_\Ann (S)$, and---proceeding by contradiction---we assume that $a^{*}_\Bob \notin A^{\ell}_\Bob [n]$ for every $s_\Bob \in \Und^{\ell}_\Bob (S)$ such that $s_b (n) := a^{*}_\Bob$. Now, letting $\overline{s}_\Bob$ be any such a strategy, from \Mref{claim:mb1}, it follows that there exists a $k \in \bN$ with $k < \ell$ such that $\overline{s}_\Bob \notin \Und^{k}_\Bob (S)$, with $a^{*}_\Bob \notin A^{k}_\Bob [n]$.  In particular, this implies---by definition---that for every $Q^k_\Ann [n] \subseteq A^{k}_\Ann [n]$ there exists an $a_\Bob \in A^{k}_\Bob [n]$ such that $a_\Bob$ weakly dominates $a^{*}_\Bob$ relative to $Q^k_\Ann [n]  \times A^{k}_\Bob [n]$, where the statement is \emph{a fortiori} true relative to every singleton ${\overline{Q}}{\vphantom{Q}}^{k}_\Ann [n]$. However, this leads to a contradiction in light of \Mref{rem:singleton}, the fact that $\{a^{*}_\Ann\} \subseteq A^{k}_\Ann [n]$, and the assumption that $v_\Bob (a^{*}_\Ann , a^{*}_\Bob) > v_\Bob (a^{*}_\Ann , a_\Bob)$, for every $a_\Bob \in A_\Bob \setminus \{a^{*}_\Bob\}$.
\end{proofclaim}

\begin{claim}
\label{claim:mb3}
Given an $\ell \in \bN$, if $(n, a^{*}_\Ann) \notin \Und^{\ell+1}_\Ann (S)$, then there exists a $(\overline{n}, a_\Ann) \in \Und^{\ell}_\Ann (S)$ such that
\begin{equation*}
v_\Ann \big((\overline{n}, a_\Ann) , s_\Bob (\overline{n}) \big)  \geq %
v_\Ann \big( (n , a^{*}_\Ann ) , a^{*}_\Bob \big) ,
\end{equation*}
for every $s_\Bob \in \Und^{\ell}_\Bob (S)$.
\end{claim}

\begin{proofclaim}
We fix an $\ell \in \bN$ and we assume that $(n, a^{*}_\Ann) \notin \Und^{\ell+1}_\Ann (S)$. Thus,  from \Mref{lem:fundamental_CBD_WD}, there exists a strategy $(\overline{n}, a_\Ann) \in \Und^{\ell}_\Ann (S)$ that weakly dominates $(n , a^{*}_\Ann)$ relative to $\Und^{\ell} (S)$. Now, from \Mref{claim:mb2}, there exists a $\widetilde{s}_\Bob \in \Und^{\ell}_\Bob (S)$ such that $\widetilde{s}_b (n) := a^{*}_\Bob$. Also, from \Mref{claim:mb1}, for every $s_\Bob \in \Und^{\ell}_\Bob (S)$ there exists a strategy $\widehat{s}_\Bob \in \Und^{\ell}_\Bob (S)$ such that
\begin{equation*}
\widehat{s}_\Bob (m) :=
\begin{cases}
\widetilde{s}_\Bob (m), & \text{ if } m = n, \\%
s_\Bob (m), & \text{ otherwise, }
\end{cases} 
\end{equation*}
where it has to be observed that $\widehat{s}_\Bob (n) = \widetilde{s}_\Bob (n) = a^{*}_\Bob$, which implies that
\begin{align*}
v_\Ann \big( (\overline{n}, a_\Ann) , s_\Bob \big) = v_\Ann \big( (\overline{n}, a_\Ann) , \widehat{s}_\Bob \big) & \geq v_\Ann \big( (n , a^{*}_\Ann) , \widehat{s}_\Bob \big) \\
& = v_\Ann \big( (n , a^{*}_\Ann) , \widehat{s}_\Bob (n) \big) \\
& = v_\Ann \big( (n , a^{*}_\Ann) , a^{*}_\Bob \big) ,
\end{align*}
where the inequality comes from the fact that $(\overline{n}, a_\Ann)$ weakly dominates $(n , a^{*}_\Ann)$ relative to $\Und^{\ell} (S)$.
\end{proofclaim}

\begin{claim}
\label{claim:mb4}
Given an $\ell \in \bN$, if $(n+1 , a^{*}_\Ann) \notin \Und^{\ell}_\Ann (S)$, then  $(n, a_\Ann) \notin \Und^{\ell}_\Ann (S)$, for every $a_\Ann \in A_\Ann \setminus \{a^{*}_\Ann\}$. 
\end{claim}

\begin{proofclaim}
We fix an $\ell \in \bN$ and we assume that $(n+1 , a^{*}_\Ann) \notin \Und^{\ell}_\Ann (S)$. From \Mref{claim:mb3}, it follows that there exists a strategy $(\overline{n} , a_\Ann) \in \Und^{\ell-1}_\Ann (S)$ such that
\begin{equation*}
v_\Ann \big((\overline{n}, a_\Ann) , s_\Bob (\overline{n}) \big)  \geq %
v_\Ann \big( (n+1 , a^{*}_\Ann ) , a^{*}_\Bob \big) ,
\end{equation*}
for every $s_\Bob \in \Und^{\ell-1}_\Bob (S)$. Now, we have that
\begin{equation*}
v_\Ann \big( (n+1, a^{*}_\Ann), a^{*}_\Bob \big) - %
v_\Ann \big( (n, a_\Ann), a_\Bob \big) \geq \delta - \varep > 0 ,
\end{equation*}
for every $a_\Ann \in A_\Ann \setminus \{a^{*}_\Ann\}$ and $a_\Bob \in A_\Bob$ in light of how we defined $\delta \in \Re_{>0}$ with $\varep \in (0, \delta)$, which implies that 
\begin{equation*}
v_\Ann \big( (n+1, a^{*}_\Ann), a^{*}_\Bob \big)  \geq%
v_\Ann \big( (n, a_\Ann), a_\Bob \big) + \delta .
\end{equation*}
Thus, it follows that  $(\overline{n} , a_\Ann)$ strictly dominates $(n, a_\Ann)$, for every $a_\Ann \in A_\Ann \setminus \{a^{*}_\Ann\}$, establishing the result in light of \Mref{rem:relation_dominance}.
\end{proofclaim}

\noindent We assume that $\Und^{N} (S) = \Und^{\infty} (S)$ and we let $\overline{n} := \max \Set { m | m \in \Burnt^N }$. Thus, we proceed by contradiction by assuming that $\overline{n} \neq 0$ and we let $\ell < N$ be such that  $(\overline{n} + 1, a^{*}_\Ann) \notin \Und^{\ell}_\Ann (S)$. Now, from \Mref{claim:mb4}, for every $(\overline{n} , a_\Ann)$ with $a_\Ann \in A_\Ann \setminus \{a^{*}_\Ann\}$, we have that $(\overline{n} , a_\Ann) \notin \Und^{\ell}_\Ann (S)$. Hence, it follows that $A^{\ell}_\Ann [\overline{n}] := \{ a^{*}_\Ann \}$, which implies that $A^{\ell}_\Bob [\overline{n}] := \{ a^{*}_\Bob \}$. Now, since $\overline{n} \neq 0$, every $(\overline{n} -1 , a_\Ann)$ with $a_\Ann \in A_\Ann \setminus \{a^{*}_\Ann\}$ and every $s_\Bob \in \Und^{\ell}_\Bob (S)$ such that $s_\Bob\big(  (\overline{n}-1) \big) \neq a^{*}_\Bob$ have been deleted. Thus, it follows that $(\overline{n} - 1, a^{*}_\Ann)$ strictly dominates $(\overline{n} , a^{*}_\Ann)$. However, this contradicts the original assumption that $\overline{n} := \max \Set { m | m \in \Burnt^N }$, hence, establishing the result.
\end{proof}

%%%%%%%%%%%%%%%%%%%%%%%%%%%%%%%%%%%%%%%%%
%%%%%%%%%%%%%%%%%%%%%%%%%%%%%%%%%%%%%%%%%
%%%%%%%%%%%%%%%%%%%%%%%%%%%%%%%%%%%%%%%%%
%%%%%%%%%%%%%%%%%%%%%%%%%%%%%%%%%%%%%%%%%
%%%%%%%%%%%%%%%%%%%%%%%%%%%%%%%%%%%%%%%%%
%%%%%%%%%%%%%%%%%%%%%%%%%%%%%%%%%%%%%%%%%
%%%%%%%%%%%%%%%%%%%%%%%%%%%%%%%%%%%%%%%%%
%%%%%%%%%%%%%%%%%%%%%%%%%%%%%%%%%%%%%%%%%
%%%%%%%%%%%%%%%%%%%%%%%%%%%%%%%%%%%%%%%%%
%%% BIBLIOGRAPHY
%%%%%%%%%%%%%%%%%%%%%%%%%%%%%%%%%%%%%%%%%
%%%%%%%%%%%%%%%%%%%%%%%%%%%%%%%%%%%%%%%%%
%%%%%%%%%%%%%%%%%%%%%%%%%%%%%%%%%%%%%%%%%
%%%%%%%%%%%%%%%%%%%%%%%%%%%%%%%%%%%%%%%%%
%%%%%%%%%%%%%%%%%%%%%%%%%%%%%%%%%%%%%%%%%
%%%%%%%%%%%%%%%%%%%%%%%%%%%%%%%%%%%%%%%%%
%%%%%%%%%%%%%%%%%%%%%%%%%%%%%%%%%%%%%%%%%
%%%%%%%%%%%%%%%%%%%%%%%%%%%%%%%%%%%%%%%%%
%%%%%%%%%%%%%%%%%%%%%%%%%%%%%%%%%%%%%%%%%
% To get the backref link of the same color of the citecolor
\hypersetup{colorlinks=true,linkcolor=green!50!black}

\end{document}